\documentclass[11pt]{elsarticle}
\usepackage[utf8]{inputenc}
\usepackage{times}
\usepackage{url}
\usepackage{color}
\usepackage{hyperref}
\usepackage{multirow}
\usepackage{tabularx}
\usepackage{mathtools}
\usepackage[table]{xcolor}

\newcommand{\rl}{RL}

\newcommand{\runin}[1]{\vspace{.2em}\noindent\textbf{#1.}}

\usepackage{amssymb}

\journal{Information Systems}

\begin{document}

\begin{frontmatter}



\title{Automatically Reconciling the Trade-off between Prediction Accuracy and Earliness in Prescriptive Business Process Monitoring}



\author{Andreas Metzger*, Tristan Kley, Aristide Rothweiler, Klaus Pohl}

\address{paluno (The Ruhr Institute for Software Technology)\\
University of Duisburg-Essen; 
Essen, Germany\\
andreas.metzger@paluno.uni-due.de, tristan.kley@paluno.uni-due.de, aristide.r@hotmail.de; klaus.pohl@paluno.uni-due.de\\
* corresponding author}

\begin{abstract}
Prescriptive business process monitoring provides decision support to process managers on when and how to adapt an ongoing business process to prevent or mitigate an undesired process outcome.
We focus on the problem of automatically reconciling the trade-off between prediction accuracy and prediction earliness in determining when to adapt.
Adaptations should happen sufficiently early to provide enough lead time for the adaptation to become effective. 
However, earlier predictions are typically less accurate than later predictions.
This means that acting on less accurate predictions may lead to unnecessary adaptations or missed adaptations. 

Different approaches were presented in the literature to reconcile the trade-off between prediction accuracy and earliness. 
So far, these approaches were compared with different baselines, and evaluated using different data sets or even confidential data sets.
This limits the comparability and replicability of the approaches and makes it difficult to choose a concrete approach in practice. 

We perform a comparative evaluation of the main alternative approaches for reconciling the trade-off between prediction accuracy and earliness.
Using four public real-world event log data sets and two types of prediction models, we assess and compare the cost savings of these approaches.
The experimental results indicate which criteria affect the effectiveness of an approach and help us state initial recommendations for the selection of a concrete approach in practice.



\end{abstract}




\end{frontmatter}



\section{Introduction}
\label{sec:Introduction}

\emph{Prescriptive} business process monitoring is an important next step from predictive business process monitoring~\cite{KubrakMND22}.
Prescriptive business process monitoring provides decision support to process managers on when and how to intervene during an ongoing business process to prevent or mitigate the occurrence of an undesired process outcome.
In other words, while predictive process monitoring attempts to answer ``\emph{what will happen and when?}'', prescriptive process monitoring attempts to answer ``\emph{when to intervene and how?}''.

\subsection{Background on Predictive Business Process Monitoring }

Predictive business process monitoring forecasts the future state of an ongoing business process \emph{instance} (a.k.a. \emph{case}) by using data produced by the execution of the case so far together with historical data~\cite{FrancescomarinoG22}. 
A broad range of techniques exists that address different prediction tasks, such as predicting the next activity, predicting the remaining execution time of the case, as well as predicting the outcome of the case~\cite{NeuLF22,TeinemaaDRM19,VerenichMM19,Marquez-Chamorro_2017}.
For example, predictive process monitoring may predict whether there will be a delay in a transport process, or whether an order-to-cash process will be completed successfully.

If the predicted future state of an ongoing case indicates a deviation from the expected future state, process managers may intervene by proactively adapting the case; e.g., by re-scheduling process activities or by changing the assignment of resources~\cite{NunesSWR18,WeberSR09,CAISE2017}.
This can help prevent a deviation or at least mitigate the impact of a deviation~\cite{ParkS2019,PollPRRR18,TeinemaaTLDM18}.  
As an example, a delay in the expected delivery time for a freight transport process may incur contractual penalties~\cite{Gutierrez_2013}.
If during process execution, a delay is predicted, the execution of faster, alternative transport activities (such as air delivery instead of road delivery) can be proactively scheduled to prevent the delay and thus avoid the contractual penalty.

State-of-the-art predictive process monitoring techniques can continuously generate predictions during the execution of a business process -- typically whenever new data from the ongoing case arrives.
This means there are multiple points in time, when process managers may decide whether or not to trust the current prediction and act upon it.
One key limitation of predictive business process monitoring is that it provides limited support for process managers to decide which prediction to trust and act upon.
This decision is important.
While sophisticated prediction models (such as deep learning and ensemble models) and higher data volumes lead to impressive improvements in prediction accuracy~\cite{NeuLF22,TeinemaaDRM19,VerenichMM19}, predictions will never be 100\% accurate.
This means that some predictions will be wrong.
Deciding to adapt an ongoing case based on wrong predictions has negative consequences.
On the one hand, false positive predictions may lead to unnecessary process adaptations.
On the other hand, false negative predictions may mean that necessary process adaptations are missed.

\subsection{Problem Statement}
\emph{Prescriptive} business process monitoring provides decision support for business process managers, helping them to decide when and how to adapt an ongoing case~\cite{BozorgiEA2023,FahrenkrogTTD19,LeoniDR20,WeinzierlZSMP20,MehdiyevF20}.

We address the specific problem of \emph{when} to intervene, i.e., whether and when to adapt an ongoing case.
Answering this question provides the backbone of a prescriptive process monitoring system~\cite{ShoushD2022}.
More precisely, we focus on how a \emph{prescriptive} process monitoring technique can generate \emph{alarms}~\cite{TeinemaaTLDM18,FahrenkrogTTD19,ShoushD2022}. 
An alarm suggests to a process manager to take action or it may directly feed into an automated decision-making process that adapts the running process instance.
By addressing the question ``\emph{when to intervene?}'', we thereby complement related work on prescriptive process monitoring that answers ``\emph{how to intervene?}" (such as presented in~\cite{BozorgiEA2023,DonadelloEA2022,LeoniDR20,MehdiyevF20}).

Generating alarms entails reconciling a fundamental trade-off between prediction accuracy and prediction earliness~\cite{CAISE2019,TeinemaaTLDM18,TeinemaaDRM19}.
On the one hand, alarms should ideally be based on \emph{accurate} predictions.
As motivated above, if decisions are taken based on inaccurate predictions, this may imply unnecessary adaptations or missed adaptations.
On the other hand,  alarms should ideally be raised early.
The later an alarm is raised, the less time and options remain for proactively addressing process deviations~\cite{LeitnerFHD13,TeinemaaDRM19,MorenoCGS18}.
However, earlier predictions typically have a lower prediction accuracy than later predictions, because less information about the ongoing case is available than for later predictions~\cite{TeinemaaDRM19,MetzgerF021}.

Different approaches were presented in the BPM literature to reconcile this trade-off between prediction accuracy and earliness.
They include:
\begin{itemize}
\item Using a \emph{static prediction point }chosen by using the average accuracy of the underlying prediction model~\cite{CAISE2017,Metzger_et_al_2015};
\item Considering the first prediction with a sufficiently high reliability~\cite{Francescomarino_Dumas16,CAISE2019} -- typically expressed in terms of a reliability threshold chosen via \emph{empirical thresholding}~\cite{TeinemaaTLDM18,FahrenkrogTTD19}; 
\item Dynamically deciding which prediction point to consider using \emph{online reinforcement learning} (\emph{Online RL})~\cite{BPM2020,CAISE2020}.
\end{itemize}

So far, these approaches were evaluated using different evaluation setups.
The approaches were compared with different baselines; e.g., while the approach in~\cite{BPM2020} is compared with the approach in~\cite{TeinemaaTLDM18}, the approach in~\cite{CAISE2017} is only compared with the na{\"i}ve baselines of never or always adapting.
Different data sets were used for evaluating the approaches; e.g.,~\cite{Francescomarino_Dumas16} used the BPIC11\footnote{BPIC stands for Business Process Intelligence Challenge.} and BPIC15 data sets, while~\cite{BPM2020} used the BPIC12 and BPIC17 data sets.
Data sets were pre-processed differently; e.g., in~\cite{FahrenkrogTTD19}, the BPIC17 event log data was subdivided into two sub-processes, while in~\cite{CAISE2019,CAISE2020} the whole event log data set was used.
Different splits into training and testing data were used; e.g., a 67\%-33\%-split in ~\cite{BPM2020,CAISE2019}, 80\%-20\%-split in~\cite{FahrenkrogTTD19}, and a 90\%-10\%-split in~\cite{Francescomarino_Dumas16}.
Also, confidential data was sometimes used, such as in~\cite{FahrenkrogTTD19,TeinemaaTLDM18}, which limits the replicability of results.
These differences in evaluation setups make it difficult to use evaluation results from the literature to perform a fair comparison of state-of-the-art approaches in terms of their cost savings. 
In turn, it is difficult to identify the relative strengths and weaknesses of existing approaches and how to choose a concrete approach in practice. 

\subsection{Paper Contribution}

This paper provides the following main contributions:

\runin{Comparative evaluation}
We perform a comparative evaluation of the main alternative approaches for automatically reconciling the trade-off between prediction accuracy and earliness in prescriptive business process monitoring.
Using four real-world event log data sets and two types of prediction models, we assess and compare the potential cost savings of these approaches.
The experimental results show that the more recent approaches outperform the simpler, older approaches in terms of cost savings.
Our experimental results thereby sustain the individual experimental results of previous work.
Yet, our results also show that no single approach works best in all situations, but that whether an approach outperforms the others depends, among other factors, on the concrete characteristics of the data and cost structure. 

\runin{Relating to research on early time series classification} 
The early classification of time series (ECTS) also faces the problem of how to reconcile the trade-off between accuracy and earliness.
ECTS aims to predict the final label of a temporally-indexed set of, typically real-valued, data points with sufficiently high accuracy by using the lowest number of data points~\cite{GuptaGBD20}. 
While differing in terms of the type of the underlying data (event logs vs.~time series), the approaches proposed by the BPM and ECTS communities exhibit conceptual similarities.
As an example, Mori et al. use probabilistic classifiers to produce a class label for a time series as soon as the probability at a time step exceeds a class-dependent threshold~\cite{Mori_ea_17-mpl}, which thus is similar to considering the first prediction with a sufficiently high reliability as introduced above.
We discuss and make explicit the commonalities and differences between these approaches and provide links between BPM and ECTS; e.g., for what concerns the conceptual ideas behind the approaches or the used cost models.

\runin{Artificial curiosity-driven online reinforcement learning}
In simple terms, Online RL learns to balance accuracy and earliness by receiving rewards that quantify whether the chosen balance was a good one or not.
We relax a fundamental assumption of Online RL, which limits our earlier work~\cite{BPM2020,CAISE2020} but is also a limitation of RL approaches used for ECTS~\cite{BonduABCCGHLM22,MartinezRPR20}.
The assumption was that to determine rewards, one can determine the process outcome if an adaptation were not executed (i.e., to know the true process outcome without intervention).
Yet, knowing such alternative process outcome once the process has been adapted is not feasible in general, as it would require an accurate and reliable what-if business process analysis~\cite{Dumas21}.
This poses an important limitation for the practical application of these earlier Online RL approaches.
We overcome this limitation by leveraging the concept of artificial curiosity~\cite{PathakAED17}.
The principal idea of artificial curiosity is that instead of only using feedback received via the  system's environment, we also use feedback generated internally by the system.

\runin{Initial practical recommendations}
Based on our theoretical insights and experimental results, we formulate a set of initial recommendations for selecting a concrete approach in practice.
These initial recommendations are based on key aspects, such as the amount of process data available, the overall accuracy of the used prediction models, and potential concept drifts that may affect the reliability of individual predictions.
In addition, we provide suggestions on how to practically determine these key aspects. 

\vspace{1em}

Overall, our work contributes to the emerging research area of AI-Augmented Business Process Management -- ABPM~\cite{ABPM}.
We demonstrate how the ‘adaptation’ characteristic in ABPM can be automated and empowered by AI (i.e., reinforcement learning) to facilitate real-time adaptation.
In addition, by connecting with the work on ECTS, we provide input to machine-learning-based early decision-making research~\cite{BonduABCCGHLM22}.

\subsection{Paper Organization}

Section~\ref{sec:fundamentals} provides relevant fundamentals. 
Section~\ref{sec:approaches} elaborates the problem and introduces state-of-the-art approaches for reconciling the trade-off between prediction accuracy and earliness.
Section~\ref{sec:approach} describes the enhancements of our Online RL approach.
Section~\ref{sec:Experiments} reports on the evaluation setup.
Section~\ref{sec:predmodels} characterizes the data used to compare the approaches.
Section~\ref{sec:ExperimentalResults} presents the results of our comparative evaluation.
Section~\ref{sec:disc} provides initial practical recommendations.
Section~\ref{sec:discussion} discusses validity risks and directions for future work.
Section~\ref{sec:sota} analyses related work.

\section{Fundamentals}
\label{sec:fundamentals}

This section introduces fundamental concepts of predictive business process monitoring, explains the relevance of prediction accuracy and earliness, and finally presents a cost model for prescriptive business process monitoring.

\subsection{Predictive Business Process Monitoring}
\label{sec:prbpm}

Predictive business process monitoring forecasts how an ongoing business process instance, aka.~\emph{case}, will unfold.
A case $k$ is characterized by a finite sequence $\sigma_k = \langle e_1, \ldots, e_l \rangle$ of events, with $l$ being the \emph{length} of the case.
An \emph{event} $e_i$ represents the execution of an activity~\cite{BozorgiEA2023,Aalst12} and has at least two attributes: a categorical attribute \emph{event type} describing the type of activity that was executed and a numeric attribute \emph{timestamp} recording when the event occurred.
Note that timestamps are typically not equidistant in contrast to time series~\cite{GuptaGBD20}.
An event may have further attributes, such as the resources or people that carried out the activity.
An \emph{event log }is a set of $\sigma_k$~\cite{BozorgiEA2023}.

Predictive business process monitoring utilizes prediction models trained on event logs~\cite{Francescomarino18,Marquez-Chamorro_2017}.
One may, e.g., predict the next activity~\cite{TaxTZ20}, the remaining time~\cite{VerenichDRMT19}, or the outcome of an ongoing case~\cite{TeinemaaDRM19}.
In this paper we focus on prescriptive business process monitoring approaches for outcome prediction, i.e., we aim to predict the label $\hat{y}$ associated to the complete sequence of events of a case.

\begin{figure}[ht]
	\centering
		\includegraphics[width=.9\textwidth]{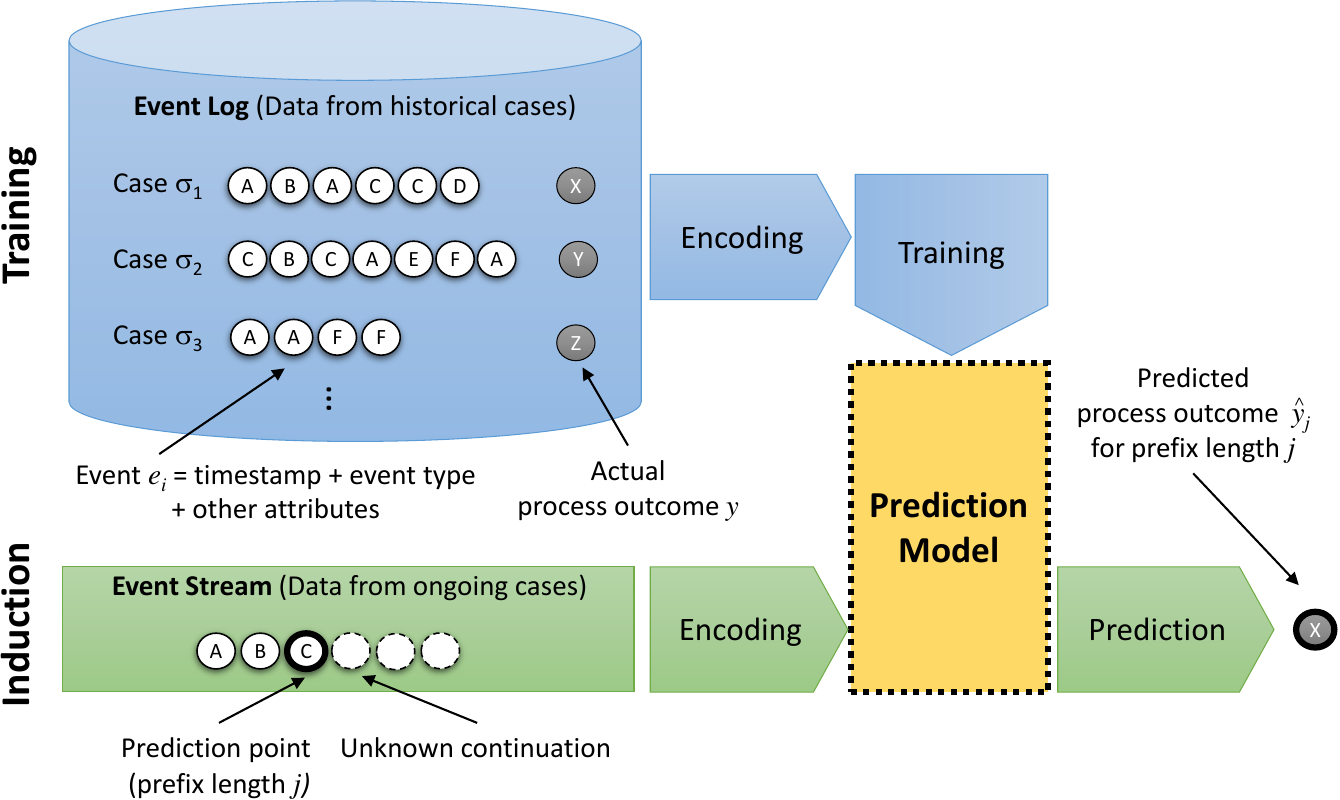}
	\caption{Conceptual view on predictive business process monitoring}
	\label{fig:concept}
\end{figure}

Fig.~\ref{fig:concept} depicts the two main phases and the key steps of predictive business process monitoring.
During the \emph{training} phase, a prediction model is trained using event log data.
During the \emph{induction} phase, the prediction model is used to generate predictions about the ongoing case by using data from the ongoing case as input, typically in the form of a \emph{sequence of events} executed up to the prediction point.
For both training and induction, the process data typically needs to be encoded such as to serve as suitable input for a prediction model (e.g., see~\cite{TeinemaaDRM19}).

\subsection{Prediction Accuracy}
\label{sec:accuracy}

Various types of prediction models were proposed for predictive business process monitoring.
Increasingly, we see the use of sophisticated prediction models, such as random forests (e.g., gradient boosted trees~\cite{TeinemaaDRM19,VerenichMM19}), and deep artificial neural networks (e.g., LSTMs~\cite{LeoniDR20}).
Compared to simple prediction models, such as decision trees or linear regression, these sophisticated prediction models achieve consistently better prediction accuracy in various types of predictive process monitoring problems~\cite{TaxTZ20,VerenichDRMT19}.

Informally, prediction accuracy characterizes the ability of a prediction model to predict as many true deviations from expected process outcomes as possible, while predicting as few false deviations as possible~\cite{SalfnerLM10}.
Obviously, predictions generated by a predictive business process monitoring system should be accurate in order to be useful~\cite{TeinemaaDRM19}.
When used for proactive process adaptation, this is especially important, because adaptation decisions are based on these predictions~\cite{PollPRRR18,TeinemaaTLDM18,AschoffZ11,ASAS_Book}.

To elaborate the need for accurate predictions, Table~\ref{tab:Situations} depicts the four cases that can result from using predictions to decide on proactive process adaptation.
The two cases in boldface are the critical ones.

\begin{table}[htb]
	\centering
	\small
		\begin{tabular}{c|c|c}
			 & Prediction $\hat{y}_j$ =  & Prediction $\hat{y}_j$ = \\ 
			 & \emph{deviation  }& \emph{no deviation}\\ 
	\hline
	
			\multirow{2}{*}{Actual $y$ = \emph{deviation}} & True Positive (\emph{TP})  & False Negative  (\emph{FN})\\ 
      & $\Rightarrow$ Necessary adaptation & $\Rightarrow$  \textbf{Missed} adaptation  \\ 
	\hline
			\multirow{2}{*}{Actual $y$ = \emph{no deviation}} & False Positive (\emph{FP}) & True Negative (\emph{TN})\\ 
			 & $\Rightarrow$ \textbf{Unnecessary} adaptation & $\Rightarrow$ No adaptation \\ 
		\end{tabular}
	\caption{Prediction contingencies and adaptation decisions based on predictions}
	\label{tab:Situations}
\end{table}

\runin{Unnecessary adaptation} If the prediction model falsely predicts that there is a deviation, such a false positive prediction implies an unnecessary adaptation.
An adaptation typically entails costs, e.g., due to executing of additional or different process activities or due to adding more resources to the case.
Therefore, unnecessary adaptations incur additional costs, while not addressing actual deviations.

\runin{Missed adaptation} If the prediction model falsely predicts that there is no deviation, an opportunity for adaptation is missed.
As a result, the case will face a deviation during its execution, which may imply, among others, penalties in case of contractual violations. 
Each  required adaptation that is missed means one less opportunity for preventing or mitigating a deviation.

\subsection{Prediction Earliness}
\label{sec:earliness}

Predictions can be made at different prediction points during the execution of a case.
Predictions are early if they are made toward the beginning of a case.
Ideally, one would like to use early predictions, as this gives more time and options for proactively addressing predicted process deviations~\cite{LeitnerFHD13,TeinemaaDRM19,MorenoCGS18}.

There are different ways to define prediction points.
As an example, prediction points can be explicitly defined by determining relevant activities or milestones from a process model.
Such a process model may already exist or may be generated using process mining~\cite{Aalst2011process}. 
Such explicitly defined prediction points are often called \emph{checkpoints}~\cite{LeitnerFHD13,Metzger_et_al_2015} or \emph{decision points}~\cite{TeinemaaDRM19}.
Prediction points may also be defined using equidistant points in time, i.e., using prediction windows of a given duration~\cite{FolinoGP15}.

More typically, predictions are generated after each event $e_j$ of an ongoing case $\sigma_k$.
Such prediction points can be characterized by their \emph{prefix length}, which gives the number of events that were produced up to the point of the prediction~\cite{TeinemaaDRM19,LeontjevaCFDM15}. 
In this paper we characterize prediction points by their prefix length by giving the index $j$ of event $e_j \in \sigma_k$ for which a prediction is made. 

\subsection{Cost Model for Prescriptive Business Process Monitoring}
\label{sec:cost-model}
A proactive adaptation decision entails different costs, which depend on whether an alarm is based on a false positive or a false negative prediction (see Section~\ref{sec:accuracy}), and when the prediction was made during process execution (see Section~\ref{sec:earliness}).
This leads to different costs $C(j)$ depending at which prefix length $j$ an alarm is raised\footnote{If no alarm is raised, we model this as $j = 0$.}.

We use a cost model based on previous work in BPM~\cite{FahrenkrogTTD19,BPM2020,CAISE2020,CAISE2019} and ECTS~\cite{BonduABCCGHLM22}.
This cost model is shown in Table~\ref{tab:ContingencyTableCosts} and follows the structure of Table~\ref{tab:Situations}.
The cost model considers the following parameters:

\runin{Penalty costs $C_\mathrm{p}$} This parameter (a.k.a. cost of undesired outcome~\cite{FahrenkrogTTD19}) models the costs for violating an expected process outcome. 
As an example, a penalty may have to be paid for late deliveries in a transport process.
Penalties may be faced in two situations.
First, a necessary proactive adaptation may be missed.
Second, a proactive adaptation may not have been effective (see below) and thus the deviation remains after adaptation.

\runin{Adaptation costs $C_\mathrm{a}$} This parameter (a.k.a. cost of intervention~\cite{FahrenkrogTTD19}) models the costs of performing the actual adaptation, because adapting an ongoing case typically requires effort and resources. 
As an example, adding more staff to speed up the execution of a case incurs additional personnel costs.

\runin{Adaptation effectiveness $\alpha$} This parameter (a.k.a. mitigation effectiveness~\cite{FahrenkrogTTD19}) models the probability that an adaptation is effective.
To model the fact that earlier prediction points provide more options and time for proactive adaptations than later prediction points, earlier proactive adaptations are given a higher $\alpha$ than later ones.

\runin{Compensation costs $C_\mathrm{c}$} This parameter (a.k.a. cost of compensation~\cite{FahrenkrogTTD19}) models the costs resulting from unnecessary adaptations.
Unnecessary adaptation may require roll-back or compensation activities that result in additional process execution costs.
	As an example, if a credit was falsely issued to a client, this may entail additional costs for recovering the money or compensating the client for this error.

\begin{table}[htb]
	\centering
	\footnotesize
			\resizebox{1\textwidth}{!}{%

		\begin{tabular}{r r|ccc|c}
		
			 & Costs $C(j) = $& \multicolumn{3}{c|}{Prediction $\hat{y}_j$ = } & Prediction $\hat{y}_j$ = \\ 
			 			 &  & \multicolumn{3}{c|}{\emph{deviation }} & \emph{no deviation  }\\ 

	   &     & \emph{with probability $\alpha$:} & \vline &\emph{with probability $1-\alpha$:}& \\
	   &     & \emph{Effective Adaptation} & \vline &\emph{Non-effective Adaptation}& \\
	  \hline 
			 &  Actual $y$ = \emph{deviation}  & $C_\mathrm{a}$ &\vline  & $C_\mathrm{a}$ & $C_\mathrm{p}$ \\ 
& & & \vline & + $C_\mathrm{p}$ & \\
\hline
			& Actual $y$ = \emph{ no deviation} & $C_\mathrm{a}$ & \vline &$C_\mathrm{a}$ & 0\\ 
			&   & + $C_\mathrm{c}$ & \vline &  &  
		\end{tabular}}
	\caption{Cost Model}
	\label{tab:ContingencyTableCosts}
\end{table}

\section{Trading-off Prediction Accuracy and Earliness}
\label{sec:approaches}

Below, we first elaborate our problem formulation, followed by the description of different approaches for reconciling the trade-off between prediction accuracy and earliness.

\subsection{Problem Formulation}
\label{sec:probl}
In general, the later an alarm is raised the less time remains for proactively addressing a deviation via proactive process adaptation~\cite{LeitnerFHD13,TeinemaaDRM19,TeinemaaTLDM18,CAISE2019}.
This can be important as adaptations typically have non-negligible latencies, i.e., it may take some time until they become effective~\cite{MorenoCGS18}.
As an example, dispatching additional personnel to mitigate delays in container transports may take several hours. 
Also, the later a case is adapted, the fewer adaptation options are available.
As an example, while at the beginning of a transport process one may be able to transport a container by train instead by ship, once the container is on board the ship such an adaption is no longer possible.
Finally, if an adaptation is performed late and turns out to be ineffective (i.e., not preventing the predicted deviation), not much time may remain for any remedial actions or further adaptations.
This means one should favor earlier prediction points for raising alarms, thereby leaving sufficient time and options for process adaptation.

However, there is a conflict between generating accurate alarms and generating early alarms~\cite{LeitnerFHD13,VerenichMM19,CAISE2019}.
Typically, prediction accuracy increases as the case unfolds, because more information about the ongoing case becomes available.
While early predictions tend to exhibit low prediction accuracy, later predictions typically exhibit higher prediction accuracy.
This  means later predictions have a higher chance of being accurate, thus one should favor later prediction points for raising alarms.
However, later alarms leave less time and options for process adaptation.

Considering this trade-off means determining a prediction point $j^*$, which balances accuracy and earliness in such a way as to minimize costs $C(j^*)$.
On this conceptual level, the problem matches the problem of ECTS\footnote{With the exception that ECTS forces a decision, i.e., classification, while in prescriptive process monitoring we may not raise an alarm at all; modeled as $j=0$ as mentioned above.} and can be expressed in the below equation~\cite{achenchabe2021early}:

\begin{equation}
j^* = \textrm{argmin}_{j \in \{0, \ldots, l\}} C(j)
\end{equation}

Obviously, we do not know the resulting costs for future timesteps and thus solving this optimization problem is not straightforward.

\subsection{Approaches for Trading-off Prediction Accuracy and Earliness}
Below, we introduce three competitive state-of-the-art approaches and and one simple one to address the above problem.
The simple one, we introduce first, may be the straightforward choice of a practitioner, and we include it to assess its limitations.

\runin{First Positive Prediction} 
One simple approach is to act on the first positive prediction as the case unfolds.
This means an alarm is raised for the first prefix length, for which the prediction model gives a positive prediction (i.e., forecasts a deviation).
Acting on the first positive prediction provides the earliest point for intervention.

While this approach is simple to implement and thus may appear attractive from a practical point of view, it ignores the fact that early predictions may not be as accurate as later predictions, and thus may lead to many false alarms.

\runin{Static Prediction Point / Minimal Prefix Length} 
One principle approach is to use the predictions of a well-chosen, static prediction point at prefix length $j^*_\mathrm{fix}$.
An alarm is generated if an ongoing case reaches prefix length $j^*_\mathrm{fix}$ and at the same time if the prediction generated at this prefix length forecasts a deviation.
There are different ways to choose such a static prediction point.
Like in~\cite{CAISE2017,Metzger_et_al_2015}, such a static prediction point may be determined by analyzing the average prediction accuracy measured for each of the different prediction points (using some test data set; e.g., as we do in Section~\ref{sec:predacc}).
Using this average accuracy information, one may choose the earliest point that exhibits sufficiently high prediction accuracy.

Similar approaches were proposed for ECTS and categorized as prefix-based early classification, where the idea is to learn a minimum prefix length using training data instances~\cite{GuptaGBD20,XingPY12} (also see Section~\ref{sec:ects}).

Using a fixed prediction point has two main shortcomings.
First, no alarms will be raised for cases that have a length that is shorter than $j^*_\mathrm{fix}$.
While one may choose a very early prediction point that captures many or most of the cases, this comes with potentially lower prediction accuracy.

Second, the average accuracy of a prediction model does not provide direct information about the accuracy of an \emph{individual} prediction for a concrete case~\cite{CAISE2019}.
In particular, for a given prediction model, the accuracy of individual predictions may differ  across prediction points and cases~\cite{TeinemaaDLM18}.
These differences in prediction accuracy are not taken into account when choosing a fixed, static prediction point.

\runin{(Empirical) Thresholding} 
To address the second shortcoming of using a static prediction point, one emerging approach is to use \emph{reliability estimates} (aka. posterior probabilities of the underlying prediction model~\cite{MoriMDL18-probab}).
A reliability estimate quantifies the likelihood that an individual prediction is correct~\cite{FahrenkrogTTD19}.
A typical example of a reliability estimate is the class probability generated by a random forest.

As a straightforward approach to leverage reliability estimates, one may use the earliest prediction with sufficiently high reliability to raise an alarm~\cite{TeinemaaDRM19,CAISE2019,Francescomarino17,TeinemaaDMF16,TeinemaaTLDM18}.
To determine whether the reliability is sufficiently high, one can set a concrete reliability threshold; e.g., 95\%.
Depending on how this threshold is set, one can trade earliness against accuracy~\cite{CAISE2019,TeinemaaTLDM18}.
If earlier predictions are preferred, a lower threshold may be chosen, which raises alarms more speculatively at the  risk that predictions are not very accurate.
If accurate predictions are required, a higher threshold may be chosen, which raises alarms more conservatively and thus poses the risk that alarms may be raised too late as to be effective or that no alarm may be raised.

Similar approaches were proposed for ECTS and categorized as model-based approaches using discriminative classifiers, i.e., classifiers that provide a probability together with the actual prediction~\cite{GuptaGBD20,Mori_ea_17-mpl,HatamiC13} (also see Section~\ref{sec:ects}).

It can be difficult for a process manager to define a threshold that is optimal for a given situation, because, for instance, the optimal threshold depends on the actual type of business process and costs entailed in process execution.
A recent BPM approach to address the problem of determining a threshold is \emph{Empirical Thresholding}~\cite{TeinemaaTLDM18,FahrenkrogTTD19}.
In Empirical Thresholding, a dedicated training process -- involving a dedicated training data set -- is used to determine a suitable threshold.
In the basic variant of Empirical Thresholding, a cost model, which defines adaptation, compensation and penalty costs, is used together with a dedicated training data set to compute the optimal threshold.

In~\cite{FahrenkrogTTD19}, two variants of this basic Empirical Thresholding approach are suggested.
First, it is suggested to introduce a so-called firing delay $p$.
This means that an alarm is only raised if the reliability was above the threshold for $p$ consecutive prefix lengths.
Second, it is suggested to train different thresholds for different (groups of) prefix lengths.
While the rationale behind these variants is convincing, experimental results in~\cite{FahrenkrogTTD19} indicate that they only provide limited improvements.
Both the firing delay approach and the multiple threshold approach only provided additional cost savings in around 8\% of the experimental situations.
We thus focus on the basic variant of Empirical Thresholding in the remainder of this paper.

One general concern of Empirical Thresholding is that while the threshold is optimal for the training data, it may not remain optimal over time.
Concept drifts of the process environment and data~\cite{MaisenbacherW17,OstovarLR20} may impact on prediction model accuracy and thus affect the reliability of individual predictions.

\runin{Online Reinforcement Learning (Online RL)}
An approach to address concept drifts is to employ \emph{online }reinforcement learning  (RL)~\cite{BPM2020,CAISE2020}.
Online RL means that learning happens at runtime during the actual execution of the business processes.
Based on the process predictions and their reliability estimates, Online RL decides for each prediction individually whether to raise an alarm.
In turn, Online RL learns from whether such an alarm was correctly raised to improve the RL decision-making policy.
Thereby, Online RL avoids the need to determine an optimal threshold.
Also, for Online RL we do not have to calibrate the reliability estimates like it may be done for Empirical Thresholding~\cite{TeinemaaTLDM18,FahrenkrogTTD19} or when human decision-makers need to take a decision based on the reliability estimate.

Similar RL-based approaches were proposed for ECTS~\cite{GuptaGBD20,BonduABCCGHLM22,MartinezRPR20} (also see Section~\ref{sec:ects}).

There is one critical assumption encoded in earlier Online RL approaches, such as presented in our previous work~\cite{BPM2020,CAISE2020} and in ECTS research~\cite{BonduABCCGHLM22,MartinezRPR20}.
The assumption is hat one can assess whether an adaptation was correct by determining the alternative process outcome if that adaptation were not executed.
Yet, knowing such alternative process outcome once the process has been adapted may not be feasible in general, as it would require an accurate and reliable what-if business process analysis~\cite{Dumas21}.
This posed an important limitation for the practical application of our earlier approach.

Here, we eliminate this assumption by introducing the concept of artificial curiosity into  the reinforcement learning process as explained below.

\section{Online RL with Artificial Curiosity}
\label{sec:approach}
\label{sec:overview}

\begin{figure}[htbp]
	\centering
		\includegraphics[width=.8\textwidth]{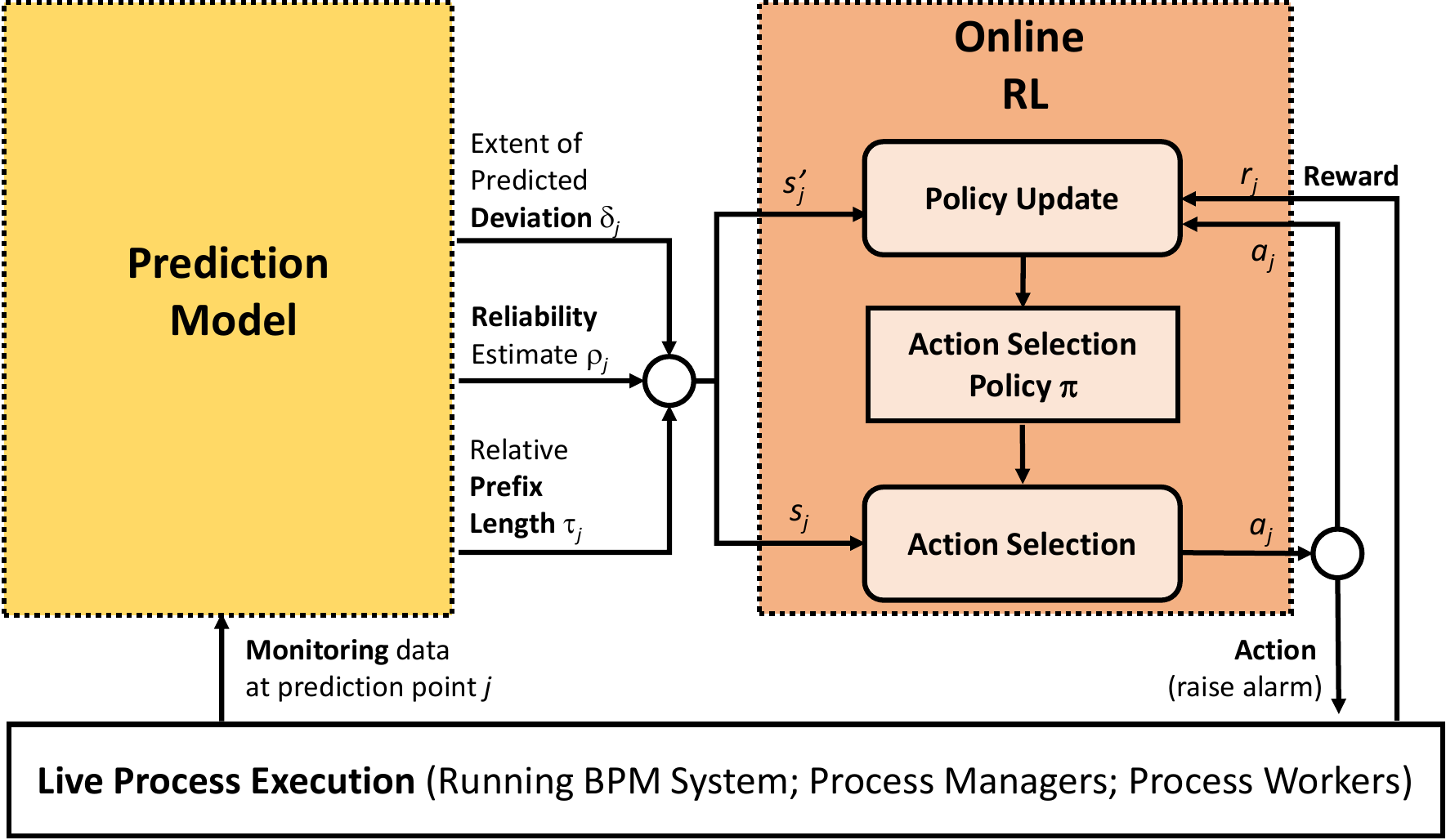}
	\caption{Conceptual overview of Online RL approach}
	\label{fig:approach}
\end{figure}

Figure~\ref{fig:approach} provides an overview of our Online RL approach and how it connects to the prediction models from predictive business process monitoring (as introduced in Section~\ref{sec:fundamentals}).
Below, we explain the basic Online RL approach and its extension with artificial curiosity, as well as the requirements for a prediction model so it can be used as input for the Online RL approach.

\subsection{Online RL}
\label{sec:RL}

In general, RL aims at finding an optimal action selection policy $\pi$ for a given environment by interacting with this environment. 
A policy $\pi$ defines a mapping from environment states to actions.
Upon executing an action $a$ in a state $s$, the environment transitions to the next state $s'$ and awards a specific numeric reward based on a reward function $r(s, a)$.
An optimal policy $\pi$ is a policy that optimizes the cumulative rewards received~\cite{sutton2018reinforcement}.

\runin{Action Selection and Policy} 
To capture concept drifts that affect the accuracy and reliability of the prediction model, we employ \emph{policy-based} deep RL~\cite{NachumNXS17}.
The fundamental idea of policy-based RL is to directly use and optimize a parametrized stochastic action selection policy $\pi_\theta$.
The action selection policy maps states to a probability distribution over the action space (i.e., set of possible actions).
Formally, $\pi_\theta : S \times A \to [0,1]$, giving the probability of taking adaptation action $a$ in state $s$, i.e., $\pi_\theta(s,a) = \mathrm{Pr}(a|s)$.
Policy-based \emph{deep} RL means that the policy function $\pi_\theta$ is represented as a deep artificial neural network.
The policy's parameters $\theta \in {\rm I\!R}^d$ are the weights of the artificial neural network.
Policy-based RL stochastically selects actions by sampling from the probability distribution $\pi_\theta(s,a)$~\cite{SuttonMSM99}.

Policy-based RL offers several benefits.
First, it can cope with multi-dimensional, continuous state spaces, which we face due to the input for the Online RL approach including the continuous variables $\rho_j$ and $\delta_j$~\cite{BPM2020}.
Second, it can generalize over unseen neighboring states.
Third, and most importantly, it can readily cope with non-stationarity via the aforementioned stochastic action selection and thereby can address concept drifts of the underlying prediction model.

Deep neural networks are also used in another class of deep RL, which is\emph{ value-based }deep RL~\cite{sutton2018reinforcement}.
Here, the neural network is used to represent the action-value-function $Q(s, a)$, which gives the expected reward when taking action $a$ in state $s$.
Value-based RL requires explicitly implementing a policy function to determine whether -- in given state -- the agent should exploit the current policy (i.e., choose the best action based on $Q$) or to explore new actions (e.g., by randomly selecting an action).
As we discussed in our earlier work, this imposes the additional engineering challenge on determining  how to balance exploitation and exploration, in particular in the presence of concept drifts~\cite{CAISE2020,BPM2020}.

\runin{Policy Update}
A learning episode (see definition below) consists of $l$ time steps. 
At the end of each episode, the trajectory of $l$ actions, states and rewards are used for
a policy update. 
During a policy update, the weights of the neural network are updated via so-called policy gradient methods.
These methods update the policy to optimize the gradient of a given objective function, such as average rewards over some time horizon~\cite{sutton2018reinforcement}.

\runin{Rewards} The reward function $r(s, a)$ specifies the numeric reward received for executing action $a$ in state $s$. 
The reward function thereby quantifies the learning goal to achieve.
The reward function has to be designed such as to optimize cumulative rewards.
As an example, a simple reward function may provide a positive reward $r = 1$ if an action $a$ has the desired effect (i.e., leading to a desired new environment state), and a negative reward $r = -1$ if it did not.
We elaborate on the concrete reward function for  Online RL in Section~\ref{sec:curious}.

\runin{States and Actions} 
The environment states refer to the  output of the prediction model $\delta_j$, $\rho_j$, and $\theta_j$ (see Section~\ref{sec:ensemble}).
Actions are binary and refer to raising resp. not raising an alarm at a given prediction point $j$, i.e., $a_j \in \{\textrm{alarm}, \textrm{no alarm}\}$. 

\begin{figure}[htbp]
	\centering
	\includegraphics[width=.75\textwidth]{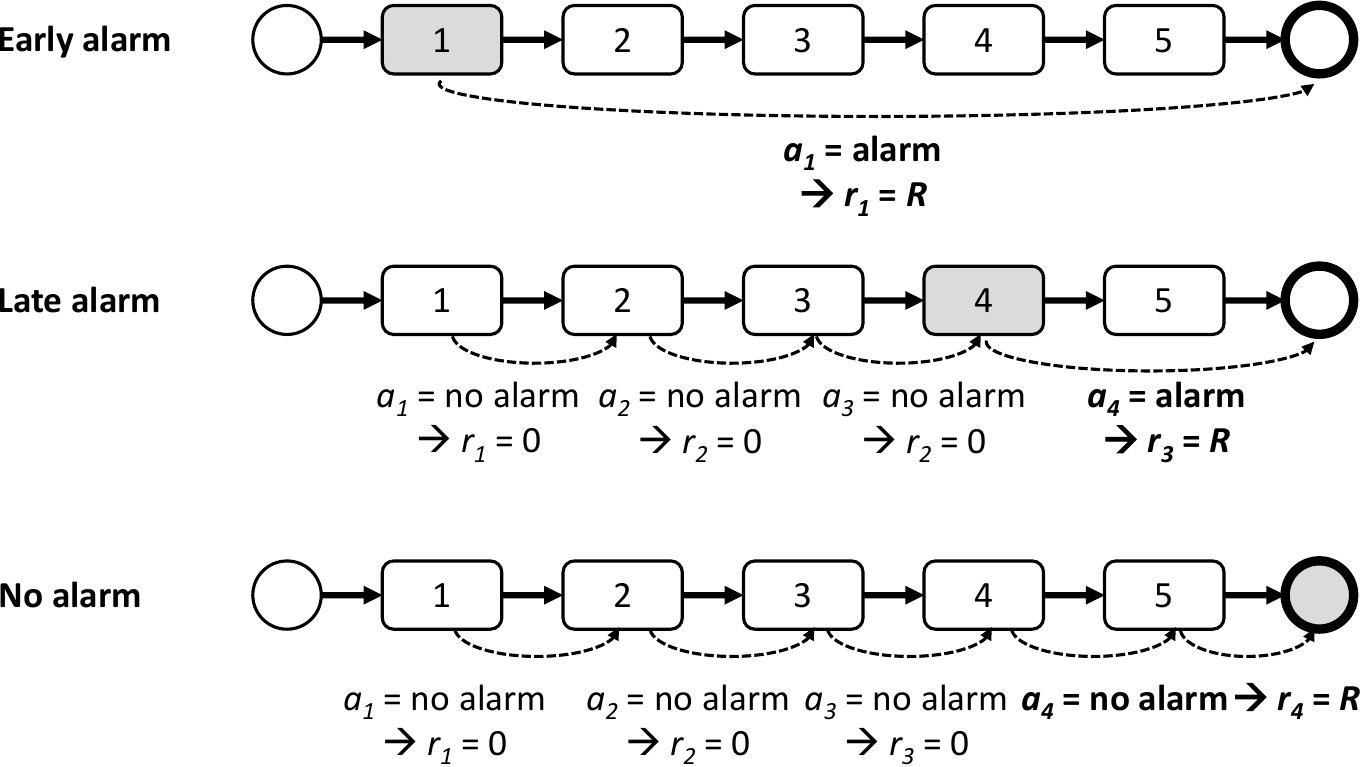}
	\caption{End states and episodes for Online RL with a case length and thus episode of length $l = 5$}
	\label{fig:episode}
\end{figure}

\runin{Learning Episodes}
We break down the Online RL process into suitable learning episodes.
A learning episode matches the execution of a single case.
Whenever Online RL raises an alarm or when the end of the case is reached, the episode ends and we provide a numeric reward $r = R$ as described below.
Otherwise, for any actions that do not lead to the end of such an episode, we provide zero rewards, i.e., $r = 0$.
Fig.~\ref{fig:episode} illustrates the three main types of end states that result from raising alarms at different points along process execution.

\subsection{Artificial Curiosity for RL}
\label{sec:curious}

The successful application of RL depends on how well the learning problem, and in particular the reward function, is defined~\cite{Dewey14}.
By defining a reward function, one expresses the learning goal in a declarative fashion.
As mentioned above, the goal of RL is to maximize cumulative rewards.
Therefore, designing a suitable reward function is key to successful learning.

Note that the definition of the reward function is an inherent and fixed element of our Online RL approach, and there is no need to fine-tune it for specific data sets.

When designing a suitable reward function for reconciling the trade-off between prediction accuracy and earliness, the particular challenge is that we need to assess whether raising an alarm was the right decision.
This requires determining the alternative process outcome if such an alarm were not raised and consequently no process adaptation were performed.
As motivated in Sections~\ref{sec:Introduction} and~\ref{sec:approaches}, knowing such alternative process outcome once the process has been adapted is not feasible in general.
Not knowing the alternative process outcome after an adaptation means that we lack an essential element of the reward function that would punish false alarms; e.g., by awarding negative reward values for false alarms.

We address this lack of an essential element of the reward function by leveraging the concept of artificial curiosity~\cite{PathakAED17}.
The principal idea of artificial curiosity is that instead of only using \emph{extrinsic} rewards, we also use \emph{intrinsic} rewards in the definition of the reward function.
While an extrinsic reward has to be provided externally by the environment (i.e., computed from environment states), an intrinsic reward can be created from information available internally to the RL algorithm.
In the reward function for Online RL, we include intrinsic rewards that positively reward transitioning to previously unexplored states. 
This enables the RL system to explore its environment even in the absence of extrinsic rewards.

As shown in Fig.~\ref{fig:episode}, we differentiate between three kinds of end states that result from taking an adaptation decision and thus should be explored.

\begin{table}[htb]
	\centering
	\small
	\begin{tabular}{c|c|c}
		 & Adaptation      & No adaptation                                                  \\ \hline
		Actual = Deviation      & \multirow{2}{*}{$R = b (1-c) - 2d$} & \begin{tabular}[c]{@{}c@{}}$R = -1$ \end{tabular}  \\ \cline{1-1} \cline{3-3} 
		Actual = No deviation   &                                      & \begin{tabular}[c]{@{}c@{}}$R = +1.5$ \end{tabular}
	\end{tabular}
	\caption{Reward function for Online RL}
	\label{tab:rewards}
\end{table}

\runin{End states resulting from "no adaptation"}
For these two end states, we provide extrinsic rewards.
In particular, we provide strong reward signals by rewarding a correct decision with $+1.5$ and by punishing a wrong decision with $-1$.
One may consider using actual costs (see Section~\ref{sec:cost-model}) as a more fine-grained reward function.
However, as we have shown in our earlier work~\cite{BPM2020} and as indicated in~\cite{SatyalWPCM19}, doing so does not provide a strong enough reward signal and slows down convergence of the learning process. 

\runin{End state resulting from "adaptation"}
For the end state, we provide intrinsic rewards that rely on the three parameters $b, c,$ and $d$ as follows.

The parameter $d\in [0,1]$ is the rate of adaptations among the last seen 30 cases.
As it has the negative coefficient $-2$, $d$ punishes high adaptation rates and thus fosters exploring not raising an alarm. 
We choose to average over 30 cases, as this provides a working compromise between the parameter changing quickly enough, yet not too erratically. 
When the RL policy update leads to a lower adaptation rate, the parameter $d$ changes quickly enough to prevent the policy update from further lowering the adaptation rate. 
On the other hand, the parameter does not change too quickly to prevent the policy update from effectively lowering the adaptation rate. 

The parameters $b \in [\frac{1}{2},1]$ and $c \in [0,3]$ foster exploring different degrees of the earliness of alarms.
The parameter $b$ decreases linearly with the prefix length $j$, being equal to $1$ on the first prediction point and $\frac{1}{2}$ on the last.
Thereby, $b$ facilitates that early alarms should be preferred to late alarms.

The parameter $c$ is defined to be bi-linearly dependent on $d$ and $v$ as follows: 

\small
\begin{equation}
c=\mathrm{max}\left(3, \mathrm{min}(0, (-30 v+21)\cdot (d-\frac{1}{2}))\right)
\label{eq:curiosity}
\end{equation}
\normalsize

The parameter $v$ is the negative predictive value computed for the last $100$ non-adapted cases.
We use the negative predictive value $v$ as an estimate for the accuracy of raising alarms and consider the last 100 cases to get a stable enough accuracy estimate.
The negative predictive value $v$  is computed as follows:

\small
\begin{equation}
v = \frac{TN}{TN+FN}
\end{equation}
\normalsize

The reward function parameter $c$ thereby encodes two assumptions we make about the learning process. 
First, when the current RL policy leads to a high negative predictive value $v$, indicating  high accuracy in raising alarms, there is no longer a need to explore raising alarms later.
Concretely, once $v$ nears or exceeds 70\%, we assume that intrinsic rewards are no longer necessary, which results in $c = 0$. 

Second, when the adaptation rate $d$ is small, the extrinsic rewards will suffice to facilitate learning. 
Thereby, the parameter $c$ is gradually reduced to $0$, effectively leaving only the extrinsic rewards.
In the overall reward function, if $c$ is equal or close to $3$, this reinforces late adaptations, if $c$ is equal to $0$ or at least smaller than 1, this reinforces early adaptations.

\begin{figure}[htbp]
	\centering
	\includegraphics[width=.65\textwidth]{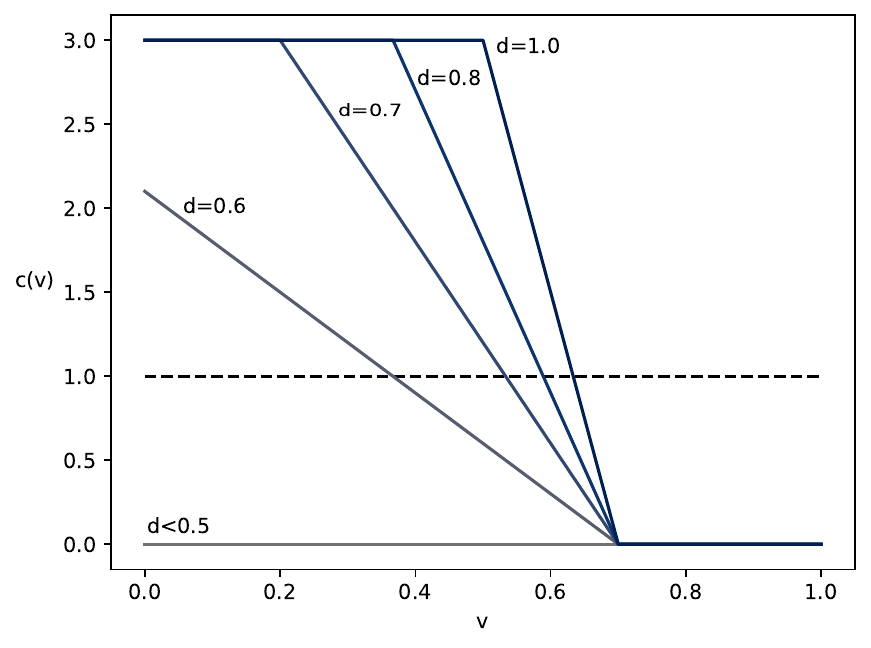}
	\caption{Curiosity modifier $c$ over negative predictive value $v$, values greater than 1 reinforce curiosity about later adaptations}
	\label{fig:c(v)}
\end{figure}

Figure \ref{fig:c(v)} plots the curiosity modifier $c$ over the negative predictive value $v$ with regard to several adaptation rates $d$.
It demonstrates visually the relationship between the three variables as defined by Equation~\ref{eq:curiosity}: only for low value of $v$ and high values of $d$ is the curiosity modifier $c$ higher than the threshold of 1.

\subsection{Prediction Model Requirements}
\label{sec:ensemble}

The Online RL approach requires the following inputs that form the state space for the RL algorithm.

\runin{Predictions for each prefix length $j$}
The prediction model should be able to generate a prediction after each event of the ongoing case. 
As mentioned in Section~\ref{sec:fundamentals}, we characterize these prediction points by their \emph{prefix length} $j$, which gives the number of events that were produced up to the prediction point.

\runin{Relative predicted deviation $\delta_i$}
Online RL requires a numeric input from the prediction model that quantifies the relative predicted deviation.
Typically, this can be generated as follows.
First, one may use regression models to generate real-valued predictions $\hat{y}_{j} \in \mathbb{R}$ (e.g., see~\cite{ICSOC2017}).
Given $A$ as the expected process outcome\footnote{Section~\ref{sec:predmodels} explains how $A$ can be determined.}, the relative predicted deviation $\delta_j$ is computed as:

\small 
\begin{equation}
\delta_j = \frac{\hat{y}_j  - A}{A}
\label{eq:dev}
\end{equation}
\normalsize

\runin{Reliability estimate $\rho_i$}
The Online RL approach requires as input continuous reliability estimates $\rho_i \in [0,1]$ for each prediction point $j$.
A useful reliability estimate should be a good indicator of actual prediction accuracy.

Typically, reliability estimates computed from ensembles of prediction models (a typical example are random forests) can provide good estimates of the probability that an individual prediction is correct~\cite{BosnicK08}.
Ensemble prediction is a meta-prediction technique where the predictions of $m$ base prediction models are combined. 
While the main aim of ensemble prediction is to increase prediction accuracy, ensemble prediction also allows computing reliability estimates~\cite{CAISE2017,CAISE2019}.

Assuming that for a prediction point $j$ we are given the predictions of each base prediction model $i = 1, \ldots, m$ of the ensemble as $\delta_{j,i}$. 
One straightforward yet effective way to determine the reliability estimate $\rho_j$ is to compute the fraction of the predictions of the individual prediction models $i$ as follows~\cite{CAISE2017,CAISE2019}:

\small
\begin{equation}
 \rho_j = max(\frac{|\{i = 1, \ldots, m: \delta_{j,i} > 0\}|}{m},
\frac{|\{i = 1, \ldots, m: \delta_{j,i}  \leq  0\}|}{m})
\label{eq:rel}
\end{equation}
\normalsize

This way of computing reliability estimates facilitates the practical application of Online RL.
First, many ensemble prediction models directly provide such an estimate (e.g, in form of class probabilities of random forests).
Second, compared to other reliability approaches, it does not require additional tuning steps.
For example, "conformal prediction" requires defining a suitable non-conformity measure and calibration~\cite{PapadopoulosVG07,ShoushD2022}, while "corrected variance" requires hyper-parameter tuning~\cite{CarneyCB99}.

\runin{Relative prefix length $\tau_i$}
In addition to the extent of the deviation $\delta_j$, we also compute the relative prefix length $\tau_j$ as input.
Using $\tau_j$ provides an important signal to the Online RL approach about the earliness of the respective ensemble prediction $\delta_j$.
This relative prediction point can be computed by dividing the prediction point $j$ by the case length $l$.

\small
\begin{equation}
\tau_j = \frac{j}{l}
\label{eq:len}
\end{equation}
\normalsize

As we are dealing with an episodic RL problem (see Figure~\ref{fig:episode}), $l$ is known at the end of the episode and can be used to compute the reward signal.

\section{Evaluation Setup and Execution}
\label{sec:Experiments}

This section describes the setup and execution of a series of controlled experiments to comparatively evaluate the approaches from Section~\ref{sec:approaches} and ~\ref{sec:approach}.
Concretely, we seek to answer:
\begin{center}
\emph{"How do the approaches compare in terms of cost savings?"}
\end{center}
This means, like in~\cite{FahrenkrogTTD19}, we analyze and characterize the reduction of the average process execution costs for different situations.
In addition, we quantify and analyze the extent of these cost savings.

All artifacts, including code, data sets, prediction results, as well as experimental outcomes, are publicly available to facilitate replicability\footnote{\url{https://git.uni-due.de/abpm/isj}}.

\subsection{Realization of Prediction Models}
\label{sec:realization1}

We experiment with two alternative ways of realizing the prediction models.
On the one hand, we use ensembles of random forests (concretely regression forests), as a representative of rule-based learning~\cite{Kotsiantis07}.
On the other hand, we use ensembles of deep artificial neural networks, concretely LSTMs, as a representative of perceptron-based learning~\cite{Kotsiantis07}.
Both random forests and LSTMs are widely used for predictive process monitoring~\cite{TeinemaaDRM19,ParkS20}.

These two realizations exhibit different advantages and shortcomings.
On the one hand, LSTMs can directly handle arbitrary-length sequences of events, while random forests require the encoding of such event sequences into fixed-length input vectors~\cite{TeinemaaDRM19,MetzgerN18,Marquez-Chamorro_2017}.
On the other hand, random forests can be trained rather efficiently~\cite{TeinemaaDRM19}, while LSTMs require significant time and resources for training~\cite{CAISE2019}.

\runin{Random Forests}
Random forests were introduced as an ensemble of decision-tree-based prediction models by Breiman~\cite{Breiman01}. 
They work by growing an ensemble of decision trees using the observations from the training data. 
To benefit from using an ensemble, the decision trees must be diverse, i.e., they should make different prediction errors.
To achieve diversity, each decision tree is grown by using the principle of bagging (bootstrap aggregating~\cite{Dietterich2000}).
This means, data from the original training data set are drawn with replacement and put into bags. 
Each bag is then used to train one decision tree. 
In addition,  a random subset of features is considered to determine the best split at each node of a decision tree. 
This contributes to the diversity of the trees even if the bags are similar to each other.
The predictions of all decision trees are combined. 
For regression trees, predictions are combined by computing the average value of the individual predictions. 

\runin{LSTM Ensembles}
LSTMs are a popular class of deep learning models, introduced by Hochreiter and Schmidhuber~\cite{HochreiterS96}.
They are an extension of Recurrent Neural Networks (RNNs).
One key benefit of using LSTMs for predictive process monitoring is that they can handle arbitrary length sequences of input data~\cite{Goodfellow-et-al-2016}.
Thus, a single LSTM can be employed to make predictions for business processes that have an arbitrary length in terms of process activities.
LSTMs can be trained by directly feeding the event sequences of historical process executions.
In contrast, other prediction models, including random forests, require the special encoding of the input data to train the models~\cite{TeinemaaDRM19,TaxTZ20}.

While research in predictive process monitoring used individual LSTM models~\cite{LeoniDR20}, in our earlier work we introduced ensembles of LSTMs for the sake of generating reliability estimates~\cite{CAISE2019,MetzgerF021}, which we also use for the experiments reported below.
We use bagging as a concrete ensemble technique.

\subsection{Training of Prediction Models}
\label{sec:training}

We executed the following steps for both types of prediction models:
To train a prediction model, we used 67\% of the event log as training data, while we used the remaining 33\% as "test" data, i.e., as data used as input to the approaches being compared.

For the sake of generalizability, we used a na{\"i}ve approach to select the input features from the respective event logs.
This means we used all input features that were available (provided an input feature did not reveal the process outcome) and did not perform any additional feature engineering or selection. 
We used one-hot encoding to treat categorical (i.e., non-numeric) features such as event labels.

For the size of the ensemble, we chose an ensemble size of $m = 100$.
In earlier experiments, we varied ensemble sizes from 2 to 100. 
The size of the ensemble did not lead to different principal findings~\cite{CAISE2017}. 
There was a trend that larger ensembles generally delivered predictions with higher accuracy. 
More importantly, larger ensembles delivered more fine-grained reliability estimates.

As for bootstrap size (i.e., the size of the bags), we used 60\%.
Earlier experiments did not show a clear trend that larger bootstrap sizes would perform better than smaller ones and different bootstrap sizes did not impact the general shape of the experimental results~\cite{CAISE2017}. 

As a principle, we used the standard hyper-parameters as provided by the respective implementations of the machine learning models.
One main reason was that we were not interested in finding a prediction model with the highest accuracy, but wanted to evaluate how the approaches work for predictions even if they are of lower accuracy.
Also, ensembles in general provide good prediction accuracy even if composed of weak individual prediction models~\cite{zhou2012ensemble}.
This reduces the relevance of finding optimal hyper-parameter settings.
Finally, hyper-parameter tuning for LSTMs can become prohibitively expensive, because the training of an individual LSTM model takes considerable resources and time.
This is exacerbated when using $n$-fold cross-validation, a typical approach to reliably estimate the model performance for a given set of hyper-parameters.

\runin{Random Forests}
The random forest prediction models used in our experiments were trained using the statistical computing environment \emph{R}, using the ‘randomForest’ package. 

As mentioned in Section~\ref{sec:realization1}, random forests require the encoding of sequential event sequences into fixed-length inputs.
To facilitate as much comparability with the LSTM realization as possible, we encoded the event sequences in such a way as to retain the same amount of information as given to the LSTM models.
Also, we chose an encoding that does not require tuning of additional parameters, such as cluster size or aggregation functions\footnote{Note that experimental results suggest that choosing an optimal encoding is challenging and may require choosing a different encoding and parameterization for each data set~\cite{TamaCK20}.}.

Following~\cite{TeinemaaDRM19}, we performed the following two-step encoding:
First, event sequences were divided into buckets, a strategy called ``trace bucketing''.
We used prefix-length bucketing, which means that each bucket $p$ contains the (partial) event sequences of length $p$. 
For each of the buckets, a random forest model was trained.

Second, we transformed the event sequences within each of the buckets into fixed-length input vectors, a strategy called ``sequence encoding''.
We use index encoding, which generates one feature per event and class attribute.
Each feature that encodes an event or class attribute includes an index $(1, \ldots, p)$ that specifies where the event occurred in the case. 
Index encoding retains all the information in the event sequence and we thereby use the same information as for the LSTM models.
Index encoding requires that all event sequences in a bucket have the same length, thus the decision to use prefix-length bucketing in the first step.

\runin{LSTM Ensembles}
For our experiments, we reused the prediction models from our earlier work~\cite{CAISE2019}.
They were trained by building on the implementation of individual LSTM models presented in~\cite{TaxVRD17}.
Different from~\cite{TaxVRD17}, which incrementally predicts the next activities until the process end is reached, we directly predict the process outcome following the approach from~\cite{NavarinVPS17}.
To this end, we modified the LSTM architecture and implementation such that for any point in time, the LSTM directly predicts the process outcome and not the next activity.
Earlier work has indicated that directly predicting the process outcome delivers better accuracy than incremental prediction~\cite{MetzgerN18}.

\subsection{Realization and Configuration of Online RL}
\label{sec:realization2}

We realize the Online RL approach described previously in the same way we did in our earlier work~\cite{CAISE2020,BPM2020}.
An exception is the reward function, which is realized as explained in Section~\ref{sec:curious}.
As a concrete deep RL algorithm, we use PPO (proximal policy optimization~\cite{schulman2017proximal}), a state-of-the-art policy-based actor-critic algorithm.
One main advantage of PPO is that it avoids too large policy updates by using a so-called clipping function. 
A too-large policy update could mean that Online RL misses the global optimum and remains stuck in a local optimum, or even that the policy is destroyed.

Also, compared with other RL algorithms, the PPO algorithm is rather robust for what concerns selecting hyper-parameter settings that facilitate stable learning. 
Thereby, we avoid extensive hyper-parameter tuning compared to similar algorithms and basically can use standard hyper-parameter settings.
One exception is the discount factor $\gamma$, which defines the relevance of future rewards.
Given the way we define the learning episodes (see Section~\ref{sec:RL}), we set the discount factor to $\gamma = 1$ in  order not to discount the reward received for the end state of each case. 

To represent the actor and the critic parts of PPO, we use two multi-layer perceptron networks with two hidden layers of 64 neurons each.
The input layers of both networks consist of five neurons representing the three state variables ($\delta_j$, $\rho_j$, $\tau_j$), from which extrinsic rewards are computed, as well as the two variables adaptation rate $d$ and the negative predictive value $v$ (see Section~\ref{sec:curious}), from which intrinsic rewards are computed.
The output layer of the critic consists of one neuron representing the estimated value of a state, while the two neurons of the actor's output layer represent the probability distribution for each of the two available actions "alarm" and "no alarm".
The concrete action is chosen via sampling from these probability distributions.

\subsection{Chosen Cost Model Parameters}
\label{sec:metrics}
For our experiment, we parametrize the cost model introduced in Section~\ref{sec:cost-model} as follows.
As we are interested in comparing the relative cost savings of the approaches, it is sufficient to work with normalized cost model parameters for determining $C(j)$.
As a basis, we use normalized penalty costs of $C_\textrm{p} = 100$.
Like in~\cite{FahrenkrogTTD19}, we express $C_\mathrm{a}$  and $C_\mathrm{c}$ relative to $C_\textrm{p}$.
Concretely, we model this as:

\small
\begin{equation}
C_\mathrm{a} = \lambda \cdot C_\mathrm{p}; 
C_\mathrm{c} = \kappa \cdot  C_\mathrm{p}
\end{equation}
\normalsize

By varying $\lambda$ and $\kappa$, 
we can reflect different situations that may be faced in practice concerning how costly a process adaptation and compensation may be.
	
We also vary $\alpha$ in our experiments such that it linearly decreases from $\alpha_\mathrm{max} = 1$ for the first prediction point to $\alpha_\mathrm{min}$ for the last prediction point.
We vary $\alpha_\mathrm{min}$  between two extreme settings to cover different possible circumstances:
$\alpha_\mathrm{min} = 1$ models that late adaptations are always feasible, while $\alpha_\mathrm{min} = 0$ models that  late adaptations are never feasible.

We vary the cost model parameters for the experiment as follows: $\lambda, \kappa, \alpha_\mathrm{min}  \in \{0, 0.25, 0.75, 1.0\}$ and consider all combinations of these values, leading to a total of 64 combinations.

Of the approaches, only Empirical Thresholding requires a cost model as input to determine the optimal threshold.
As such a cost model may not be precisely known in practice or a-priori, we model the uncertainty of knowing the actual cost model parameters using the parameter $\xi \in \{0.025, 0.1, 0.175, 0.25\}$.
The parameter $\xi$ gives an envelope around the actual cost model parameters from which the actual cost model parameters that are used as input for Empirical Thresholding are uniformly sampled.
As an example, if $\lambda_\mathrm{actual}$ is the actual value for the adaptation costs, then we sample from $[\lambda_\mathrm{actual} - \xi, \lambda_\mathrm{actual} + \xi]$.

\subsection{Experiment Execution}
\label{sec:exec}

\runin{Use of data} In general, we use the "test" data as described in Section~\ref{sec:training} as input for our experiment.
From this "test" data we use 33\% to determine the threshold for empirical thresholding.
We also initially train the RL approach using this 33\% of the "test" data.
The costs of each of the approaches  are measured using the remaining 67\% of the "test" data.

\runin{Acting on alarms} We assume that a process manager acts on each alarm generated.
Due to the nature of the data sets used (they do not reflect the continuation of the cases after an adaptation), we can only measure the effect of raising at most one alarm per each ongoing case (like in~\cite{FahrenkrogTTD19,ShoushD22}).

\runin{Randomization} As explained above, by sampling within the $\xi$-envelope, we introduce randomness into Empirical Thresholding.
Similarly, Online RL is intrinsically stochastic given the way that the chosen deep RL algorithm works (see Section~\ref{sec:RL}).
To account for these random effects, we repeat the experiment execution for Empirical Thresholding, as well as for Online RL 10 times and report average costs as well as their standard deviation.

\section{Selection and Characterization of Experimental Data Sets}
\label{sec:predmodels}

To facilitate the replicability of our experiments, we use four public real-world event log data sets.
We characterize these data sets and the trained prediction models below.

\subsection{Data Sets}
\label{sec:ds}
The data sets used exhibit different characteristics as shown in Table~\ref{tab:DataSets}.
They cover different application domains: finance (BPIC12 and BPIC17), government (Traffic), and transport (Cargo), and also exhibit different kinds of 
deviations\footnote{Note that while Cargo has a real-valued deviation (i.e., delay in terms of hours), BPIC12, BPIC17, and Traffic have categorical deviations (violation or non-violation).
To provide the numeric reward signal (relative predicted deviation) for our RL agent, we map a non-violation to 0.0 and a violation to 1.0, and train regression prediction models on these numeric outcome labels.
To compute $\delta_j$, we set $A = 0.5$ in Equation~\ref{eq:dev}.}. 
They differ in terms of the rate of actual deviation as well as in the size of the data set (note that we report the size of the "test" data set, i.e., 67\% of the actual data set as explained above).
Also, they differ in terms of complexity and length of the process instances\footnote{Like in~\cite{TeinemaaDRM19}, we only consider prediction points up to a certain prefix length in order not to bias the results toward extremely long cases. Concretely, we consider the 99\% quantile of all prefix lengths of the respective data set.}.

\begin{table}[htbp]
	\centering
	\resizebox{.95\textwidth}{!}{%
		\begin{tabular}{l|l|r|r|r|r}
     &   \emph{} & \emph{Rate of} & \emph{Size of} & \emph{Process} & \emph{Max. pre-}\\
    \emph{Name } & \emph{Type of Deviation } & \emph{ Deviations} &  \emph{"Test" Data } & \emph{Variants} & \emph{fix length}\\
		\hline

\textbf{BPIC12} & Unsuccessful loan application  &  25\% &   4,361 & 3,587 &  48 \\

\textbf{BPIC17} & Unsuccessful loan application & 41\% &  10,500 & 2,087 &  71 \\
\textbf{Traffic} & Unpaid traffic fines  &  58\% &  50,117 & 185 &  5 \\
\textbf{Cargo} & Delay in cargo delivery  &  31\% &  1,313 & 144 &  21 \\

		\hline
		\end{tabular}}

	\caption{Data sets used in experiments}

	\label{tab:DataSets}
\end{table}

\runin{BPIC12 and BPIC17} These data sets entail process monitoring data of a loan application process.
Both BPIC data sets concern the processes of the same financial institution, but differ in the form of data collection and process variants.
These data sets are frequently used in research on predictive process monitoring (e.g., see~\cite{VerenichDRMT19,TeinemaaDRM19,Marquez-Chamorro_2017}).

\runin{Traffic} This data set entails the process monitoring data of a traffic fine process.
Again, this data set is frequently used in research on predictive process monitoring (e.g., see~\cite{VerenichDRMT19,TeinemaaDRM19,Marquez-Chamorro_2017}).

\runin{Cargo} This data set covers five months of air cargo processes of an international freight forwarding company. 
We used this data set extensively in our previous research (e.g., see~\cite{ICSOC2017,CAISE2017,CAISE2019,Metzger_et_al_2015}).

For each of the four data sets, we trained an LSTM ensemble and an RF prediction model, considering the successful and unsuccessful process outcomes as indicated in Table~\ref{tab:DataSets}.
Below, we provide a characterization of the prediction models in terms of their prediction accuracy and analyze potential concept drifts in terms of the prediction models' accuracy.
We provide this characterization as relevant context for explaining the evaluation results in Section~\ref{sec:ExperimentalResults}.

\subsection{Average Prediction Accuracy}
\label{sec:predacc}
\begin{figure}[hbtp]
\centering

		$
		\begin{array}{cccc}
		\textrm{BPIC12} & \textrm{BPIC17} \\
		\includegraphics[width=.31\textwidth]{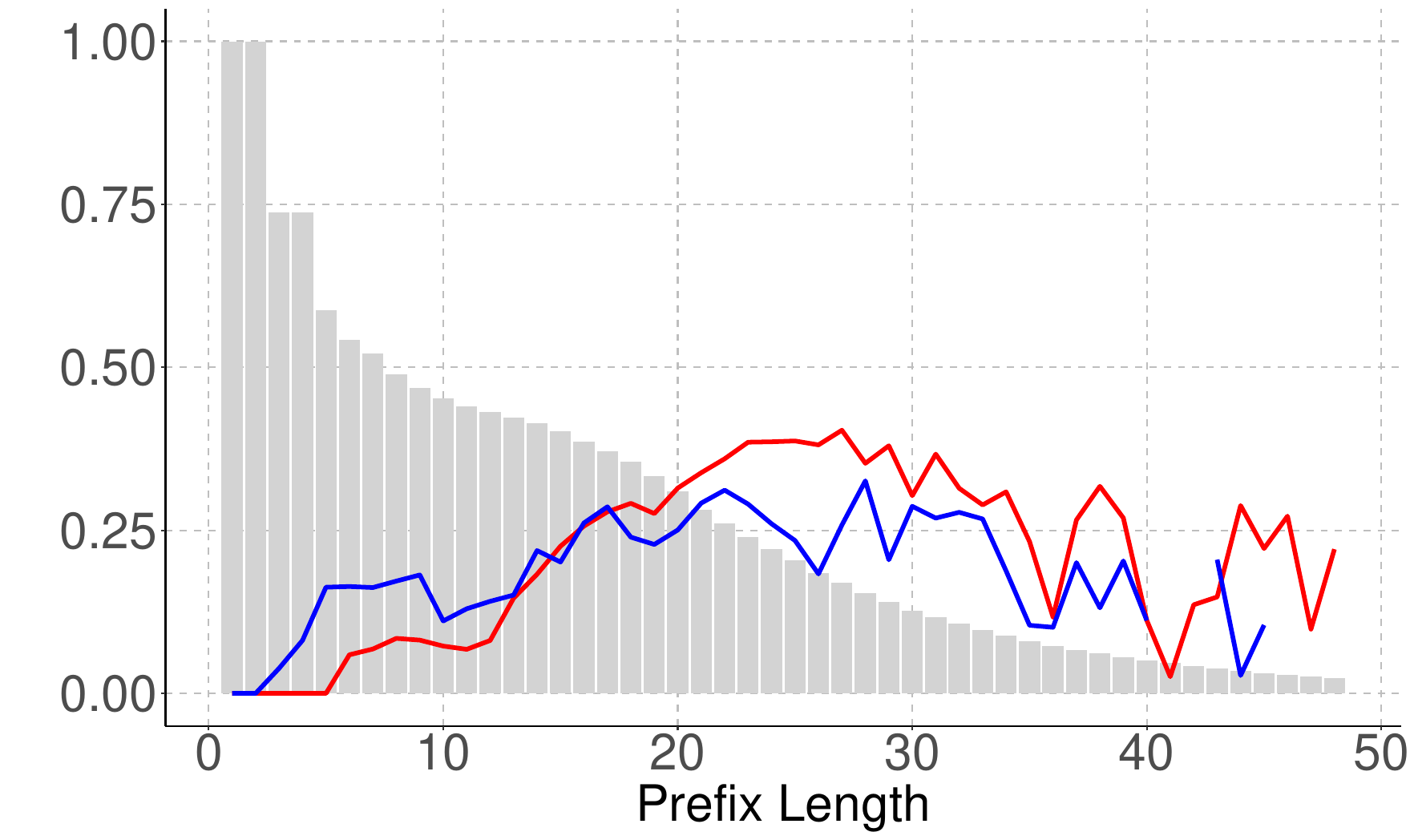} &
     	\includegraphics[width=.31\textwidth]{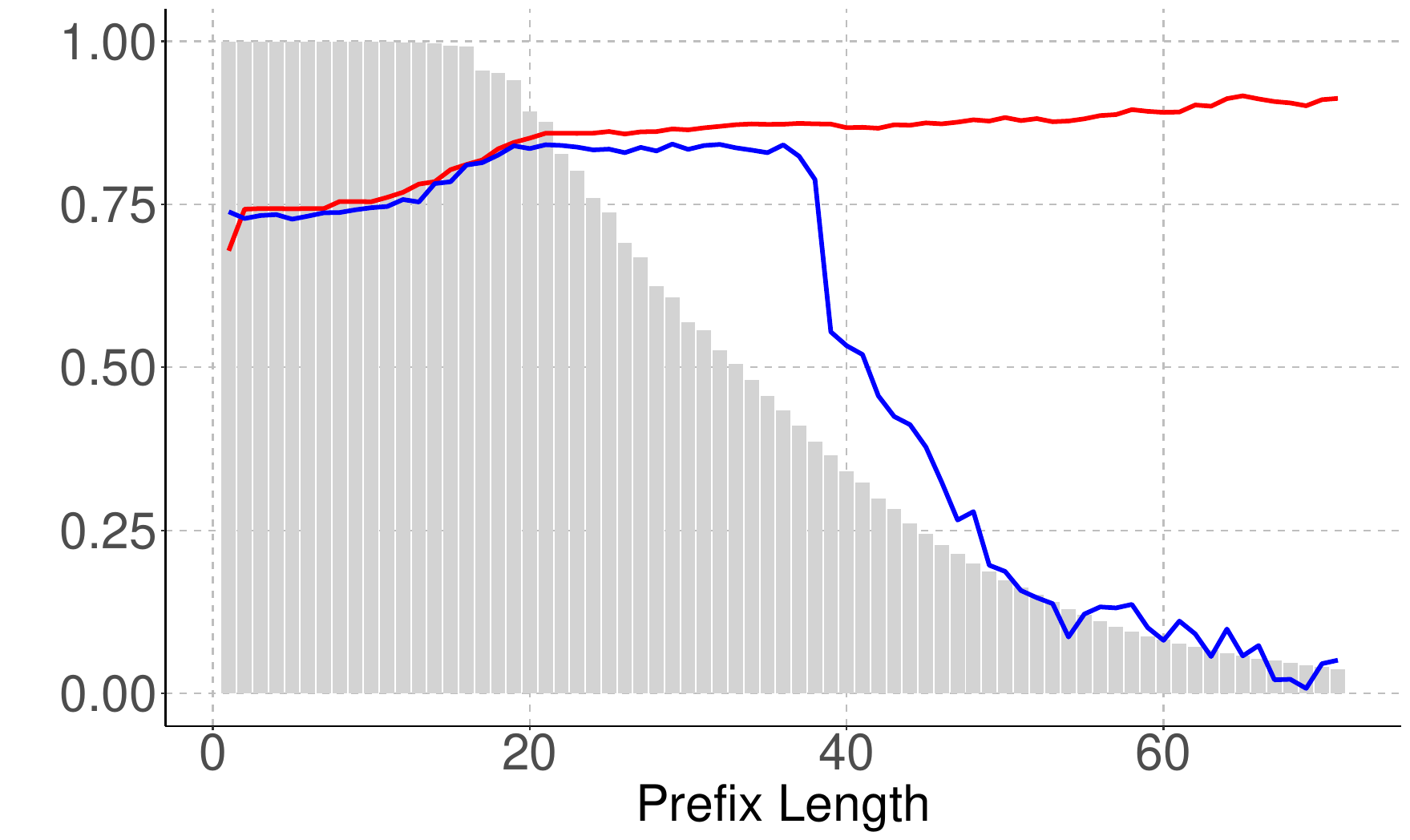} \\ 
		\textcolor{red}{\varnothing = 0.242} & \textcolor{red}{\varnothing = 0.804}  \\
		\textcolor{blue}{\varnothing = 0.309} & \textcolor{blue}{\varnothing = 0.721} \\
\\
		 \textrm{Traffic} & \textrm{Cargo}\\
	    \includegraphics[width=.31\textwidth]{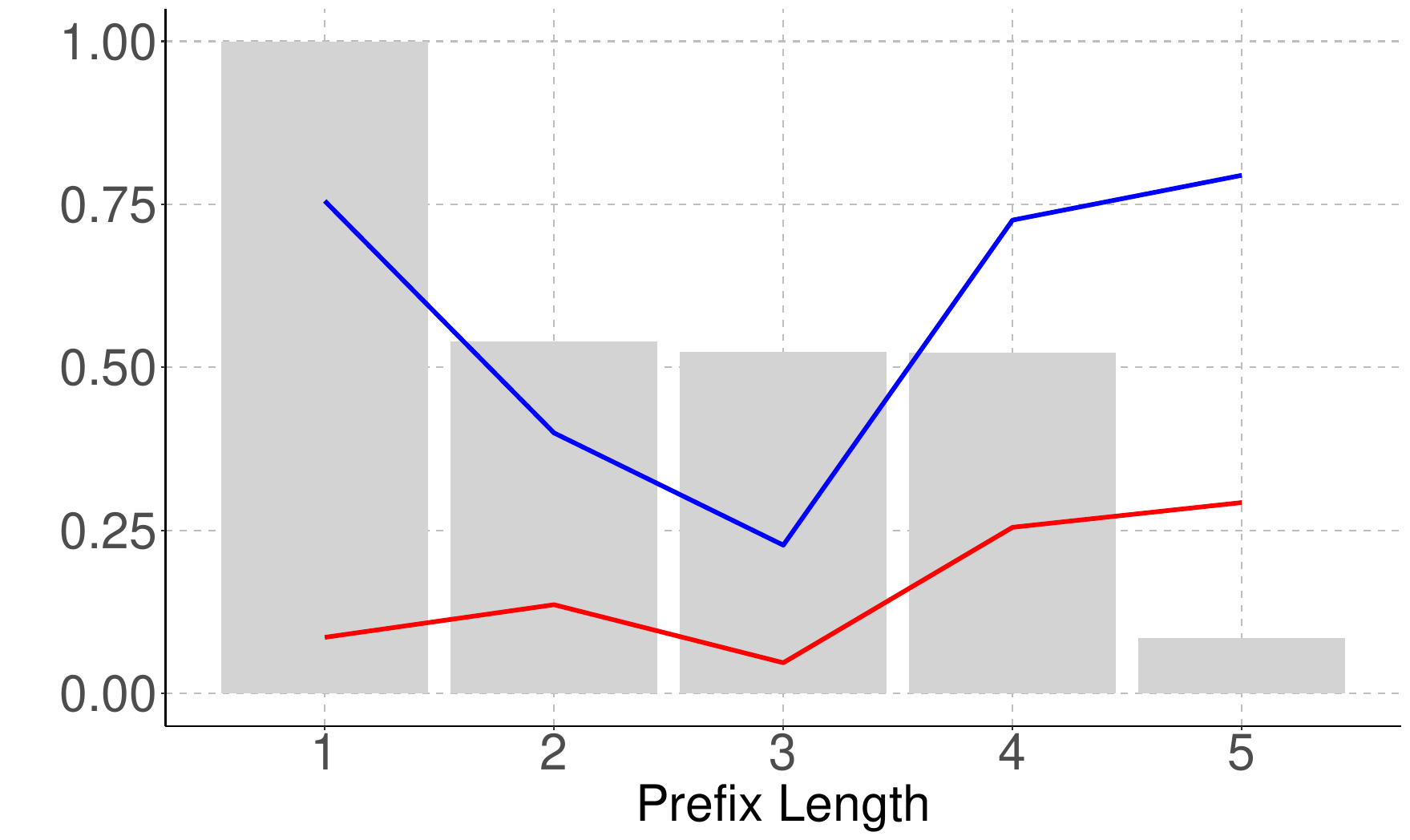} &
	    \includegraphics[width=.31\textwidth]{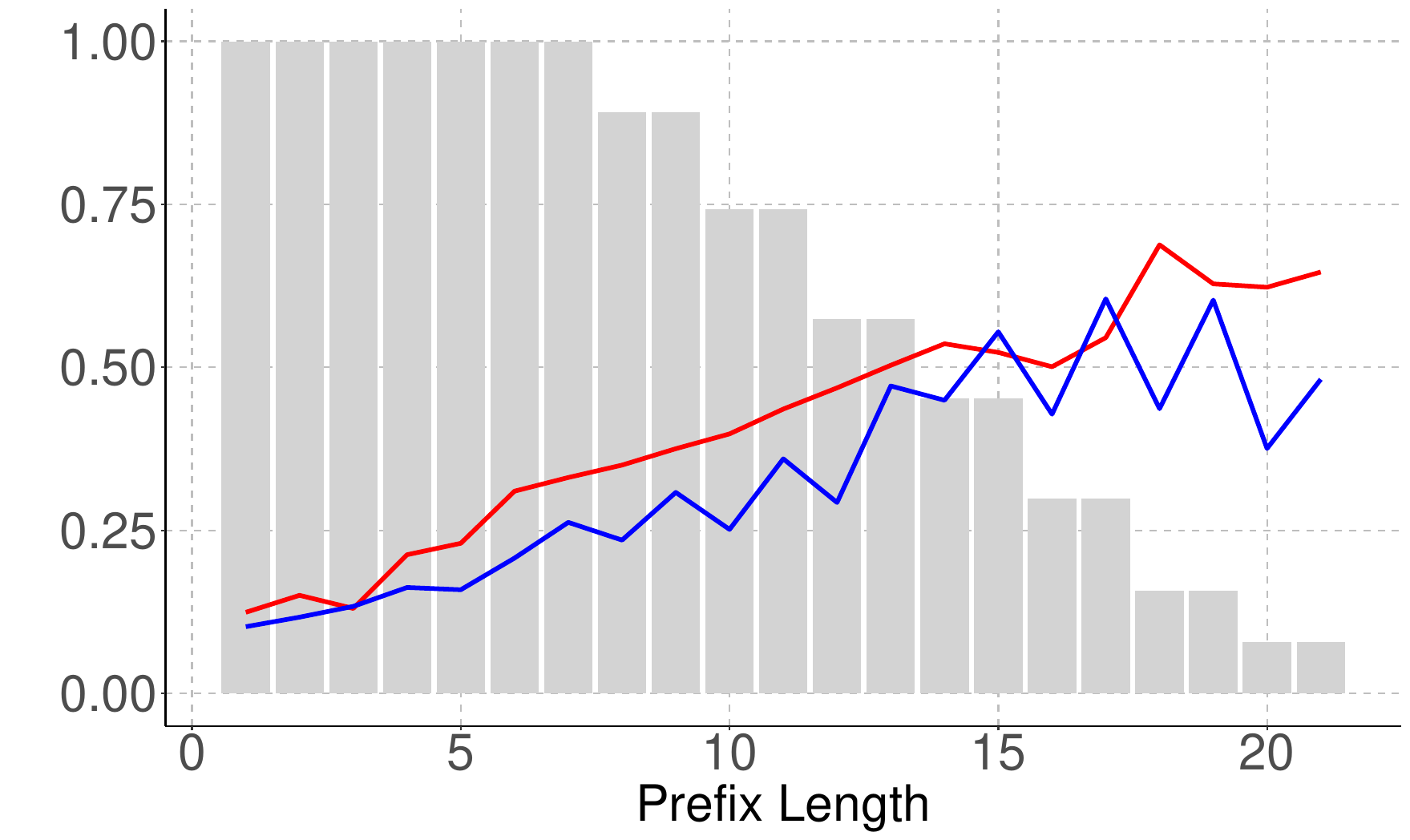} \\        
		\textcolor{red}{\varnothing  = 0.185} & \textcolor{red}{\varnothing = 0.327} \\
		\textcolor{blue}{\varnothing = 0.648} & \textcolor{blue}{\varnothing = 0.226}\\    
		\end{array}$
        
\caption{Characteristics of prediction models and data sets: \textcolor{red}{\textbf{red}}: LSTM prediction accuracy [MCC];
\textcolor{blue}{\textbf{blue}}: RF prediction accuracy [MCC];
 bars depict relative number of cases that reach prefix length $j$
}

\label{fig:mcc}
\end{figure}

For each of the data sets, Fig.~\ref{fig:mcc} shows the average prediction accuracy for each possible prediction point (i.e., prefix length) as well as the average accuracy ($\varnothing$) across all prediction points.
We measure the accuracy of the prediction models using the "test" data (i.e., 33\% of the overall data of the respective data set).
Also, the figure shows the relative number of cases that reach the respective prefix length (grey bars underlying the curves).

We use contingency table (binary) metrics, because this conveys a better impression of the performance of the prediction model.
The numeric prediction error of the regression model does not directly indicate whether the process outcome may violate the expected process outcome or not~\cite{ASAS_Book}. 
As contingency table metric, we use the Matthews Correlation Coefficient (MCC), which is robust against class imbalances~\cite{boughorbel2017optimal}.
Also, MCC is a more challenging metric to score high on and therefore provides more realistic estimates of real-world model performance.
Using the four prediction contingencies (see Section~\ref{sec:accuracy}), MCC $\in [-1, 1]$ is defined as follows (1 indicating a perfect predictor; $-1$ indicating the opposite):

\small
\begin{equation}
\text{MCC} = \frac{ TP \cdot TN - FP \cdot FN } {\sqrt{ (TP + FP) \cdot ( TP + FN ) \cdot ( TN + FP ) \cdot ( TN + FN ) } }
\end{equation}
\normalsize

As one can observe in Fig.~\ref{fig:mcc}, in most data sets there is a tendency that prediction accuracy continuously increases as the cases unfold.
There are exceptions though as discussed below.

\runin{Traffic} We can observe one exception for the RF prediction model of the Traffic data set (Traffic-RF).
Here, prediction accuracy for prefix length 1 is almost as high as towards the end of  the case, and there is a large drop in accuracy for prefix lengths 2 and 3.
For this data set, there is a high probability that deciding to delay the adaptation decision may result in higher costs, as one may hit the area of low accuracy and will pay for it with a penalty for adapting later\footnote{
Note that~\cite{TeinemaaDRM19} shows the same behavior for the Traffic data set when using gradient boosted trees (GBT), a different variant of random forests.}.

\runin{BPIC17} We observe a further exception for BPIC17-RF, which shows a visible drop in accuracy after around prefix length 40, which never recovers.
This is due to the way we perform case encoding as explained in Section~\ref{sec:realization1}.
For longer cases, the size of the buckets gets smaller and thus RF has not sufficient training data (in contrast to LSTM, which can leverage data across the whole case).
Different encodings can lead to better accuracy curves as explored in ~\cite{TeinemaaDRM19}.
Even though cases that reach this point already are through at least around $1/2$ of their execution, a relatively high number of cases (32\%) are longer than prefix length 40.
As such, this may have an impact on how -- on average -- the approaches work for this data set.

\runin{BPIC12} Similarly, but less pronounced, the prediction accuracy for BPIC12-LSTM and BPIC12-RF starts dropping at around prefix length 35.
One reason may be the lower data quality of BPIC12 when compared with BPIC17 (as indicated by the clear difference in $\varnothing$ prediction accuracy) combined with the low number of training data for late prediction points. 
As can be seen from the bars, there is less data to train the prediction models for late prediction points, thus these prediction models may not have sufficiently generalized over the data.
For the BPIC12 data set, only 7\% of the processes reach the point where prediction accuracy starts dropping and then already have reached at least around $2/3$ of their execution.
It thus should not have a major impact on how the approaches work for this data set.

\runin{Cargo} Finally, while overall exhibiting a continuous increase in prediction accuracy, Cargo-RF shows a visible zig-zag pattern.
The reason is that all cases of the Cargo data set have equal process lengths, and thus training data for odd process lengths is missing.
Where the LSTM prediction model can handle such gaps in data, the RF prediction model is more strongly affected by these gaps.

\subsection{Concept Drifts} 
\label{sec:drifts}
\begin{figure}[hbtp]
\centering

		$
		\begin{array}{cccc}
		\textrm{BPIC12-LSTM} & \textrm{BPIC12-RF} 	&	\textrm{BPIC17-LSTM} & \textrm{BPIC17-RF} \\

		\includegraphics[width=.22\textwidth]{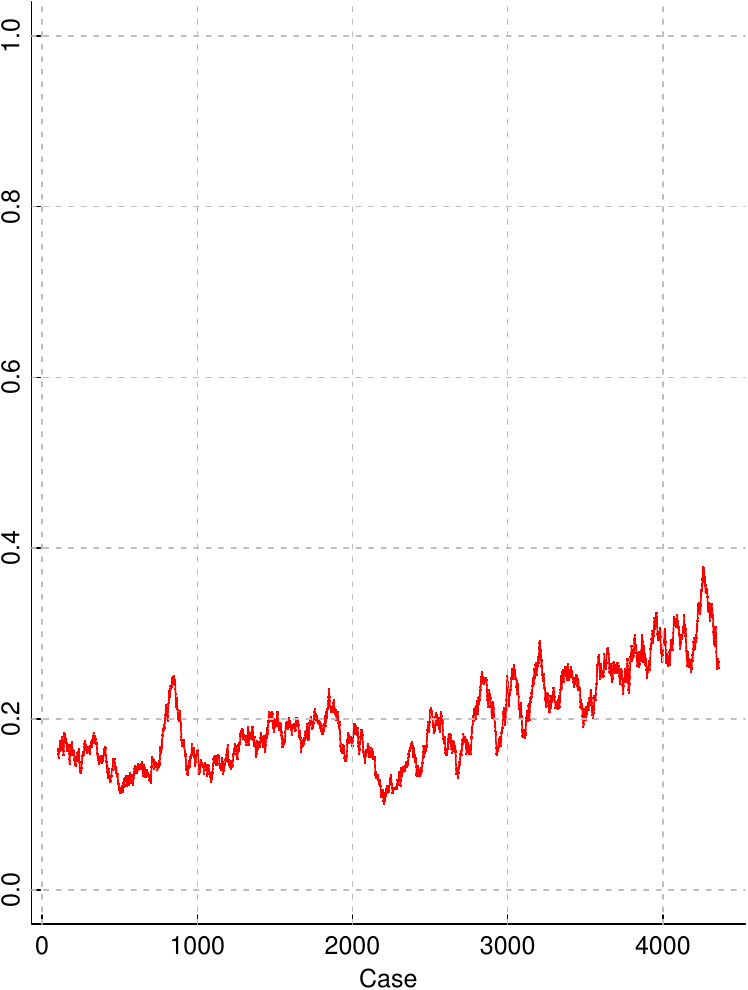} &
		\includegraphics[width=.22\textwidth]{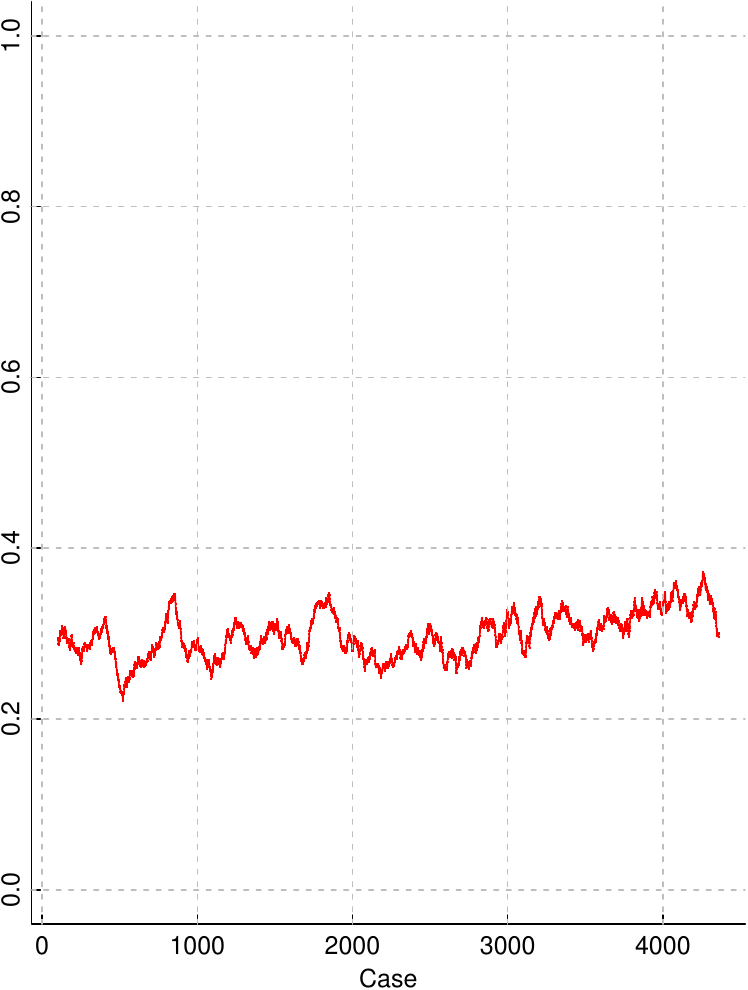} & 
     	\includegraphics[width=.22\textwidth]{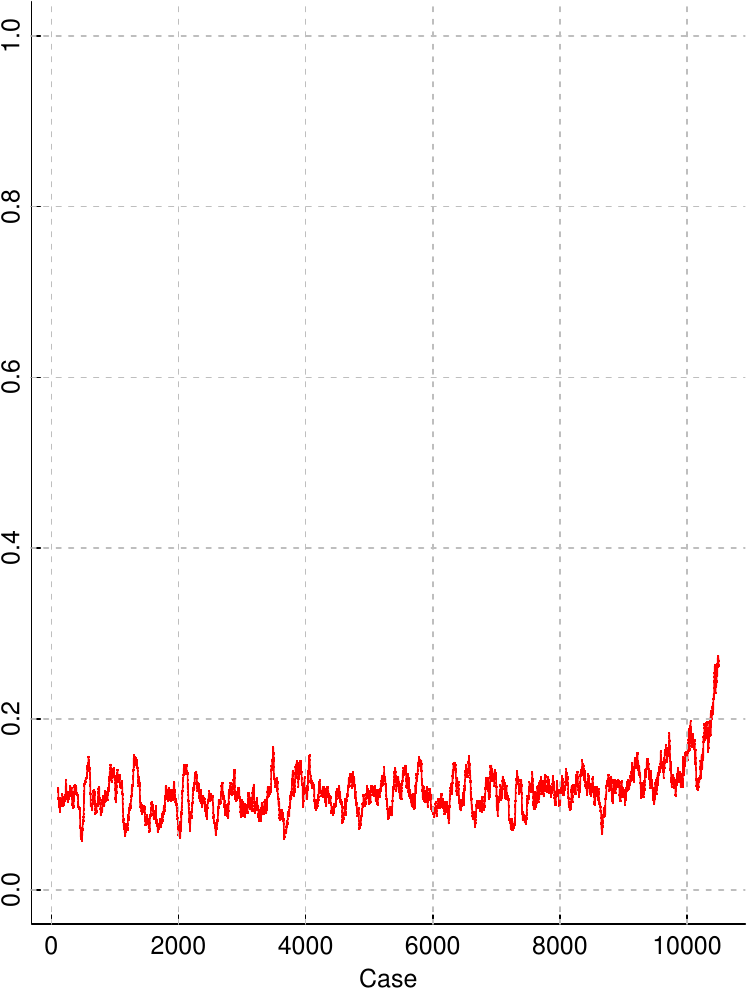} &
	    \includegraphics[width=.22\textwidth]{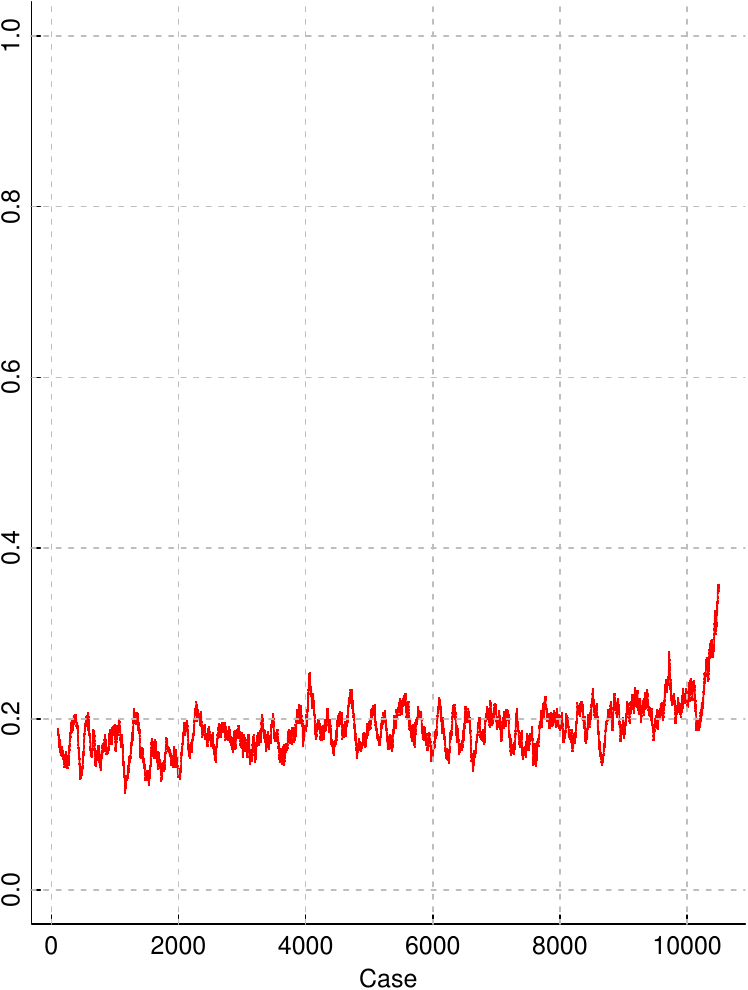} \\

		\textrm{Traffic-LSTM} & \textrm{Traffic-RF} 	&	\textrm{Cargo-LSTM} & \textrm{Cargo-RF} \\

		\includegraphics[width=.22\textwidth]{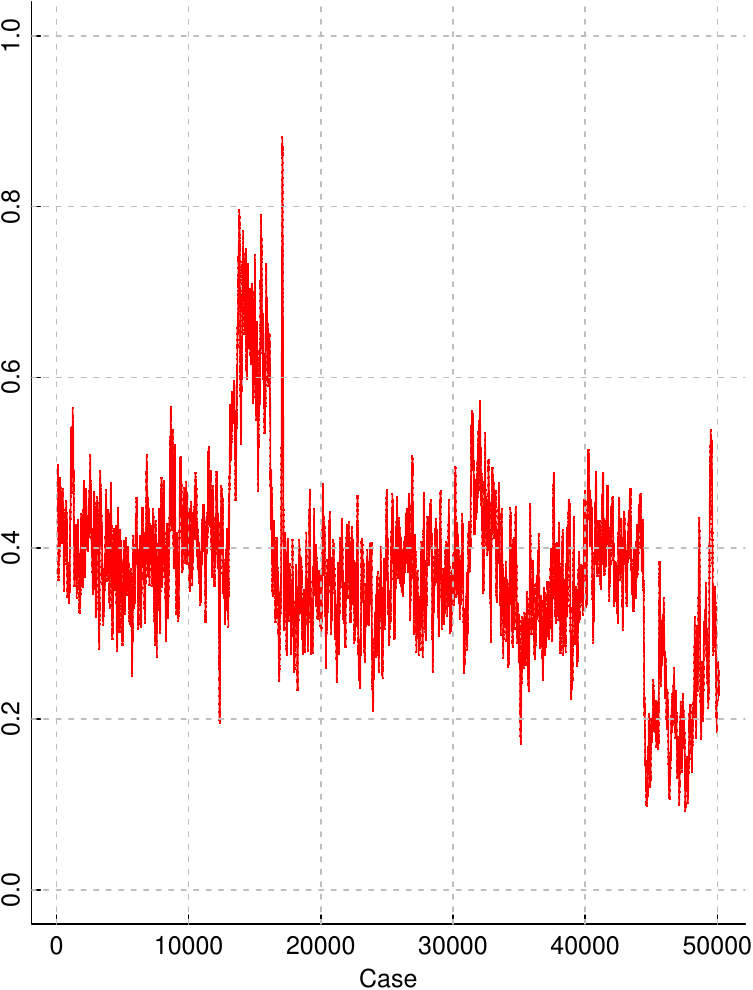} &
		\includegraphics[width=.22\textwidth]{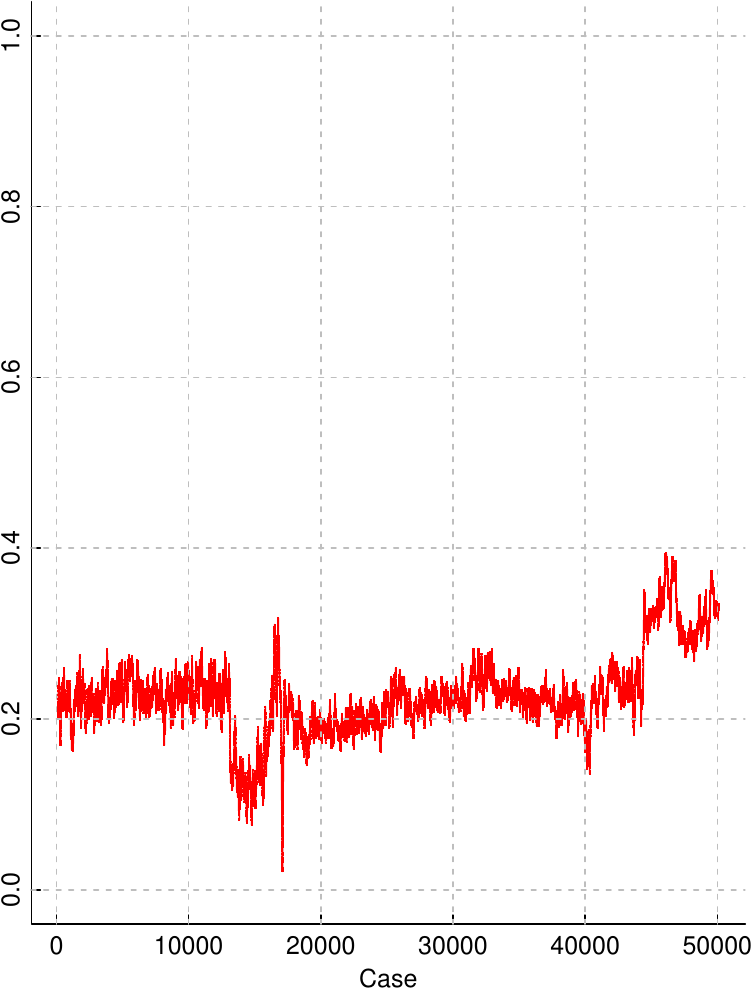} & 
     	\includegraphics[width=.22\textwidth]{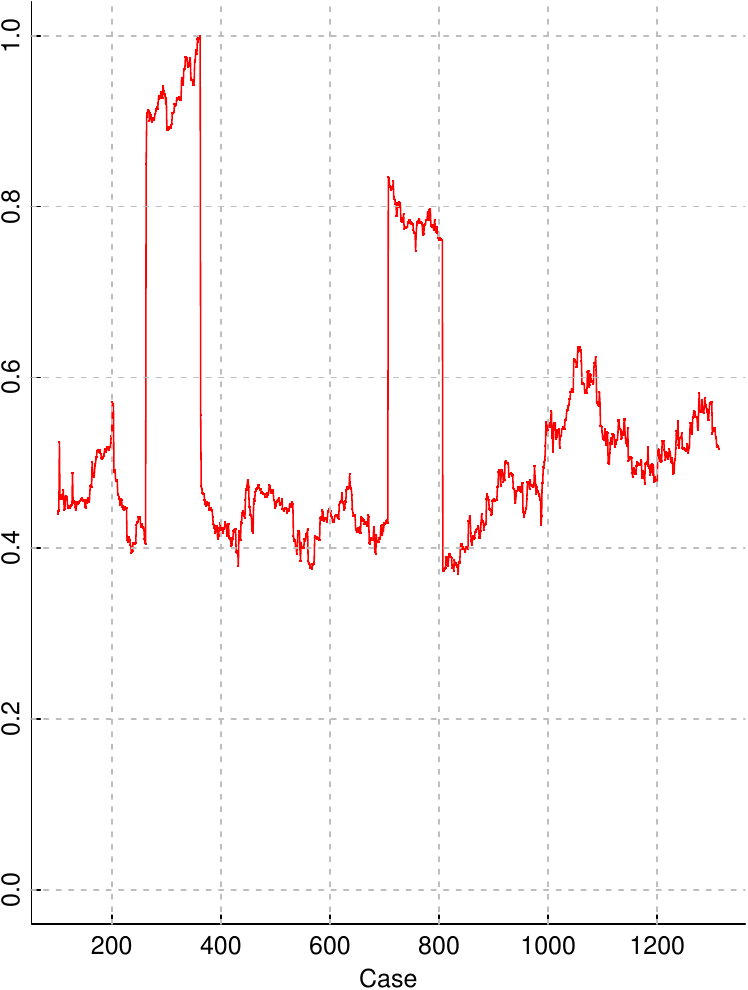} &
	    \includegraphics[width=.22\textwidth]{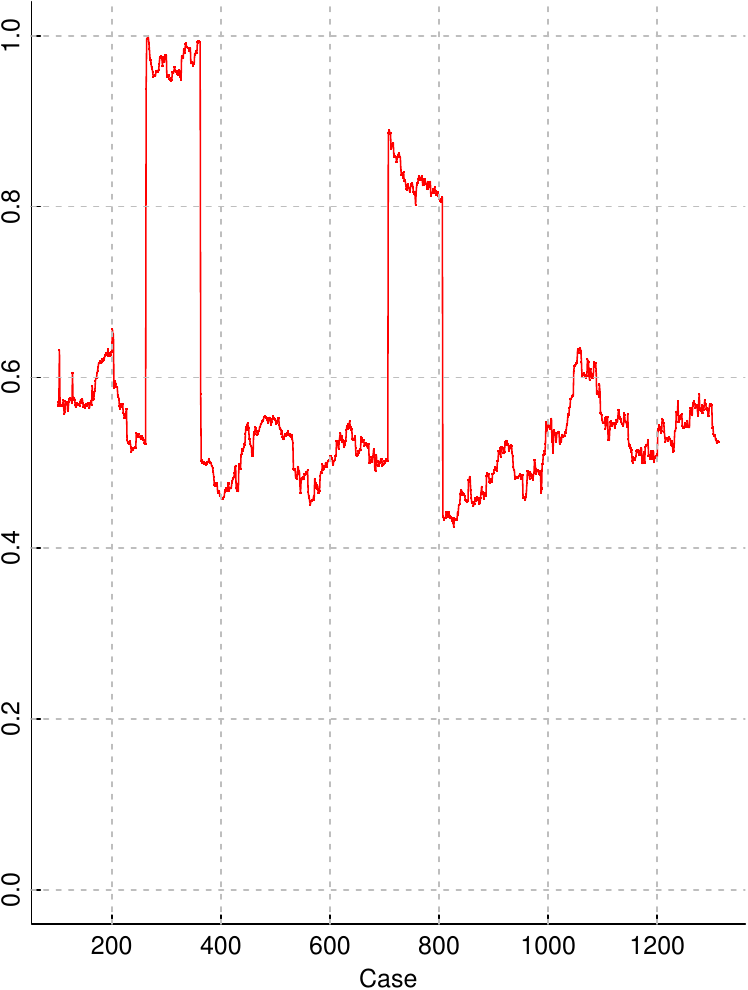} \\

		\end{array}$

\caption{Prediction accuracy [MAE] per case as an indicator for concept drifts}


\label{fig:nonstat}
\end{figure}

Figure~\ref{fig:nonstat} shows the prediction accuracy per each case in the "test" set.
This shows how prediction accuracy may fluctuate over longer periods, thereby indicating concepts drifts.
One reason for such fluctuations may be that the prediction models are presented with unseen and out-of-sample process monitoring data.
We measure the prediction accuracy for each case in terms of the Mean Absolute Error (MAE) computed for all predictions $\hat{y}_{j}$ of a case;  $j = 1, \ldots, l$, where $l$ is the case length, and $y$ the actual process outcome:

\small
\begin{equation}
\text{MAE} = \frac{|\hat{y}_{1} - y | + \cdots + |\hat{y}_{l} - y |}{l}
\end{equation}
\normalsize

\runin{Traffic}
The Traffic data set (LSTM as well as RF) shows the most visible change of prediction accuracy, especially between around case \# 14,000 and \#~16,000, and again after around case \# 45,000.
The prediction accuracy for cases \# 14,000 to \# 16,000 is 65\% higher than the average accuracy for the whole data set, while for cases after case \# 45,000 the average accuracy is 51\% lower than the average accuracy for the whole data set.

\runin{Cargo} Similarly, the Cargo data set (LSTM as well as RF)  shows a visible change in prediction accuracy at around case \# 300 and case \# 700.

\runin{BPIC12} The BPIC12-LSTM data set shows a change in prediction accuracy after around case \# 3,500, while for BPIC12-RF there is no discernible change in prediction accuracy.

\runin{BPIC17}
Finally, BPIC17 -- both LSTM and RF -- show the smallest change in prediction accuracy, for both data sets happening towards the end of the data set.
\section{Evaluation Results}
\label{sec:ExperimentalResults}

To answer the research question posed in Section~\ref{sec:Experiments}, we measured in which situations (i.e., for which cost model parameters) and how often the approaches provide the highest cost savings.
In addition, we quantified the extent of these cost savings.
Below, we present our experimental results by first presenting high-level observations and then providing a more in-depth analysis of the observations.

\subsection{High-level Observations}
\label{sec:results-1}

\begin{figure}[hbtp]
\centering

		\includegraphics[width=1\textwidth]{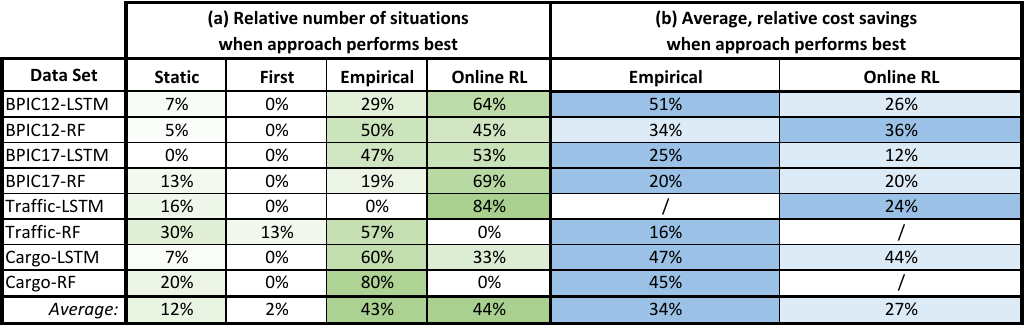} 

\caption{Overview comparison of different approaches to trade-off prediction accuracy and earliness (\textbf{bold} = approach performs best)}
\label{fig:RQ1}
\end{figure}

Figure~\ref{fig:RQ1} provides an overview of the experimental results.
Considering only situations in which proactive adaptation offers cost benefits, the figure provides (\emph{a}) the relative number of situations in which an approach performs best, and (\emph{b}) the average, relative cost savings per case in such situations.

The results in Figure~\ref{fig:RQ1}(\emph{a}) show that the more recent and advanced techniques -- Empirical Thresholding and Online RL --  outperform the more simplistic approaches.
Comparing Empirical Thresholding and Online RL, we can observe that both tend to work in many situations, albeit with some exceptions that we discuss as part of our detailed analysis in Section~\ref{sec:results-2} as well as in Section~\ref{sec:disc}.

Figure~\ref{fig:RQ1}(\emph{b}) quantifies the extent of the cost savings for the two best-performing approaches, i.e., Empirical Thresholding and Online RL.
To this end, the average, relative cost savings $c_\mathrm{rel}$ of the respective approach $x$ are computed using the costs of the approach $c_x$ and the costs of never adapting $c_\mathrm{never}$:

\small
\begin{equation}
c_\mathrm{rel} = \frac{c_\mathrm{never} - c_x}{c_\mathrm{never}}
\end{equation}
\normalsize

We can observe that both approaches consistently deliver cost savings.
On overage, Empirical Thresholding delivers cost savings of 34\% and Online RL delivers cost savings of 27\%.
The reasons for the smaller cost savings of Online RL are discussed as part of our detailed analysis below.

\subsection{Detailed Analysis}
\label{sec:results-2}

\begin{figure}[hbtp]
\centering
		$
		\begin{array}{cc}
	
		\textrm{BPIC12-LSTM} & 		\textrm{BPIC12-RF} \\
		
		\includegraphics[width=.48\textwidth]{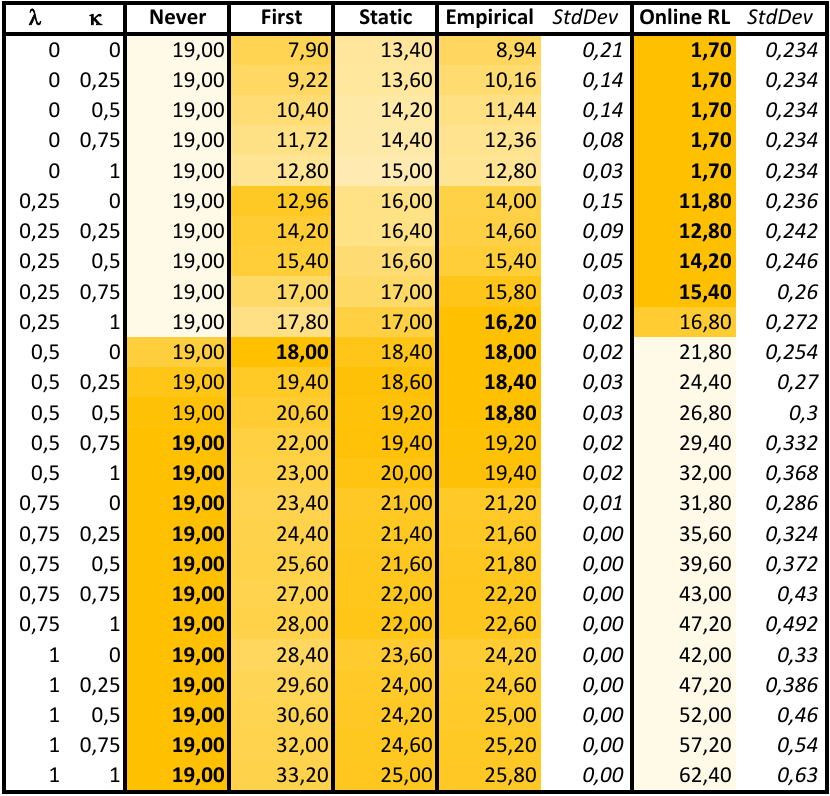} & 
    	\includegraphics[width=.48\textwidth]{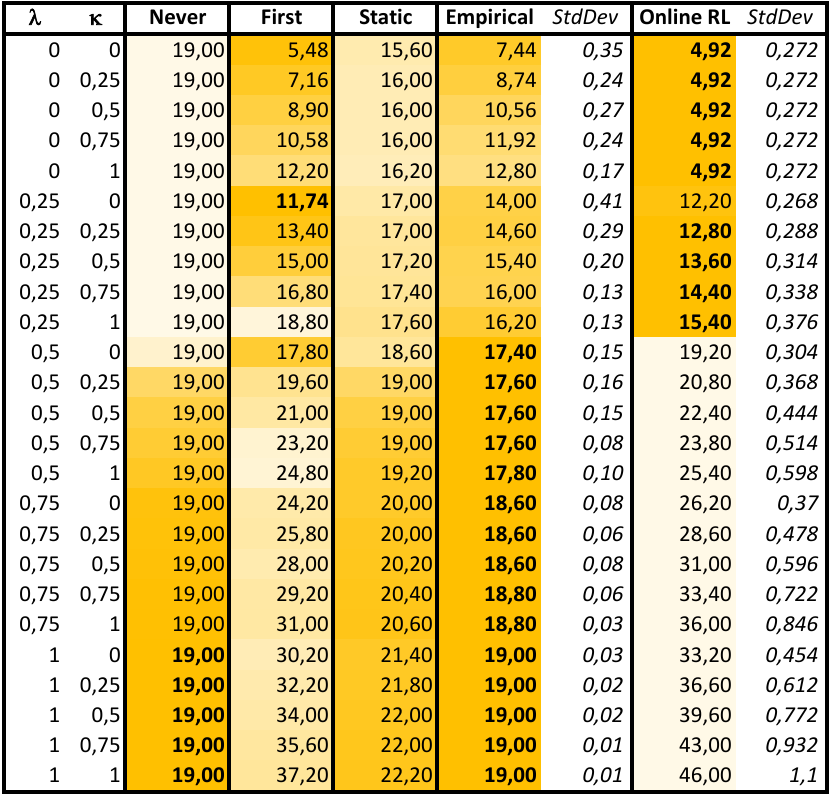} \\

 		\textrm{BPIC17-LSTM} & 		\textrm{BPIC17-RF} \\
    	\includegraphics[width=.48\textwidth]{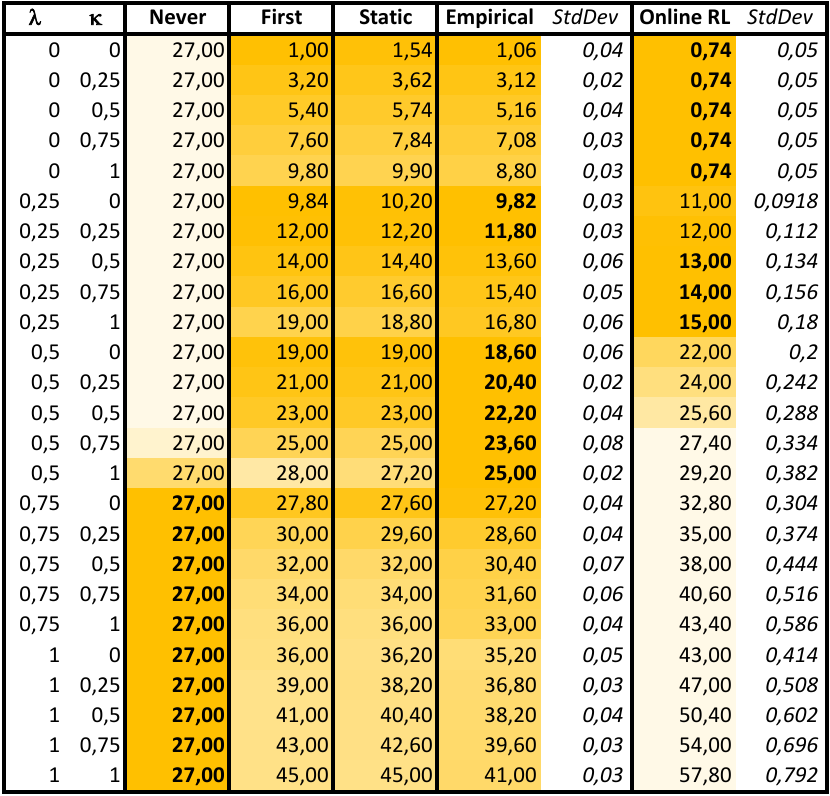} & 
    	\includegraphics[width=.48\textwidth]{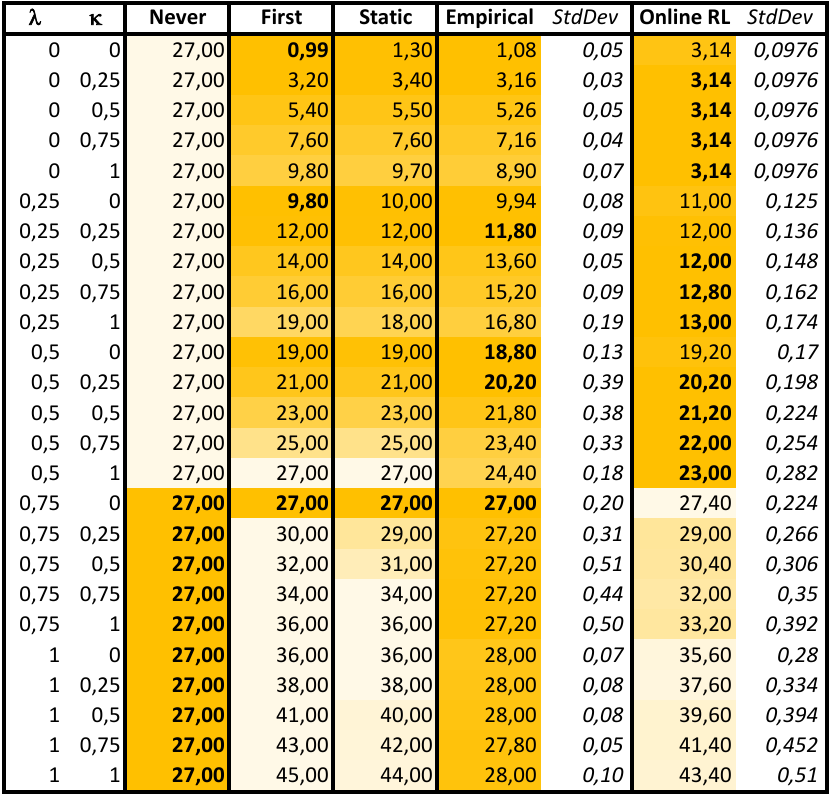} \\
    	
		\end{array}$

\caption{Average process execution costs per case for the different approaches to trade-off prediction accuracy and earliness (\textbf{bold} = approach leading to lowest cost)}
\label{fig:RQ1a}
\end{figure}

\begin{figure}[hbtp]
\centering
		$
		\begin{array}{cc}

\textrm{Traffic-LSTM} & 		\textrm{Traffic-RF} \\ 		

		\includegraphics[width=.48\textwidth]{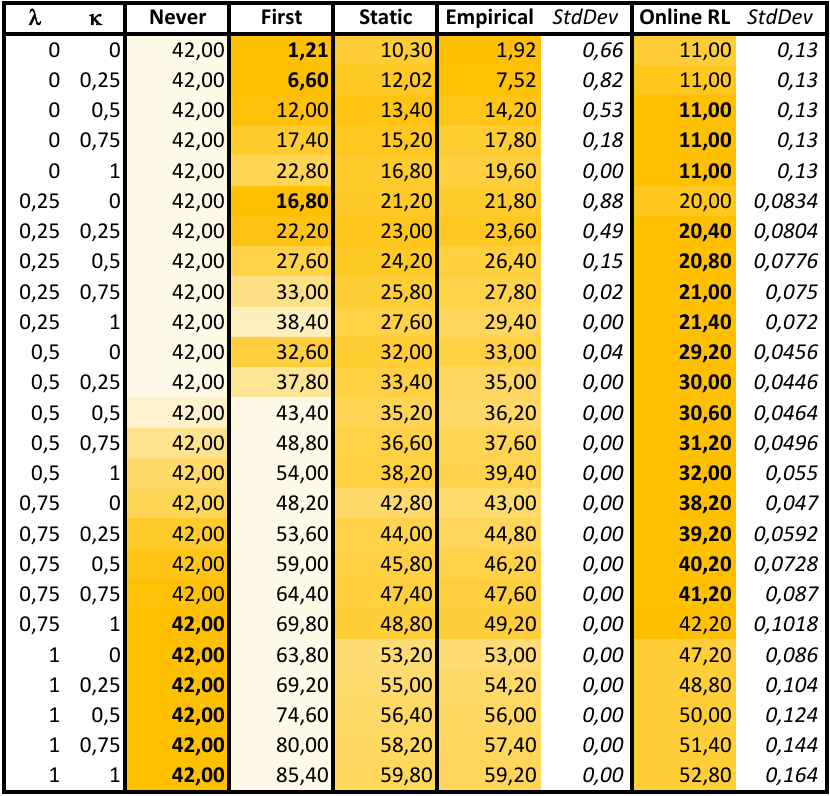} & 
    	\includegraphics[width=.48\textwidth]{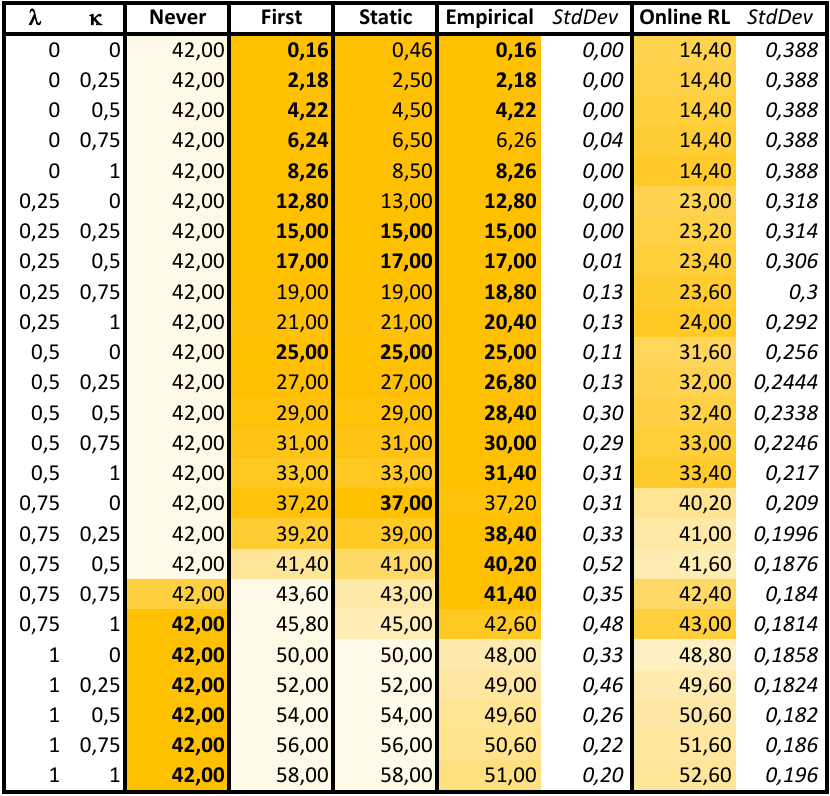} \\
    	
\textrm{Cargo-LSTM} & 		\textrm{Cargo-RF} \\    	
    	
    	\includegraphics[width=.48\textwidth]{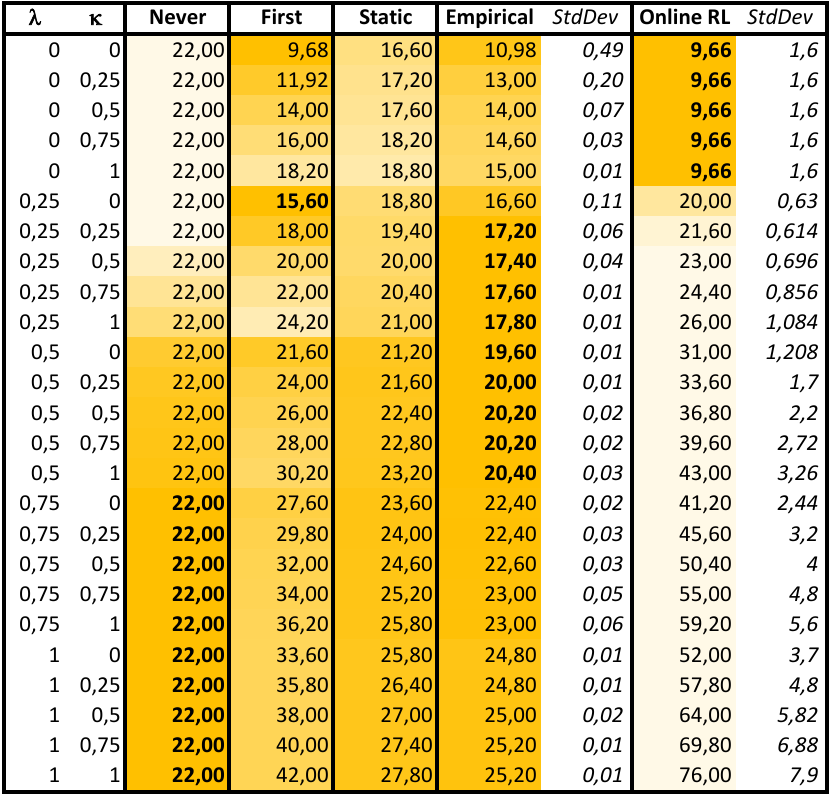} & 
    	\includegraphics[width=.48\textwidth]{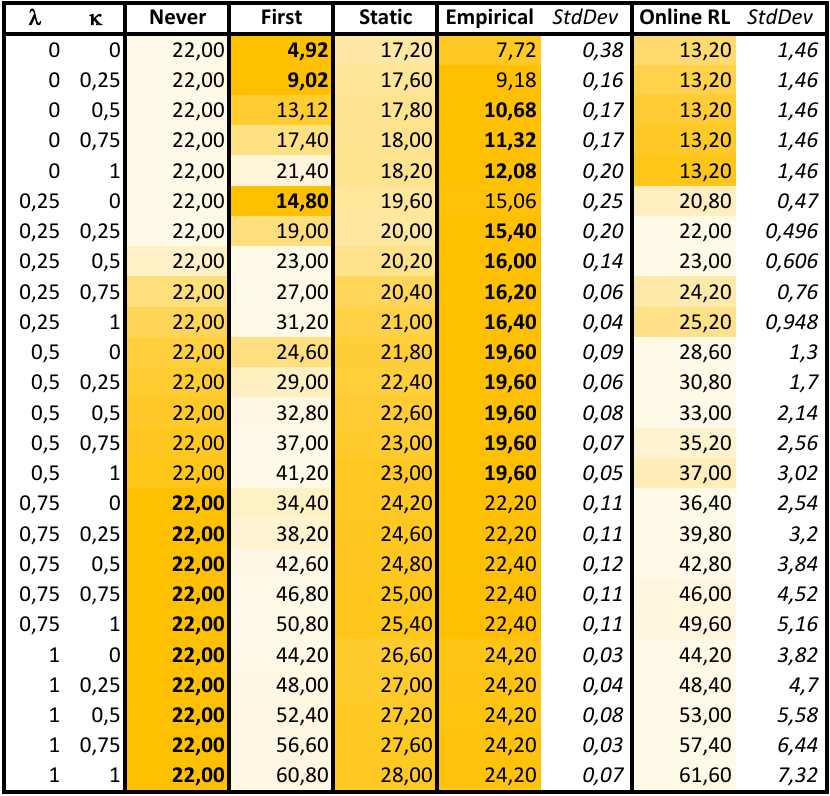} \\
		\end{array}$

\caption{Average process execution costs (cont'd)}
\label{fig:RQ1b}
\end{figure}

As a basis for a detailed analysis of the experimental results, Figures~\ref{fig:RQ1a} and~\ref{fig:RQ1b} show the average process execution costs per case for the different approaches and data sets (remember these are normalized costs as explained in Section~\ref{sec:metrics}).
The figure shows the costs clustered by cost model parameters $\lambda$ (adaptation costs) and $\kappa$ (compensation costs), averaged over $\alpha$ (adaptation effectiveness).
In addition, given the stochastic nature of Empirical and Online RL, we also report the standard deviation. 
Below, we refer to the different lines in the tables of Figure~\ref{fig:RQ1a} as $(\lambda, \kappa)$.

In addition to the four state-of-the-art approaches (as introduced in Section~\ref{sec:approaches}), we also report the results when never adapting the case, i.e., the costs entailed if process execution would commence without intervention.
This provides us an important baseline to understand under which situations proactive adaptation may not lead to cost savings.
As can be observed, in situations where adaptation and/or compensation costs are high, proactive adaptation does not pay off  corroborating the results of our earlier research~\cite{CAISE2017,CAISE2019}.
For example, proactive adaptation does not pay off for BPIC12-LSTM beginning with cost model parameters $(0.5, 0.75)$, for BPIC17-RF beginning with $(0.75, 0.25)$, and for Traffic-RF beginning with $(0.75, 1)$.

We can observe that no single approach works best for all data sets and all cost model configurations.
As a general observation, Empirical Thresholding tends to work best for cost model parameters that are in the middle range of values.
For example, Empirical Thresholding works best for BPIC12-RF between $(0.5, 0)$ and $(0.75, 1)$, and for Cargo-LSTM between $(0.25, 0.25)$ and $(0.5, 1)$.
Similarly, Online RL tends to work best for smaller cost model configurations.
For example, Online RL works best for BPIC12-LSTM between $(0, 0)$ and $(0.25, 0.75)$, and in BPIC17-LSTM between $(0, 0.9)$ and $(0.25, 1)$.
However, we can also see several exceptions to this general observation.
We thus discuss the results for each data set individually, considering the data set and prediction model characteristics identified in Sections~\ref{sec:predmodels}.

\runin{BPIC12} The results for this data set follow the general observations from above.
Cost savings for Online RL are higher for BPIC12-LSTM than for BPIC12-RF.
We attribute this to the fact that for BPIC12-LSTM the concept drift is more pronounced and thus Online RL can effectively capture this.
For BPIC12-LSTM, this data set shows a visible increase in prediction accuracy after around case \# 3,500 (see Section~\ref{sec:drifts}).
As a result, Online RL reacts to this by increasing the rate and earliness of alarms.
In contrast, Empirical Thresholding keeps the rate of alarms and earliness roughly the same, thus not leveraging on the opportunity to reduce costs by raising alarms earlier.
As shown in Figure~\ref{fig:RQ1}\emph{(a)}, Online RL performs best in 64\% of the situations, while Empirical only does so in 29\% of the situations. 

For RF, the change in prediction accuracy (as analyzed in Section~\ref{sec:drifts}) is much smaller.
For Online RL, the changes in the rate and earliness of alarms are so small as to fall within the variance of these metrics.
Again, Empirical Thresholding keeps the rates of alarms and earliness roughly the same.
Taking into account the variance (standard deviation) of the approaches, Empirical Thresholding and Online RL perform roughly the same with a difference of 5\% as shown in Figure~\ref{fig:RQ1}\emph{(a)}.
Empirical Thresholding works quite well even though average accuracy drops visible after around prefix length 35.
As discussed in Section~\ref{sec:predacc}, such negligible impact was expected, because only 7\% of cases reach prefix length  35 or higher.

\runin{BPIC17} The results for this data set follow the general observations from above with some exceptions.
For RF, using a static prediction point or using the first positive prediction leads to the lowest costs for cost model parameters $(0, 0)$ and $(0.25,0)$.
However, the costs are very close to the costs of Empirical Thresholding.
Considering the standard deviation of 0.05 resp. 0.08, they fall within the variance due to the sampling of cost model parameters for Empirical Thresholding.

The BPIC17-LSTM and BPIC17-RF basically exhibit no change in prediction accuracy (see Section~\ref{sec:drifts}) and thus the rate of alarms and earliness remains roughly the same.
There is an abrupt change in accuracy at the very end, i.e., after around case \# 10,000.
Yet, the remaining cases are too few to observe how Online RL may respond to it.
As shown in Figure~\ref{fig:RQ1}\emph{(a)}, Empirical Thresholding and Online RL perform similarly for BPIC17-LSTM with a difference of 6\%.

For BPIC17-RF, Online RL outperforms Empirical Thresholding with 69\% against 19\% (see Figure~\ref{fig:RQ1}\emph{(a)}).
We attribute this to the shape of the average accuracy curve shown in Figure~\ref{fig:mcc}.
In contrast to LSTM, the accuracy radically drops after around prefix length 40.
As around 32\% of all cases have a length of 40 or longer, this means that an empirically determined threshold (which is computed considering all cases) may not be optimal for this relatively high number of remaining cases.

\runin{Traffic} The results for this data set show visible exceptions from the above general observations: (1) for LSTM, Empirical Thresholding never provides the lowest cost, while (2) for RF, Online RL never provides the lowest cost.

The reason that Online RL outperforms Empirical Thresholding for Traffic-LSTM lies in the high level of concept drift.
The data set shows visible changes in prediction accuracy, which happen between around case \# 14,000 and \# 16,000, between around case \# 31,000 and \# 33,000 and again after case \# 45,000.

\begin{figure}[hbtp]
\centering

		$
		\begin{array}{cc}
		\textrm{\textbf{Online RL}} & \textrm{\textbf{Empirical}} \\
		\multicolumn{2}{c}{\textrm{Traffic-LSTM}} \\
		\includegraphics[width=.5\textwidth]{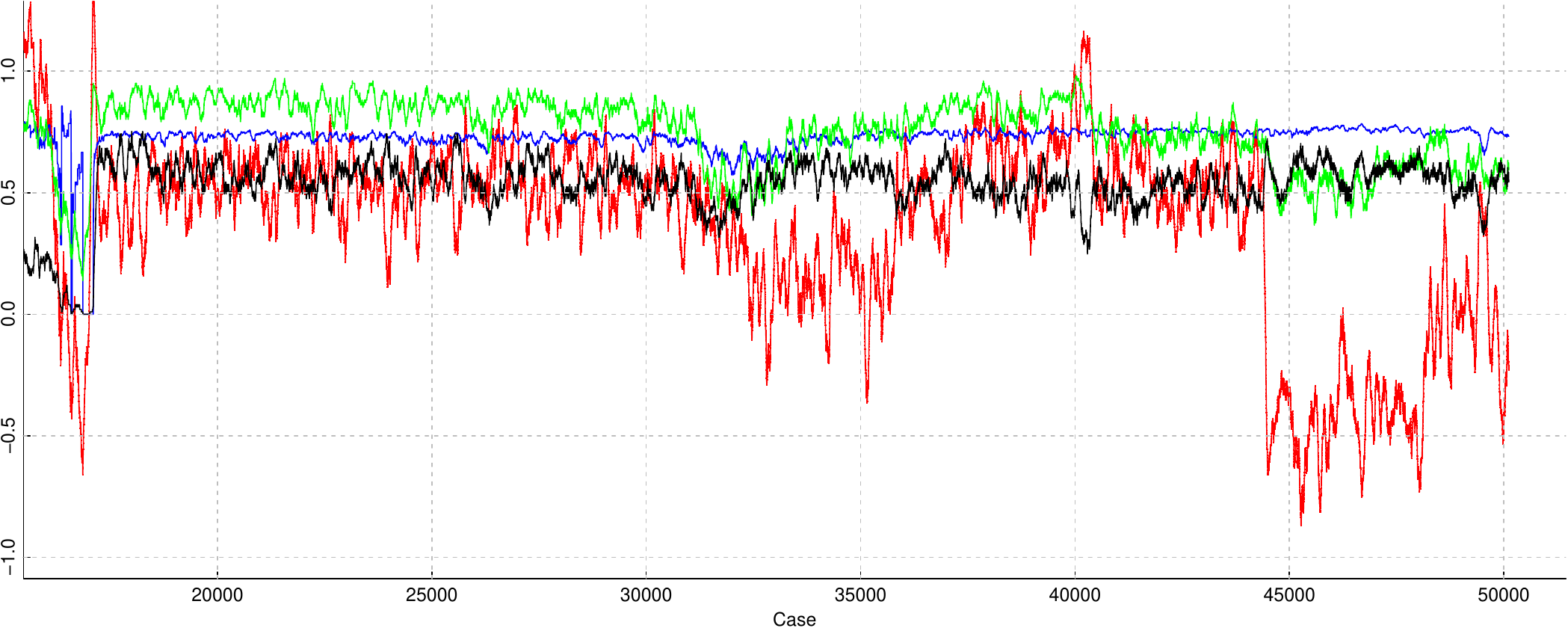} 
		& 
    	\includegraphics[width=.5\textwidth]{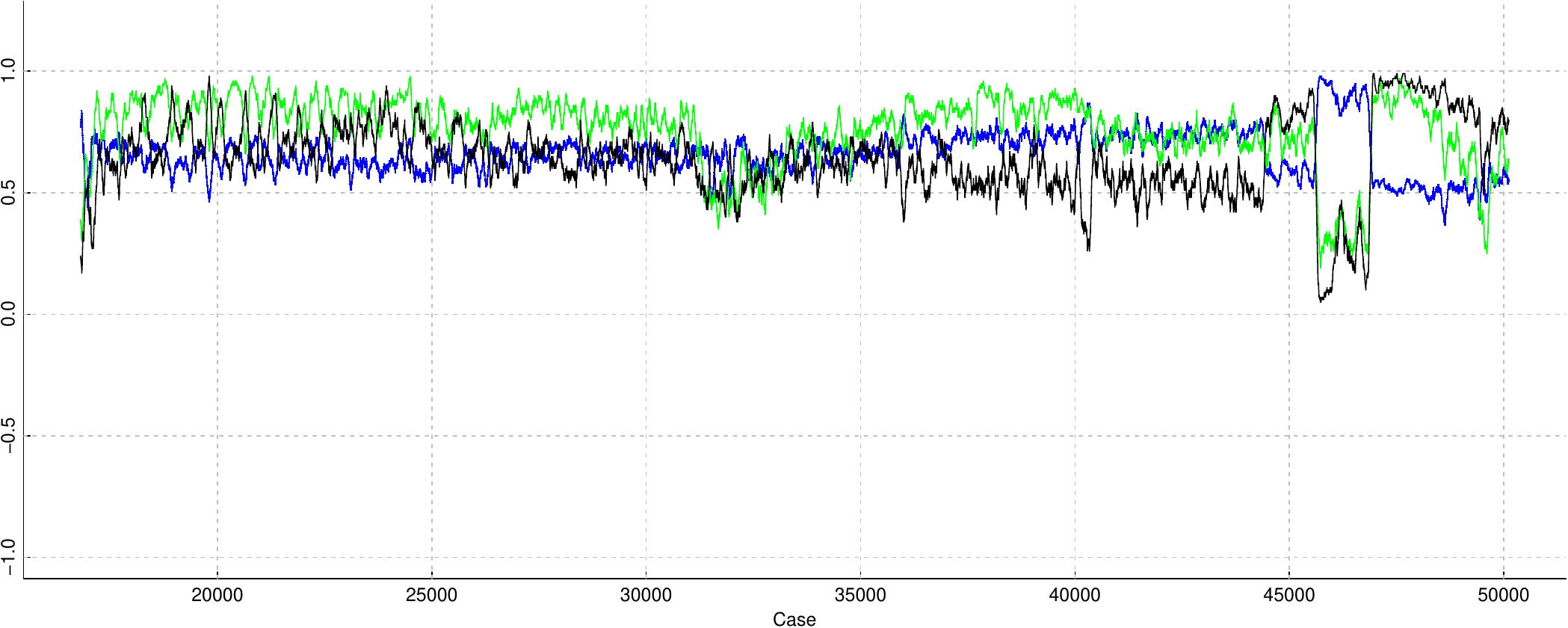} \\
	
		\end{array}$

\caption{Key characteristics of approaches: 
\textcolor{blue}{\textbf{blue}}: earliness (0 = end, 1 = beginning of process); 
\textcolor{black}{\textbf{black}}: rate of alarms; 
\textcolor{green}{\textbf{green}}: rate of accurate alarms; 
\textcolor{red}{\textbf{red}}: normalized reward (only for Online RL)}
\label{fig:RQ2a}
\end{figure}

To illustrate what happens for this specific data set, Figure~\ref{fig:RQ2a} visualizes the behavior of the two approaches\footnote{We have averaged the values for each of these metrics over the last 100 cases for reasons of visual stability and charts thus start at case \# 100.}.
For Empirical Thresholding, we chose a cost model configuration by using the average value for $\alpha_\mathrm{min}$, i.e., $0.5$.
We then use the first $\lambda$ and $\kappa$ where Empirical Thresholding outperforms the other approaches.
For Online RL, we chose the run (out of the 10 runs) that led to the highest rate of correct adaptation decisions.

As can be seen, Online RL responds to this concept drift by changing the rate of alarms and earliness.
In contrast, Empirical Thresholding responds differently.
For the first two changes in prediction accuracy, it shows a much smaller response, basically keeping earliness roughly the same, but changing slightly the rate of alarms.
For the last change in prediction accuracy, the earliness shows major fluctuations, while Online RL can keep earliness stable.
As a result, Online RL outperforms Empirical Thresholding with 84\% to 0\%.

The reason why Empirical Thresholding outperforms Online RL for Traffic-RF -- or rather why Online RL fails -- lies in the shape of the accuracy curve.
Here, after prediction point 1, the accuracy significantly decreases and only increases again after prediction point 3.
The way we formulate artificial curiosity in Online RL (see Section~\ref{sec:approach}) is that we start with the curious exploration at the later prediction points.
In this specific data set it means, that the curious exploration will be stuck in a local maximum at prefix lengths 4 and greater and will not be able to overcome the trench to reach prediction point 1.
As a result, the earliness of Online RL is low, which negatively impacts process costs. 

When analyzing the actual thresholds computed by Empirical Thresholding for LSTM-RF, it turns out that for 90\% of the cost model parameters the threshold is equal to $0.5$. 
This means that effectively there is no threshold\footnote{Due to the way that reliability estimates are computed (see Section~\ref{sec:ensemble}), all estimates are $> 0.5$.}
and that Empirical Thresholding raises an alarm for the first positive prediction in an ongoing case.
The costs for the first positive prediction approach are accordingly close to the ones of Empirical Thresholding for Traffic-RF.

\runin{Cargo} Here, Empirical Thresholding outperforms the other approaches as shown in Figure~\ref{fig:RQ1}\emph{(a)}.
The reason that Online RL works so badly for this data set lies in the small size of the data.
Cargo is only 10\% of the size of the next larger data set.
We further discuss this in Section~\ref{sec:disc}.

\section{Initial Practical Recommendations}
\label{sec:disc}

Based on the above conceptual and experimental findings, we a set of initial practical recommendations for which approach to choose in practice.
Note that these recommendations can be considered the main hypotheses supported by our comparative evaluation and thus additional empirical evidence is needed to further test the validity of these hypotheses (see Section~\ref{sec:future}).

\begin{figure}[hbtp]
\centering

		\includegraphics[width=.8\textwidth]{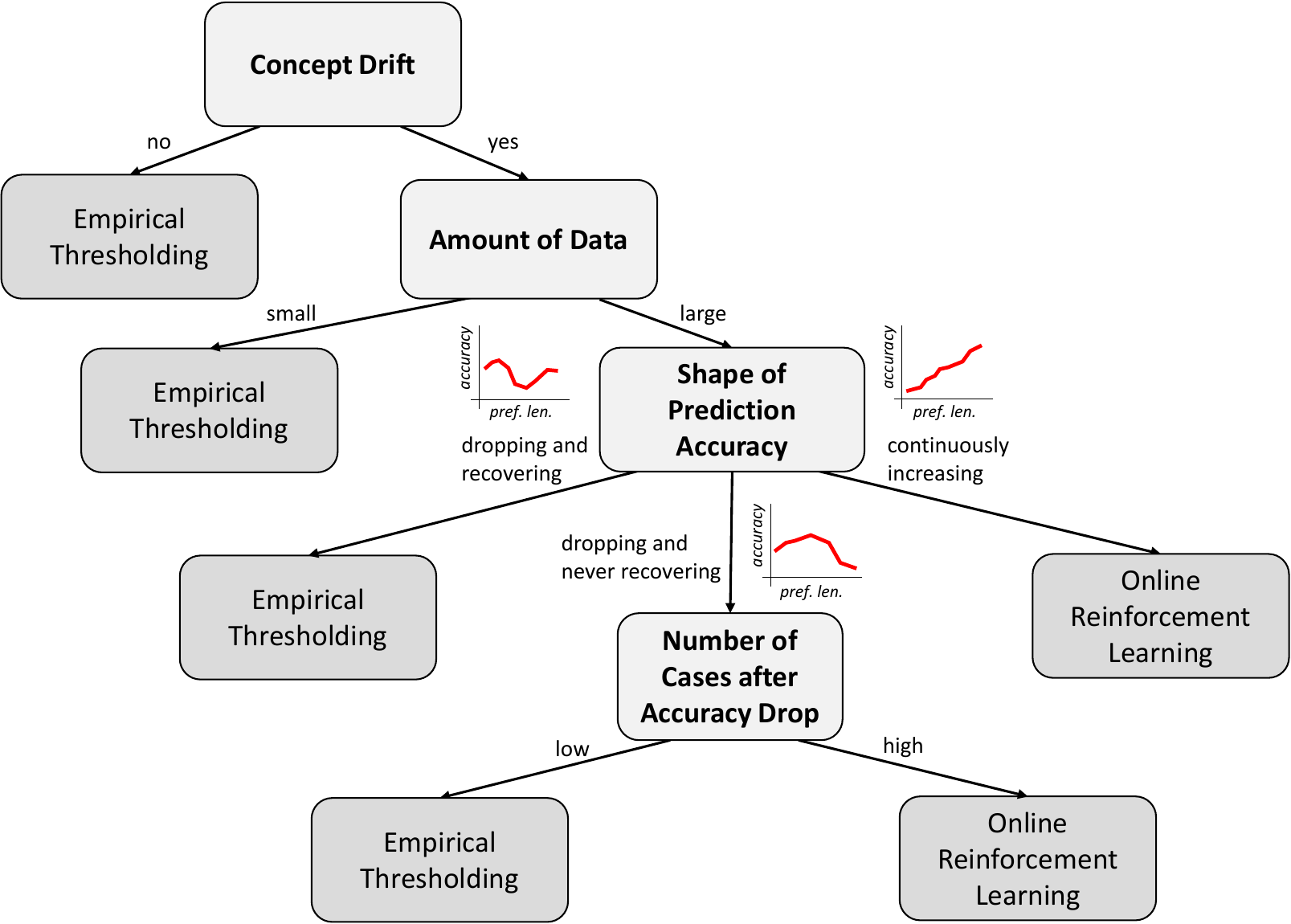} 

\caption{Decision tree providing recommendations to decide between alternative approaches}
\label{fig:rec}
\end{figure}

As Empirical Thresholding and Online RL consistently outperformed the more simplistic approaches, we provide recommendations for how to select between these two alternatives.
Figure~\ref{fig:rec} shows a decision tree, which includes key criteria to be considered when deciding which approach to deploy for a concrete process monitoring use case.
It may also be used to revisit this decision once more real-time data has been collected during process execution.

\runin{Concept Drift}
As suggested by our experimental results, Online RL should be preferred over Empirical Thresholding in the presence of concept drift and given a sufficient amount of data (see below).

One way to measure concept drift is how we did in our experiments, i.e., by considering how the prediction accuracy per case evolves along the different cases.
In a practical setting, such an analysis may be done based on a subset of the training data, or it may be done after having collected a sufficient amount of real-time data.
As prediction accuracy may be only one indicator for concept drift, another way to determine non-stationarity is to analyze the distribution of the data to determine anomalies.
One particular approach is using process drift detection, such as presented in~\cite{MaaradjiDRO15,LiuHC18}.

\runin{Amount of data}
In addition to requiring a set of training data for the prediction models, both approaches require additional data before deployment.
For Empirical Thresholding we need such additional data to determine an optimal threshold.
For Online RL we need such additional data that it can learn the principal trade-off between prediction earliness and accuracy.

Empirical Thresholding may work with a smaller amount of such additional data; e.g., in~\cite{FahrenkrogTTD19}, a training set of 20\% instead of the 33\% we used was sufficient to demonstrate cost savings for Empirical Thresholding.

\begin{figure}[hbtp]
\centering

		$
		\begin{array}{cccc}
		\textrm{BPIC12-LSTM} & \textrm{BPIC12-RF} &	 \textrm{BPIC17-LSTM} & \textrm{BPIC17-RF} \\
		\includegraphics[width=.23\textwidth]{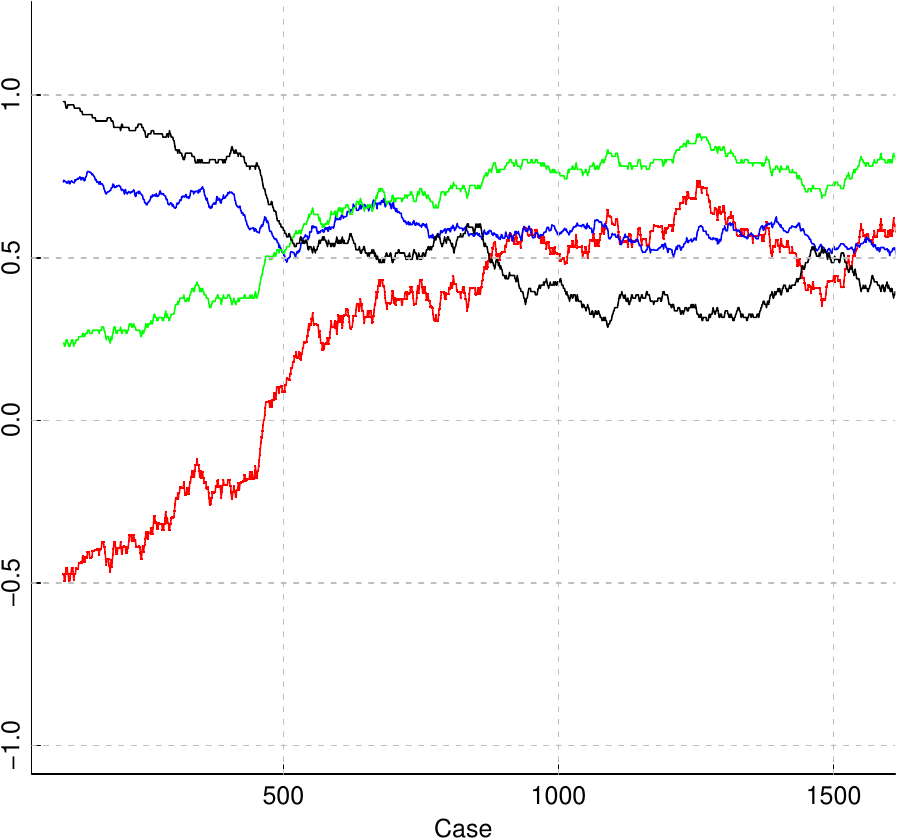} 
		& 
		\includegraphics[width=.23\textwidth]{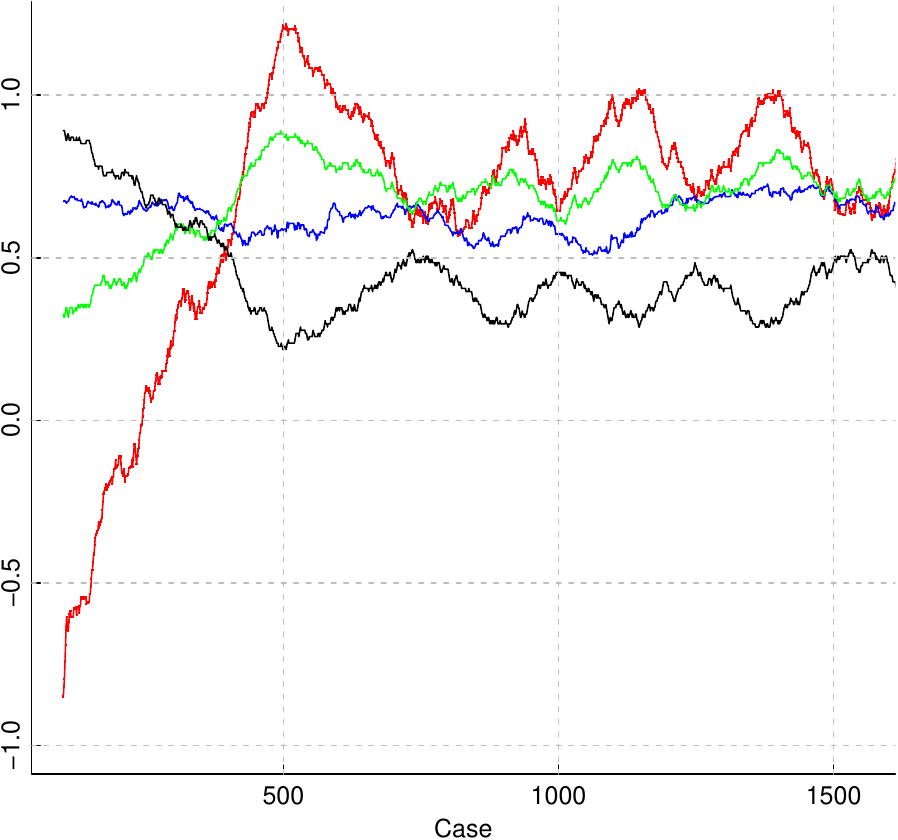} 
		&
		\includegraphics[width=.23\textwidth]{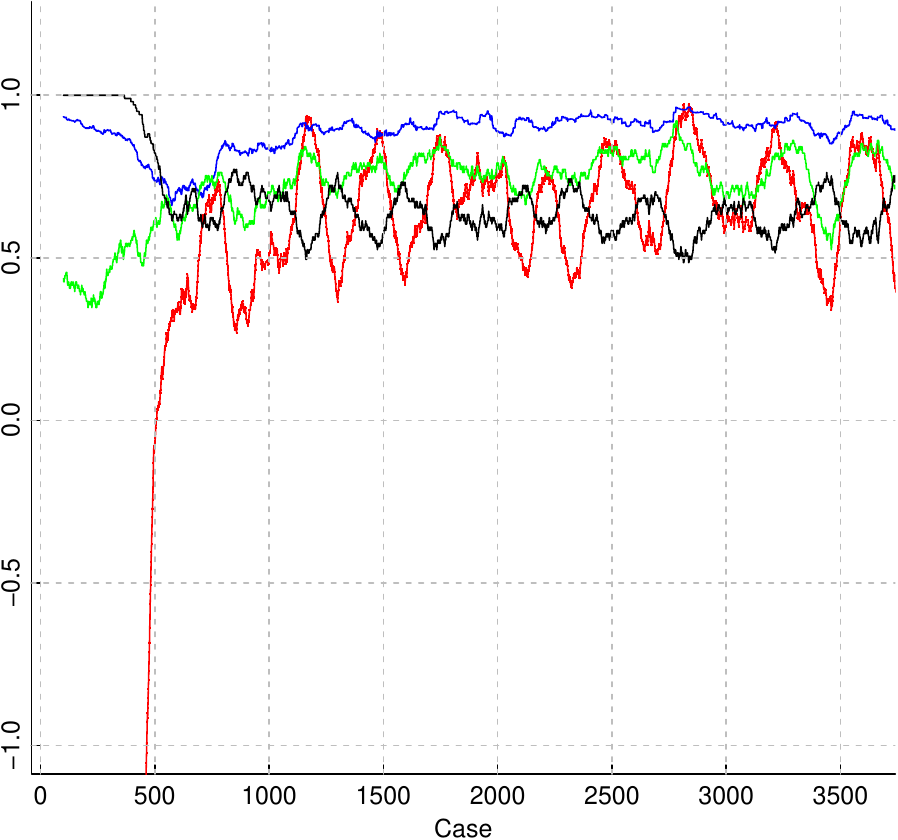} 
		&
		\includegraphics[width=.23\textwidth]{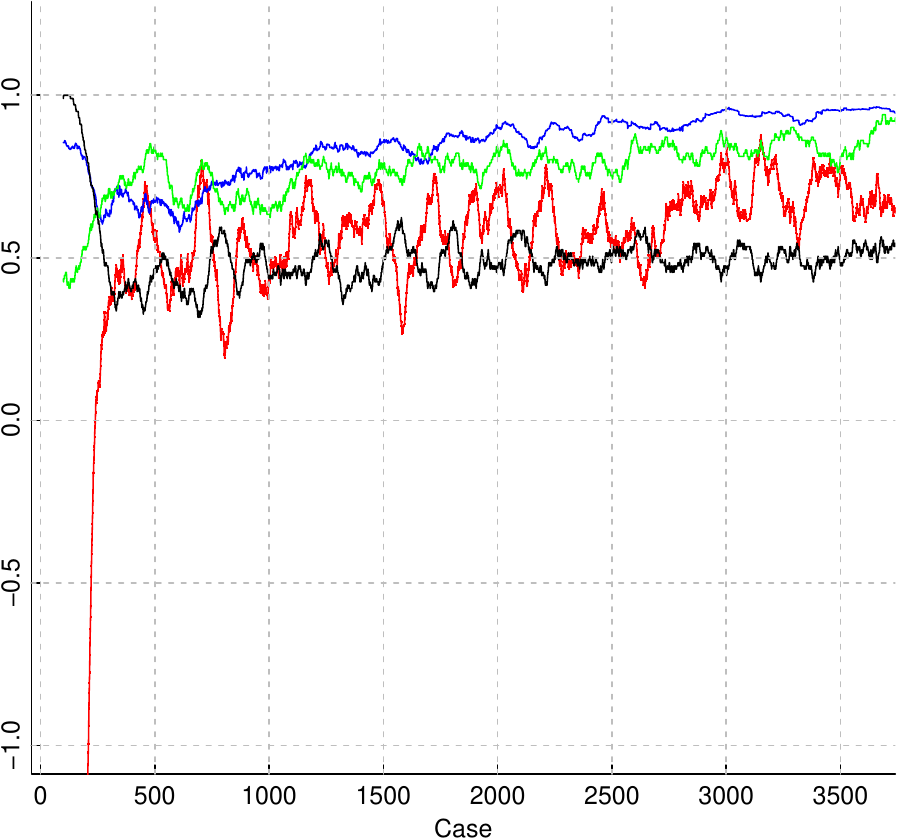} \\

		\textrm{Traffic-LSTM} & \textrm{Traffic-RF} &	 \textrm{Cargo-LSTM} & \textrm{Cargo-RF} \\
		\includegraphics[width=.23\textwidth]{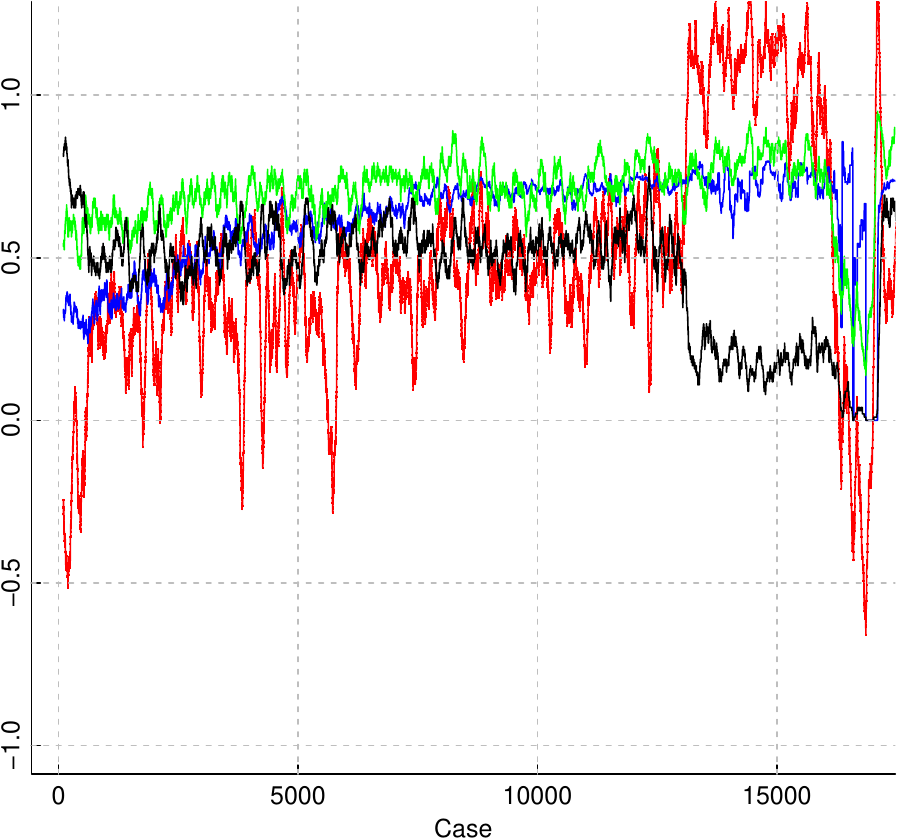} 
		& 
		\includegraphics[width=.23\textwidth]{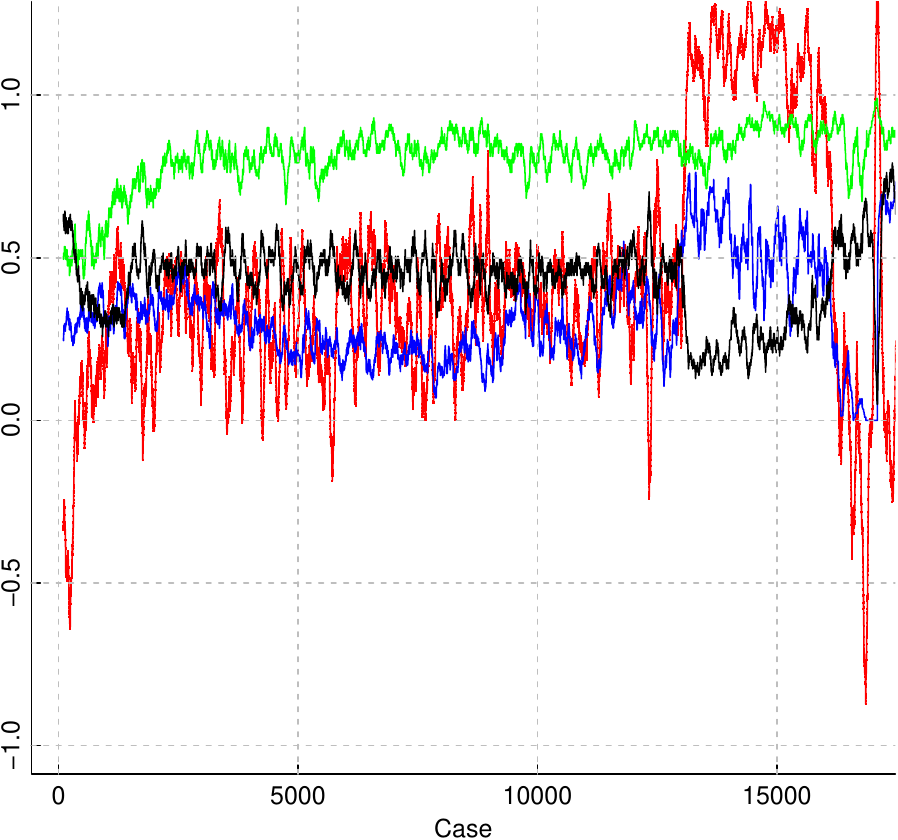} 
		&
		\includegraphics[width=.23\textwidth]{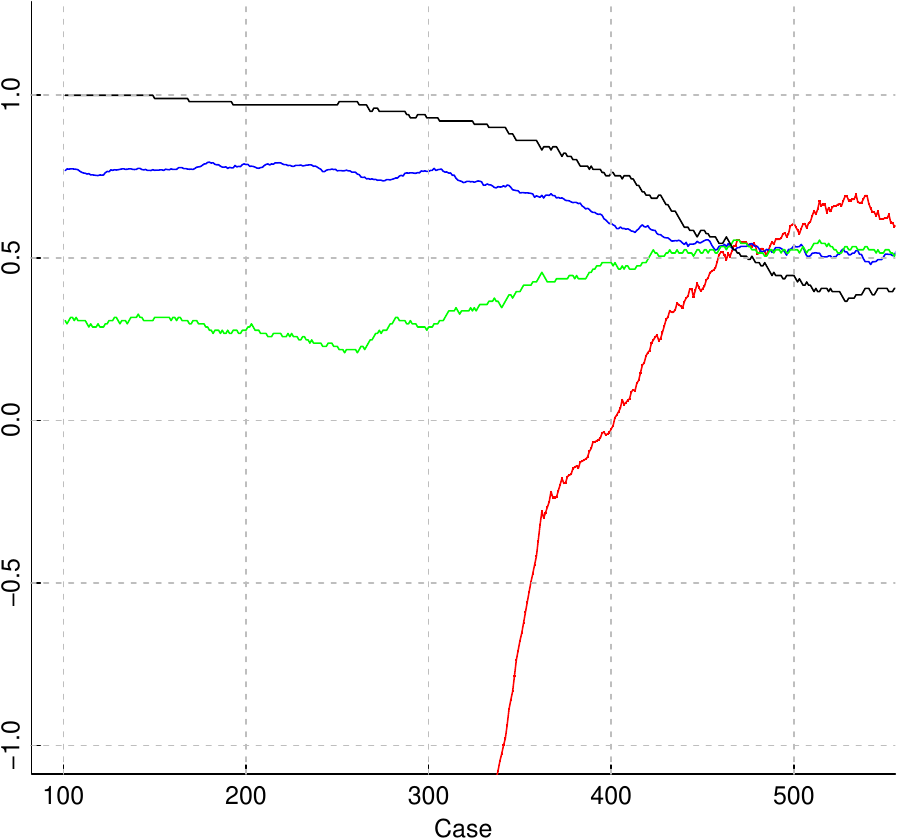} 
		&
		\includegraphics[width=.23\textwidth]{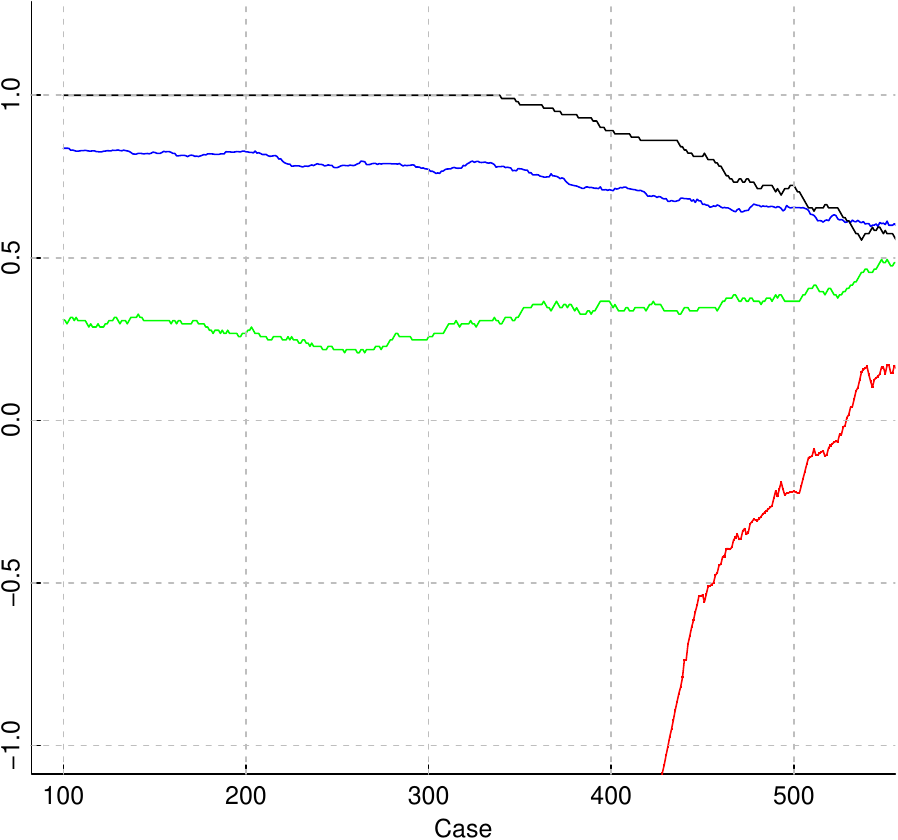} \\

		\end{array}$

\caption{Key characteristics of Online RL approach during convergence phase: 
\textcolor{red}{\textbf{red}}: normalized reward;
\textcolor{blue}{\textbf{blue}}: earliness (0 = end, 1 = beginning of process); 
\textcolor{black}{\textbf{black}}: rate of alarms; 
\textcolor{green}{\textbf{green}}: rate of accurate alarms }
\label{fig:conv}
\end{figure}

In contrast, Online RL requires a certain amount of additional data to work.
Figure~\ref{fig:conv} shows how Online RL behaves for the first 33\% of the "test" data.
Across all data sets, Online RL appears to require on average data from 600 cases to learn the basic trade-off between prediction accuracy and earliness.
In all charts, Online RL starts with a very high rate of adaptations, before learning that this may not be an optimal policy.
This means that if only a small number of cases is expected, e.g., if the business process is only very seldom invoked, Online RL most probably will not be effective.
Note that additional data requirements for Online RL may be reduced as discussed in the outlook of this paper.

The amount of data is also relevant such that Online RL can respond to concept drifts.
As we have seen, for the Cargo data set, if a concept drift affects only a small number of cases, this may not be sufficient to change the RL policy.

\runin{Shape of prediction accuracy}
If prediction accuracy continuously increases along process execution, Online RL can be applied effectively.
However, if prediction accuracy drops but never recovers, there is a high risk that Online RL will remain stuck in a local optimum.
If there is a drop in accuracy before the end of the process and accuracy remains low until the end of the process (i.e., accuracy never recovers), the choice of the approach depends on how many cases exist after the accuracy drop (see below).

Similar to measuring prediction accuracy per case (for assessing concept drift), assessing how prediction accuracy changes along process execution may be done using a subset of the training data, or using real-time data after it has been collected.

\runin{Number of cases after accuracy drop}
Whether Empirical Thresholding may be applied effectively depends on the number of cases that reach the point where accuracy drops.
If the number of cases is small, it only has a small impact on average process execution costs  even if the empirically determined threshold may not be optimal for these cases.
If the number of cases is high, Online RL may be preferred.

The number of cases can be computed, for example, from the training data set that was used for training the prediction model.
For the four data sets used in our experiments, the difference between the distribution of the training data set and the distribution of the "test" data set was only 1.5\% on average.

\runin{Computational considerations}
Deep learning, as used as part of Online RL, may raise concerns concerning computational overhead and thus infrastructure needs in practice.
However, as we are using deep RL in an online fashion, a learning step -- i.e., the update of the RL policy -- only happens once per case.
As such, the computational overhead is negligible compared to the time and resources required to execute the case.

\section{Discussion}
\label{sec:discussion}

\subsection{Validity Risks}
\label{sec:Validity}

\runin{Internal Validity} 
Concerning the extent to which our evaluation results provide a fair and unbiased comparison of the alternative approaches, we purposefully varied four independent variables in our experiments (i.e., the cost of adaptation $\lambda$, the cost of compensation $\kappa$, the effectiveness of adaptation $\alpha$ as well as the uncertainty of cost model parameters $\xi$).
Still, to keep the complexity of the experiments manageable, we did not explore all possible values and combinations of these variables.

We took great care to ensure we measure the right things.
For example, we used accuracy metrics which are robust against class imbalances.
We also made sure that our experimental results are not random.
To this end, we repeated the cost measurements for approaches that include stochastic elements (i.e., Empirical Thresholding and Online RL) $10$ times.
This number of repetitions provided a good balance between experimental effort and variance of results.


\runin{External Validity} 
Concerning the generalization of our findings, we used four real-world data sets from different application domains, which  differ in key characteristics.
In addition, we used two widely used types of predictive process monitoring techniques (LSTM and RF).

Still, the realism of our experiments and thus generalizability is limited. 
First, due to the limitations of the real-world data sets, we could only measure the effect of triggering at most one process adaptation per each ongoing case (like in ~\cite{ShoushD22}).
In reality, several adaptations may be possible during the execution of a case.

Second, we only considered a single type of alarm, i.e., we only prescribed whether to perform a proactive process adaptation or not.
In practice, one may raise different kinds of alarms that trigger different types of adaptations~\cite{FahrenkrogTTD19}.

Third, to quantify cost savings, we used a cost model from the BPM and ECTS literature.
The benefit of this cost model is that it keeps the number of independent variables in the experiments manageable.
Yet, this cost model represents an approximation of the cost situations which may be faced in practice~\cite{LeitnerHD13}.
For example, the adaptation costs may depend on the extent of the deviation, or they may depend on where in the process an adaptation is performed.
Also, we modeled adaptation effectiveness to linearly decrease as the case unfolds.
In reality, there may be different shapes; e.g., it may be that after a certain point adaptations are no longer possible~\cite{ICSOC2017}.

\subsection{Directions for Future Work}
\label{sec:future}

\runin{Handling of concept drift}
Concerning how the approaches can cope with concept drifts, one enhancement may be to exploit process drift detection and continuously re-train the prediction in an online fashion (e.g., as proposed in~\cite{BonduABCCGHLM22}), thereby continuously improving the prediction model itself.
Such an approach may then also allow empirical thresholding to better cope with concept drift.

\runin{Multiple types of alarms for Online RL} Online RL may be enhanced to raise multiple types of alarms, similar to what was proposed for Empirical Thresholding~\cite{FahrenkrogTTD19}.
This broadens the applicability of Online RL, as different types of adaptations may be triggered in different kinds of situations.
It may also be combined with causal estimations of the effect of an adaptation, as proposed by~\cite{BozorgiEA2023}, to only raise alarms if an adaptation may lead to an effective outcome at all.

\runin{Speeding up convergence of Online RL}
RL needs sufficient amount of data for convergence~\cite{BonduABCCGHLM22,BPM2020}.
As indicated by our experiments, Online RL requires the data of around 600 cases to learn how to handle the basic trade-off between prediction accuracy and earliness.
One reason is that Online RL has to learn even simple relationships from scratch; e.g.,  raising an alarm at the very end of a case may not any longer allow performing an adaptation.
As a result, it will take some time after Online RL is deployed so that it learns to accurately raise alarms.
One promising direction to address this limitation is to leverage the concept of Meta-RL~\cite{WangKSLTMBKB17}.
Meta-RL facilitates reusing knowledge from similar learning problems for the current learning problem.
Other directions are using offline pre-training using, e.g., synthetic data sets, as proposed by~\cite{BozorgiEA2023}.

\runin{Determining alternative process outcomes after adaptation}
Many of the approaches that aim to reconcile the trade-off between accuracy and earliness (also see Section~\ref{sec:sota}), need to assess whether the triggering of an adaptation was correct by determining the alternative process outcome if that adaptation were not executed (i.e., in other words by knowing the true process outcome without intervention).
As we discussed above, knowing such alternative process outcome once the process has been adapted is not feasible in general, as it would require an accurate and reliable what-if business process analysis~\cite{Dumas21}.
We proposed artificial curiosity as a solution, others suggested using causal inference~\cite{ShoushD22}.
A further idea may be to train prediction models that also receive information about adaptations of historic cases as input and use these prediction models to derive a probability distribution over possible outcomes.

\runin{Practical recommendations} 
This paper derived a set of initial recommendations for which approach to choose in practice.
These recommendations where supported by theoretical discussions as well as empirical data.
Yet, to further substantiate these recommendations, experiments with more data sets and ideally real-live case studies should be performed.
Also, additional controlled experiments are needed, e.g., to perform a sensitivity analysis on how concept drift effects the performance of the different approaches.

\runin{Explainability of alarms} 
One important direction is enriching the information presented to process managers when an alarm is raised.
While reliability estimates may serve as additional information (as reported in~\cite{MetzgerF021}), reliability estimates only provide little insight on why the alarm was raised.
As pointed out by Miller "probabilities probably don't matter"~\cite{Miller19}.
As a result, process managers may put little trust in the alarms raised.
An interesting avenue to pursue is thus enhancing prescriptive business process monitoring approaches with the capability to explain their alarms.
Here, one may build on previous work on explainable process monitoring~\cite{HuangMP21} and explainable reinforcement learning~\cite{FeitMP22}.

\section{Related Work}
\label{sec:sota}

We discuss related work along the following, complementary aspects: (1) trade-off between prediction accuracy and earliness both in BPM and in time series research, (2) prescriptive process monitoring, and (3) \rl\ for business process management.

\subsection{Trade-off between Prediction Accuracy and Earliness in BPM}
\label{sec:sota-balance}

Comprehensive overviews of predictive process monitoring approaches are gi\-ven in~\cite{NeuLF22,TeinemaaDRM19,VerenichMM19,Francescomarino18,PollPRRR18,Marquez-Chamorro_2017}.
Here we discuss the approaches that explicitly consider the trade-off between prediction accuracy and earliness.

One group of works uses prediction earliness as a dependent variable in the analysis of prediction accuracy.
This means one evaluates the prediction models separately for each prefix length. 
The prediction model is applied to a subset of prefixes of exactly the given length. 
The improvement of prediction accuracy as the prefix length increases provides an implicit notion of earliness~\cite{TeinemaaDRM19}. 
As an example, Leontjeva et al.~\cite{LeontjevaCFDM15} exploit the data payload of process events to increase prediction earliness.
Teinemaa et al.~\cite{TeinemaaDRM19}, and we in our earlier work~\cite{MetzgerN18} measured the accuracy of different prediction techniques for the different prediction points along process execution.
Results presented in the aforementioned works clearly show the trade-off between prediction earliness and accuracy.
However, it was left open how to resolve this trade-off and how to use the results to facilitate prescriptive process monitoring.

Another group of works uses reliability estimates to filter among more or less reliable predictions.
This means one keeps on monitoring each case until the prediction model gives a prediction with sufficiently high reliability.
Earliness is then measured as the average prefix length when such a prediction was made~\cite{TeinemaaDRM19}.
Maggi et al.~\cite{Maggi_Dumas14} use decision tree learning for predictive process monitoring.
They use class probabilities of decision trees and analyze how selecting predictions using class probabilities impacts  average prediction accuracy and earliness.
Di Francescomarino et al.~\cite{Francescomarino_Dumas16} employ random forests for prediction.
Similar to Maggi et al. they analyze how selecting predictions using class probabilities (of random forests) impacts average prediction accuracy.
They observe that using class probabilities may improve average prediction accuracy, but at the expedient of loosing predictions that are below a given probability threshold.
Yet, they do not analyze to what extent using these class probabilities may facilitate prescriptive business process monitoring.
Teinemaa et al. investigate whether unstructured data may increase prediction earliness and accuracy~\cite{TeinemaaDMF16}.
Francescomarino et al. investigate in how far hyper-parameter optimization~\cite{Francescomarino_Dumas16} and clustering~\cite{FrancescomarinoDMT19} can improve earliness.

\subsection{Early Time Series Classification}
\label{sec:ects}
As explained in Section~\ref{sec:Introduction}, a similar trade-off between accuracy and earliness is investigated for time series classification.
Gupta et al. provide a comprehensive review of the literature on \emph{Early Classification of Time Series} (\emph{ECTS}).

One principal difference between ECTS and process prediction is the type of the underlying data.
While time series data represent values of single or multiple variables at different points in time (such as temperatures or stock prices), process monitoring data represent sequences of process events.
A process event is characterized by a timestamp, which indicates the occurrence of the event, such as the completion of a process activity.
Like in time series, the events and thus timestamps form an ordered sequence.
However, in contrast to time series events are not typically happening at equal-space time intervals~\cite{GuptaGBD20}.
In addition to a timestamp, an event includes an event type (uniquely identifying the process step) and may include additional attributes of the event~\cite{Marquez-Chamorro_2017,TeinemaaDMF16}.
As such, there is an important semantic difference between a data point in a time series and a process event.

Recent work indicates how to generalize from ECTS to a more data-type-agnostic approach for early decision making using machine learning~\cite{BonduABCCGHLM22}.
While data streams are mentioned (which could be considered somewhat similar to event log data in BPM), the specifics of ECTS for data streams are not elaborated. 
Our paper can be considered to provides such additional specifics.

Independent from these principal differences, approaches for ECTS and prescriptive process monitoring share many similarities.
We introduced some of them already in Section~\ref{sec:approaches}, but add further ones following taxonomy provided in \cite{GuptaGBD20} below.

\runin{Prefix-based}
Prefix-based approaches conceptually follow the same solution as the static prediction point explained in Section~\ref{sec:approaches}.
A decision is made at the minimum prediction length or minimum required length for each time series, determined using dedicated training data~\cite{GuptaGBD20}.
For example, Xing et al.~\cite{XingPY12} augment unsupervised learning (1-Nearest-Neighbor) by adding an initial training process.

\runin{Shapelet-based} 
The aim of shapelet-based approaches is to find a set of key patterns in the training data set and use them as discriminatory features of the time series~\cite{GuptaGBD20}.
This approach is not directly applicable for prescriptive process monitoring, because it is very specific to time series data.
Shapelet-based approaches require a series of real-valued data to determine the different patterns on how the values develop over time~\cite{BonduABCCGHLM22}. 

\runin{Model-based} 
Model-based approaches use discriminative classifiers, i.e., classifiers that provide a probability or reliability estimate together with the actual prediction~\cite{GuptaGBD20}.
For example, Mori et al. use probabilistic classifiers to produce a class label for a time series as soon as the probability at a checkpoint exceeds a class-dependent threshold~\cite{Mori_ea_17-mpl}, and Hatami and Chira use ensemble models consisting of probabilistic classifiers~\cite{HatamiC13}.

One particular sub-class are so-called \emph{non-myopic} approaches~\cite{achenchabe2021early,DachraouiBC15}.
They work by estimating the best timestep $\tau*$ for a decision and then taking a decision when this timestep is reached. 
Non-myopic approaches do so by predicting the continuation of the time series and using this continuation to estimate $\tau*$. 
As discussed in~\cite{FahrenkrogTTD19}, non-myopic approaches assume a-priori knowledge of the length of the sequence, which is not given in BPM. 
Thereby, these approaches come with the risk that the running process will end before $\tau*$ is reached.

\runin{Miscellaneous (which includes Reinforcement Learning)} 
RL-based approaches for ECTS were presented by Bondu et al., who propose using value-based RL~\cite{BonduABCCGHLM22} to learn the trade-off between accuracy and earliness using as reward function a cost function giving costs for different contingencies.
Martinez et al. use value-based deep RL, in particular the DDQN algorithm, to learn this trade-off~\cite{MartinezRPR20}.
Both approaches assume that one can assess whether an adaptation was correct by determining the alternative process outcome if that adaptation were not executed.
As discussed throughout this paper, this poses an important limitation for the practical application of these approaches.
In addition, both approaches use value-based RL.
As we discuss in Section~\ref{sec:RL}, compared to policy-based RL (which we use in our Online RL approach), value-based RL requires determining  how to balance exploitation and exploration to capture concept drifts~\cite{CAISE2020,BPM2020}.

\subsection{Prescriptive Process Monitoring}
\label{sec:sota-prescriptive}
As introduced in Section~\ref{sec:Introduction}, existing research addresses two closely related, complementary aspects of prescriptive process monitoring.
One aspect is concerned with answering the question "\emph{how to intervene?}".
Recent work along this dimension includes~\cite{ShoushD22,BozorgiEA2023,LeoniDR20,WeinzierlZSMP20,MehdiyevF20}.
The other aspect is concerned with answering the question "\emph{when to intervene?}", thereby providing the backbone for answering the first questions~\cite{ShoushD22}.
Our contribution focuses on the question of "\emph{when}", which we thus discuss  below.

Teinemaa et al. were among the first to introduce the concept of alarm-based prescriptive process monitoring~\cite{TeinemaaTLDM18}.
They use class probabilities generated by random forests as reliability estimates to determine whether to raise an alarm to trigger a proactive adaptation.
They use Empirical Thresholding to determine reliability thresholds above which alarms are raised.
Follow-up work by Fahrenkrog-Petersen et al. extends this initial work in particular with the capability of raising multiple types of alarms~\cite{FahrenkrogTTD19}.
The benefit of Empirical Thresholding pursued by these papers is that it ensures that the threshold is optimal for the specific training data used and the given cost model.
Yet, as we have seen above, the threshold may not remain optimal over time due to concept drift.

Shoush and Dumas propose a prescriptive process monitoring approach that explicitly considers whether resources are available for performing an adaptation as well as whether it may be beneficial to delay the adaptation~\cite{ShoushD22}.
Their approach prioritizes the adaptations across a set of ongoing cases, thereby addressing the "infinite capacity" assumption concerning available resources.
As such it sketches an interesting path to enhance the approaches we analyzed in this paper.
To overcome the problem that the alternative process outcome after an adaptation is not known in general (see Section~\ref{sec:Introduction}), they use causal inference to predict the process outcome for a given adaptation as an additional input for their approach.
Thereby, they provide an alternative solution to the problem, for which we applied artificial curiosity.
In follow-up work, Shoush and Dumas introduce the use of conformal prediction to give confidence guarantees instead of only providing estimates of the prediction reliability~\cite{ShoushD22}.
While their approach is intrinsically explainable, it requires defining a suitable non-conformity measure and calibration (as we discussed in~\ref{sec:ensemble}).

In our earlier work, we used reliability estimates computed from ensembles of multi-layer perceptrons to decide on proactive adaptation~\cite{CAISE2017,ICSOC2017}.
If the reliability of a given prediction is equal to or greater than a predefined threshold, the prediction is used to trigger a proactive adaptation.
In~\cite{CAISE2017} we focused on ensembles of classification models.
In~\cite{ICSOC2017} we extended this work to ensembles of regression models, thereby also including the extent of a predicted deviation in the adaptation decision.
Yet, in this earlier work we used a static prediction point (the 50\% mark of process execution), and thus did not consider the aspect of prediction earliness.

In~\cite{MetzgerF021,CAISE2019}, we use reliability estimates computed from ensembles of LSTM models to dynamically decide on proactive adaptation.
We determine during an ongoing case the earliest prediction with sufficiently high reliability and use this prediction to trigger a proactive adaptation.
However, this previous approach required the explicit definition of a reliability threshold.


\subsection{Reinforcement Learning in BPM}
\label{sec:sota-rl}

In the literature, RL approaches in the context of BPM were proposed for different main purposes.
Huang et al. employ RL for the dynamic optimization of resource allocation in operational business processes~\cite{HuangALD11}.
However, they do not consider proactive adaptation of processes at run time.
Also, they use Q-Learning as a classical RL algorithm, and thus assume the environment can be represented by a finite, discrete set of states.
As mentioned above, Online RL does not have this limitation, as it can directly handle large and continuous environments, as well as can deal with the non-stationarity of these environments and thereby concept drifts affecting the machine learning models.

Satyal et al. propose solving the problem of deciding how new instances are assigned to a specific version of a process by modeling it as a multi-armed bandit problem~\cite{SatyalWPCM19}.
A multi-armed bandit problem can be considered a simple variant of RL, which only takes one-shot decisions and not sequential decisions.
They observe that reward engineering needs to be done carefully, and especially that the reward function must provide a strong enough signal for effective learning.
We similarly have elaborated on the reward engineering concern for Online RL.

Silvander proposes using Q-Learning with function approximation via a deep neural network (DQN) for the optimization of business processes~\cite{Silvander19}.
In contrast to policy-based RL used in Online RL, Q-Learning faces the exploration-exploitation dilemma~\cite{sutton2018reinforcement,CAISE2020}. 
To optimize rewards, RL should \emph{exploit} actions that have shown to be effective. 
However, to discover such actions in the first place, actions that were not selected before should be \emph{explored}. 
One typical solution to the exploration-exploitation dilemma is the $\epsilon$-greedy mechanism. 
During learning, $\epsilon$-greedy chooses a random action with probability $\epsilon$.
Resolving the dilemma means finding a balance between exploitation and exploration to facilitate the convergence of the learning process. 
To this end, Silvander suggests defining an $\epsilon$ decay rate, to reduce the amount of exploration over time.
However, he does not consider using RL at run time, and thus does not take into account how to increase the rate of exploration in the presence of concept drifts.

Branchi et al. propose using RL for prescriptive business process monitoring~\cite{BranchiFG22}.
As an RL algorithm they use policy iteration with Monte Carlo methods to determine the best next action to execute for optimizing process KPIs, such as revenue or costs.
The chosen RL algorithm belongs to the class of model-based algorithms, which require an explicit model of the environment.
The authors approximate such an environment model by mining it from event log data.
As they focus on \emph{which }action to execute next, they complement our work on \emph{when }to perform an adaptation.

In our previous work, we proposed a generic framework and implementation for using policy-based RL for self-adaptive information systems~\cite{CAISE2020}.
In follow-up work, we customized this framework specifically for the problem of generating alarms~\cite{BPM2020}.
In particular, we integrated this framework with our work on predictive process monitoring and presented initial promising results for typical BPM benchmark data sets~\cite{CAISE2019}.
Here, we relax the fundamental assumption about knowing the alternative process outcome after an adaptation (e.g., see discussion in Sections~\ref{sec:Introduction} and~\ref{sec:approach}).

Recent work by Bozorgi~\cite{BozorgiEA2023} enhances our previous work on Online RL ~\cite{BPM2020} along three main directions: 
First, they use causal effect estimations to determine the effectiveness of an adaptation and only raise an alarm if an adaptation may indeed lead to an effective outcome.
Second, they pre-train the RL agent in an offline setting using simulated data, thereby addressing the problem of low initial performance of online RL (see discussion in Section~\ref{sec:future}).
Third, to further speed up convergence of the learning process, they introduce conformal prediction (see Section~\ref{sec:sota-prescriptive}), such that the RL agent is able to avoid cases that most likely will end up in a successful outcome anyways.

\section{Conclusion}
\label{sec:Conclusion}

We presented a comparative evaluation of the main alternative approaches for automatically reconciling the trade-off between prediction accuracy and earliness in prescriptive business process monitoring.
The experimental results using four real-world event log data sets and two types of prediction models suggest that Empirical Thresholding and Online RL, which both leverage different forms of machine learning, outperform more traditional approaches, which do not leverage machine learning.
Yet, these approaches perform differently in different situations.
We thus provided initial recommendations for how to select a concrete approach in practice and identified several directions for future work on prescriptive business process monitoring.
Overall, our work contributes to the emerging research area of AI-Augmented Business Process Management~\cite{ABPM} by paving the way toward automated process adaptation.
Also, by connecting with the work on early classification of time series, we provide input to machine-learning-based early decision-making research~\cite{BonduABCCGHLM22}.

\vspace{2em}

\noindent\textbf{Acknowledgments.}
We cordially thank Fabiana Fournier, Rod Franklin, and Lior Limonad for their comments on earlier drafts of the paper, Stephan Fahrenkrog-Petersen and the participants of the Dagstuhl seminar on ``Software Engineering for AI-ML-based Systems`` for inspiring discussions, as well as our students Tobias Braun, Maximilian Brenke, Philip Heiber, Tobias Miosczka, Norman Reddig, and Philipp Wallutis for their support in preparing and carrying out the experiments.
Research leading to these results  received funding from the EU's Horizon 2020 R\&I programme under grant agreements 731932 (TransformingTransport), 732630 (BDVe),  and 871493 (DataPorts).





\bibliographystyle{elsarticle-num}
\bibliography{main}

\begin{thebibliography}{10}
\expandafter\ifx\csname url\endcsname\relax
  \def\url#1{\texttt{#1}}\fi
\expandafter\ifx\csname urlprefix\endcsname\relax\def\urlprefix{URL }\fi
\expandafter\ifx\csname href\endcsname\relax
  \def\href#1#2{#2} \def\path#1{#1}\fi

\bibitem{KubrakMND22}
K.~Kubrak, F.~Milani, A.~Nolte, M.~Dumas, Prescriptive process monitoring:
  \emph{Quo vadis}?, PeerJ Comput. Sci. 8 (2022) e1097.

\bibitem{FrancescomarinoG22}
C.~D. Francescomarino, C.~Ghidini, Predictive process monitoring, in: W.~M.~P.
  van~der Aalst, J.~Carmona (Eds.), Process Mining Handbook, Vol. 448 of LNBIP,
  Springer, 2022, pp. 320--346.

\bibitem{NeuLF22}
D.~A. Neu, J.~Lahann, P.~Fettke, A systematic literature review on
  state-of-the-art deep learning methods for process prediction, Artif. Intell.
  Rev. 55~(2) (2022) 801--827.

\bibitem{TeinemaaDRM19}
I.~Teinemaa, M.~Dumas, M.~L. Rosa, F.~M. Maggi, Outcome-oriented predictive
  process monitoring: Review and benchmark, {TKDD} 13~(2) (2019) 17:1--17:57.

\bibitem{VerenichMM19}
I.~Verenich, M.~Dumas, M.~L. Rosa, F.~M. Maggi, I.~Teinemaa, Survey and
  cross-benchmark comparison of remaining time prediction methods in business
  process monitoring, ACM Trans. Intell. Syst. Technol. 10~(4) (2019)
  34:1--34:34.

\bibitem{Marquez-Chamorro_2017}
A.~E. M{\'{a}}rquez{-}Chamorro, M.~Resinas, A.~Ruiz{-}Cort{\'{e}}s, Predictive
  monitoring of business processes: {A} survey, {IEEE} Trans. Serv. Comput.
  11~(6) (2018) 962--977.

\bibitem{NunesSWR18}
V.~T. Nunes, F.~M. Santoro, C.~M.~L. Werner, C.~G. Ralha, Real-time process
  adaptation: {A} context-aware replanning approach, {IEEE} Trans. Systems,
  Man, and Cybernetics: Systems 48~(1) (2018) 99--118.

\bibitem{WeberSR09}
B.~Weber, S.~W. Sadiq, M.~Reichert, Beyond rigidity - dynamic process lifecycle
  support, Computer Science - R{\&}D 23~(2) (2009) 47--65.

\bibitem{CAISE2017}
A.~Metzger, F.~F{\"o}cker, Predictive business process monitoring considering
  reliability estimates, in: E.~Dubois, K.~Pohl (Eds.), {CAiSE} 2017, Vol.
  10253 of LNCS, Springer, 2017, pp. 445--460.

\bibitem{ParkS2019}
G.~Park, M.~Song, Prediction-based resource allocation using {LSTM} and minimum
  cost and maximum flow algorithm, in: International Conference on Process
  Mining, {ICPM} 2019, Aachen, Germany, June 24-26, 2019, {IEEE}, 2019, pp.
  121--128.

\bibitem{PollPRRR18}
R.~Poll, A.~Polyvyanyy, M.~Rosemann, M.~R{\"{o}}glinger, L.~Rupprecht, Process
  forecasting: Towards proactive business process management, in: M.~Weske,
  M.~Montali, I.~Weber, J.~vom Brocke (Eds.), {BPM} 2018, Vol. 11080 of LNCS,
  Springer, 2018, pp. 496--512.

\bibitem{TeinemaaTLDM18}
I.~Teinemaa, N.~Tax, M.~de~Leoni, M.~Dumas, F.~M. Maggi, Alarm-based
  prescriptive process monitoring, in: M.~Weske, M.~Montali, I.~Weber, J.~vom
  Brocke (Eds.), Business Process Management Forum - {BPM} Forum 2018, Sydney,
  NSW, Australia, September 9-14, 2018, Proceedings, Vol. 329 of LNBIP,
  Springer, 2018, pp. 91--107.

\bibitem{Gutierrez_2013}
A.~Gutierrez, C.~{Cassales Marquezan}, M.~Resinas, A.~Metzger,
  A.~Ruiz-Cort{\'e}s, K.~Pohl, Extending {WS-Agreement} to support automated
  conformity check on transport \& logistics service agreements, in: S.~Basu,
  {et al.} (Eds.), ICSOC 2013, Berlin, Germany, Vol. 8274 of LNCS, Springer,
  2013, pp. 567--574.

\bibitem{BozorgiEA2023}
Z.~D. Bozorgi, I.~Teinemaa, M.~Dumas, M.~L. Rosa, A.~Polyvyanyy, Prescriptive
  process monitoring based on causal effect estimation, Information Systems
  (2023).

\bibitem{FahrenkrogTTD19}
S.~A. Fahrenkrog{-}Petersen, N.~Tax, I.~Teinemaa, M.~Dumas, M.~de~Leoni, F.~M.
  Maggi, M.~Weidlich, Fire now, fire later: alarm-based systems for
  prescriptive process monitoring, Knowl. Inf. Syst. 64~(2) (2022) 559--587.

\bibitem{LeoniDR20}
M.~de~Leoni, M.~Dees, L.~Reulink, Design and evaluation of a process-aware
  recommender system based on prescriptive analytics, in: B.~F. van Dongen,
  M.~Montali, M.~T. Wynn (Eds.), 2nd International Conference on Process
  Mining, {ICPM} 2020, Padua, Italy, October 4-9, 2020, {IEEE}, 2020, pp.
  9--16.

\bibitem{WeinzierlZSMP20}
S.~Weinzierl, S.~Zilker, M.~Stierle, M.~Matzner, G.~Park, From predictive to
  prescriptive process monitoring: Recommending the next best actions instead
  of calculating the next most likely events, in: N.~Gronau, M.~Heine,
  H.~Krasnova, K.~Poustcchi (Eds.), 15. Internationale Tagung
  Wirtschaftsinformatik, {WI} 2020, Potsdam, Germany, March 9-11, 2020, {GITO}
  Verlag, 2020, pp. 364--368.

\bibitem{MehdiyevF20}
N.~Mehdiyev, P.~Fettke, Prescriptive process analytics with deep learning and
  explainable artificial intelligence, in: F.~Rowe, R.~E. Amrani, M.~Limayem,
  S.~Newell, N.~Pouloudi, E.~van Heck, A.~E. Quammah (Eds.), 28th European
  Conference on Information Systems, {ECIS} 2020, Marrakech, Morocco, June
  15-17, 2020, 2020, pp. 1--20.

\bibitem{ShoushD2022}
M.~Shoush, M.~Dumas, Intervening with confidence: Conformal prescriptive
  monitoring of business processes, CoRR abs/2212.03710 (2022).

\bibitem{DonadelloEA2022}
I.~Donadello, C.~D. Francescomarino, F.~M. Maggi, F.~Ricci, A.~Shikhizada,
  Outcome-oriented prescriptive process monitoring based on temporal logic
  patterns, CoRR abs/2211.04880 (2022).

\bibitem{CAISE2019}
A.~Metzger, A.~Neubauer, P.~Bohn, K.~Pohl, Proactive process adaptation using
  deep learning ensembles, in: P.~Giorgini, B.~Weber (Eds.), CAiSE 2019, Vol.
  11483 of LNCS, Springer, 2019, pp. 547--562.

\bibitem{LeitnerFHD13}
P.~Leitner, J.~Ferner, W.~Hummer, S.~Dustdar, Data-driven and automated
  prediction of service level agreement violations in service compositions,
  Distributed and Parallel Databases 31~(3) (2013) 447--470.

\bibitem{MorenoCGS18}
G.~A. Moreno, J.~C{\'{a}}mara, D.~Garlan, B.~R. Schmerl, Flexible and efficient
  decision-making for proactive latency-aware self-adaptation, ACM Trans.
  Autonomous and Adaptive Systems 13~(1) (2018) 3:1--3:36.

\bibitem{MetzgerF021}
A.~Metzger, J.~Franke, T.~Jansen, Ensemble deep learning for proactive terminal
  process management at the port of duisburg "duisport", in: J.~vom Brocke,
  J.~Mendling, M.~Rosemann (Eds.), Business Process Management Cases Vol. 2,
  Digital Transformation - Strategy, Processes and Execution, Springer, 2021,
  pp. 153--164.

\bibitem{Metzger_et_al_2015}
A.~Metzger, P.~Leitner, D.~Ivanovic, E.~Schmieders, R.~Franklin, M.~Carro,
  S.~Dustdar, K.~Pohl, Comparing and combining predictive business process
  monitoring techniques, {IEEE} Trans. Syst. Man Cybern. Syst. 45~(2) (2015)
  276--290.

\bibitem{Francescomarino_Dumas16}
C.~D. Francescomarino, M.~Dumas, M.~Federici, C.~Ghidini, F.~M. Maggi,
  W.~Rizzi, Predictive business process monitoring framework with
  hyperparameter optimization, in: S.~Nurcan, P.~Soffer, M.~Bajec, J.~Eder
  (Eds.), {CAiSE} 2016, Vol. 9694 of LNCS, Springer, 2016, pp. 361--376.

\bibitem{BPM2020}
A.~Metzger, T.~Kley, A.~Palm, Triggering proactive business process adaptations
  via online reinforcement learning, in: D.~Fahland, C.~Ghidini, J.~Becker,
  M.~Dumas (Eds.), {BPM} 2020, Vol. 12168 of LNCS, Springer, 2020, pp.
  273--290.

\bibitem{CAISE2020}
A.~Palm, A.~Metzger, K.~Pohl, Online reinforcement learning for self-adaptive
  information systems, in: S.~Dustdar, E.~Yu, C.~Salinesi, D.~Rieu, V.~Pant
  (Eds.), {CAiSE} 2020, Vol. 12127 of LNCS, Springer, 2020, pp. 169--184.

\bibitem{GuptaGBD20}
A.~Gupta, H.~P. Gupta, B.~Biswas, T.~Dutta, Approaches and applications of
  early classification of time series: {A} review, {IEEE} Trans. Artif. Intell.
  1~(1) (2020) 47--61.

\bibitem{Mori_ea_17-mpl}
U.~Mori, A.~Mendiburu, E.~Keogh, J.~A. Lozano, Reliable early classification of
  time series based on discriminating the classes over time, Data mining and
  knowledge discovery 31~(1) (2017) 233--263.

\bibitem{BonduABCCGHLM22}
A.~Bondu, Y.~Achenchabe, A.~Bifet, F.~Cl{\'{e}}rot, A.~Cornu{\'{e}}jols,
  J.~Gama, G.~H{\'{e}}brail, V.~Lemaire, P.~Marteau, Open challenges for
  machine learning based early decision-making research, {SIGKDD} Explor.
  24~(2) (2022) 12--31.

\bibitem{MartinezRPR20}
C.~Martinez, E.~Ramasso, G.~Perrin, M.~Rombaut, Adaptive early classification
  of temporal sequences using deep reinforcement learning, Knowl. Based Syst.
  190 (2020) 105290.

\bibitem{Dumas21}
M.~Dumas, Constructing digital twins for accurate and reliable what-if business
  process analysis, in: I.~Beerepoot, C.~D. Ciccio, A.~Marrella, H.~A. Reijers,
  S.~Rinderle{-}Ma, B.~Weber (Eds.), Proceedings of the International Workshop
  on {BPM} Problems to Solve Before We Die {(PROBLEMS} 2021) co-located with
  the 19th International Conference on Business Process Management {(BPM}
  2021), Rome, Italy, September 6-10, 2021, Vol. 2938 of {CEUR} Workshop
  Proceedings, 2021, pp. 23--27.

\bibitem{PathakAED17}
D.~Pathak, P.~Agrawal, A.~A. Efros, T.~Darrell, Curiosity-driven exploration by
  self-supervised prediction, in: D.~Precup, Y.~W. Teh (Eds.), Proceedings of
  the 34th International Conference on Machine Learning, {ICML} 2017, Sydney,
  NSW, Australia, 6-11 August 2017, Vol.~70 of Proceedings of Machine Learning
  Research, {PMLR}, 2017, pp. 2778--2787.

\bibitem{ABPM}
M.~Dumas, F.~Fournier, L.~Limonad, A.~Marrella, M.~Montali, J.~Rehse,
  R.~Accorsi, D.~Calvanese, G.~D. Giacomo, D.~Fahland, A.~Gal, M.~L. Rosa,
  H.~V{\"{o}}lzer, I.~Weber, Augmented business process management systems: {A}
  research manifesto, CoRR abs/2201.12855 (2022).

\bibitem{Aalst12}
W.~M.~P. van~der Aalst, Process mining, Commun. {ACM} 55~(8) (2012) 76--83.

\bibitem{Francescomarino18}
C.~D. Francescomarino, C.~Ghidini, F.~M. Maggi, F.~Milani, Predictive process
  monitoring methods: Which one suits me best?, in: M.~Weske, M.~Montali,
  I.~Weber, J.~vom Brocke (Eds.), {BPM} 2018, Vol. 11080 of LNCS, Springer,
  2018, pp. 462--479.

\bibitem{TaxTZ20}
N.~Tax, I.~Teinemaa, S.~J. van Zelst, An interdisciplinary comparison of
  sequence modeling methods for next-element prediction, Softw. Syst. Model.
  19~(6) (2020).

\bibitem{VerenichDRMT19}
I.~Verenich, M.~Dumas, M.~L. Rosa, F.~M. Maggi, I.~Teinemaa, Survey and
  cross-benchmark comparison of remaining time prediction methods in business
  process monitoring, {ACM} Trans. Intell. Syst. Technol. 10~(4) (2019).

\bibitem{SalfnerLM10}
F.~Salfner, M.~Lenk, M.~Malek, A survey of online failure prediction methods,
  ACM Comput. Surv. 42~(3) (2010) 10:1--10:42.

\bibitem{AschoffZ11}
R.~Aschoff, A.~Zisman, {QoS}-driven proactive adaptation of service
  composition, in: G.~Kappel, Z.~Maamar, H.~R.~M. Nezhad (Eds.), ICSOC 2011,
  Paphos, Cyprus, Vol. 7084 of LNCS, Springer, 2011, pp. 421--435.

\bibitem{ASAS_Book}
A.~Metzger, O.~Sammodi, K.~Pohl, Accurate proactive adaptation of
  service-oriented systems, in: J.~Camara, R.~{de Lemos}, C.~Ghezzi, A.~Lopes
  (Eds.), Assurances for Self-Adaptive Systems, Springer, 2012, pp. 240--265.

\bibitem{Aalst2011process}
W.~Van Der~Aalst, Process mining: discovery, conformance and enhancement of
  business processes, Vol.~2, Springer, 2011.

\bibitem{FolinoGP15}
F.~Folino, M.~Guarascio, L.~Pontieri, A prediction framework for proactively
  monitoring aggregate process-performance indicators, in: S.~Hall{\'{e}},
  W.~Mayer, A.~K. Ghose, G.~Grossmann (Eds.), 19th {IEEE} International
  Enterprise Distributed Object Computing Conference, {EDOC} 2015, Adelaide,
  Australia, September 21-25, 2015, {IEEE} Computer Society, 2015, pp.
  128--133.

\bibitem{LeontjevaCFDM15}
A.~Leontjeva, R.~Conforti, C.~D. Francescomarino, M.~Dumas, F.~M. Maggi,
  Complex symbolic sequence encodings for predictive monitoring of business
  processes, in: H.~R. Motahari{-}Nezhad, J.~Recker, M.~Weidlich (Eds.), {BPM}
  2015, Vol. 9253 of LNCS, Springer, 2015, pp. 297--313.

\bibitem{achenchabe2021early}
Y.~Achenchabe, A.~Bondu, A.~Cornu{\'e}jols, A.~Dachraoui, Early classification
  of time series: Cost-based optimization criterion and algorithms, Machine
  Learning 110~(6) (2021) 1481--1504.

\bibitem{XingPY12}
Z.~Xing, J.~Pei, P.~S. Yu, Early classification on time series, Knowl. Inf.
  Syst. 31~(1) (2012) 105--127.

\bibitem{TeinemaaDLM18}
I.~Teinemaa, M.~Dumas, A.~Leontjeva, F.~M. Maggi, Temporal stability in
  predictive process monitoring, Data Min. Knowl. Discov. 32~(5) (2018)
  1306--1338.

\bibitem{MoriMDL18-probab}
U.~Mori, A.~Mendiburu, S.~Dasgupta, J.~A. Lozano, Early classification of time
  series by simultaneously optimizing the accuracy and earliness, {IEEE} Trans.
  Neural Networks Learn. Syst. 29~(10) (2018) 4569--4578.

\bibitem{Francescomarino17}
C.~D. Francescomarino, C.~Ghidini, F.~M. Maggi, G.~Petrucci, A.~Yeshchenko, An
  eye into the future: Leveraging a-priori knowledge in predictive business
  process monitoring, in: J.~Carmona, G.~Engels, A.~Kumar (Eds.), {BPM} 2017,
  Barcelona, Spain, September 10-15, 2017, Vol. 10445 of LNCS, Springer, 2017,
  pp. 252--268.

\bibitem{TeinemaaDMF16}
I.~Teinemaa, M.~Dumas, F.~M. Maggi, C.~D. Francescomarino, Predictive business
  process monitoring with structured and unstructured data, in: M.~L. Rosa,
  P.~Loos, O.~Pastor (Eds.), {BPM} 2016, Vol. 9850 of LNCS, Springer, 2016, pp.
  401--417.

\bibitem{HatamiC13}
N.~Hatami, C.~Chira, Classifiers with a reject option for early time-series
  classification, in: Proceedings of the {IEEE} Symposium on Computational
  Intelligence and Ensemble Learning, {CIEL} 2013, {IEEE} Symposium Series on
  Computational Intelligence (SSCI), 16-19 April 2013, Singapore, {IEEE}, 2013,
  pp. 9--16.

\bibitem{MaisenbacherW17}
M.~Maisenbacher, M.~Weidlich, Handling concept drift in predictive process
  monitoring, in: X.~F. Liu, U.~Bellur (Eds.), 2017 {IEEE} International
  Conference on Services Computing, {SCC} 2017, Honolulu, HI, USA, June 25-30,
  2017, {IEEE} Computer Society, 2017, pp. 1--8.

\bibitem{OstovarLR20}
A.~Ostovar, S.~J.~J. Leemans, M.~L. Rosa, Robust drift characterization from
  event streams of business processes, {ACM} Trans. Knowl. Discov. Data 14~(3)
  (2020) 30:1--30:57.

\bibitem{sutton2018reinforcement}
R.~S. Sutton, A.~G. Barto, Reinforcement learning: An introduction, 2nd
  edition, MIT press, 2018.

\bibitem{NachumNXS17}
O.~Nachum, M.~Norouzi, K.~Xu, D.~Schuurmans, Bridging the gap between value and
  policy based reinforcement learning, in: Advances in Neural Information
  Processing Systems 12 (NIPS 2017), 2017, pp. 2772--2782.

\bibitem{SuttonMSM99}
R.~S. Sutton, D.~A. McAllester, S.~P. Singh, Y.~Mansour, Policy gradient
  methods for reinforcement learning with function approximation, in: Advances
  in Neural Information Processing Systems 12 ({NIPS} 1999), 2000, pp.
  1057--1063.

\bibitem{Dewey14}
D.~Dewey, Reinforcement learning and the reward engineering principle, in: 2014
  {AAAI} Spring Symposia, Stanford University, Palo Alto, California, USA,
  March 24-26, 2014, {AAAI} Press, 2014, pp. 13--16.

\bibitem{SatyalWPCM19}
S.~Satyal, I.~Weber, H.~Paik, C.~D. Ciccio, J.~Mendling, Business process
  improvement with the {AB-BPM} methodology, Inf. Syst. 84 (2019) 283--298.

\bibitem{ICSOC2017}
A.~Metzger, P.~Bohn, Risk-based proactive process adaptation, in: E.~M.
  Maximilien, A.~Vallecillo, J.~Wang, M.~Oriol (Eds.), {ICSOC 2017}, Vol. 10601
  of LNCS, Springer, 2017, pp. 351--366.

\bibitem{BosnicK08}
Z.~Bosnic, I.~Kononenko, Comparison of approaches for estimating reliability of
  individual regression predictions, Data Knowl. Eng. 67~(3) (2008) 504--516.

\bibitem{PapadopoulosVG07}
H.~Papadopoulos, V.~Vovk, A.~Gammerman, Conformal prediction with neural
  networks, in: 19th {IEEE} International Conference on Tools with Artificial
  Intelligence {(ICTAI} 2007), October 29-31, 2007, Patras, Greece, Volume 2,
  {IEEE} Computer Society, 2007, pp. 388--395.

\bibitem{CarneyCB99}
J.~Carney, P.~Cunningham, U.~Bhagwan, Confidence and prediction intervals for
  neural network ensembles, in: International Joint Conference Neural Networks,
  {IJCNN} 1999, Washington, DC, USA, July 10-16, 1999, {IEEE}, 1999, pp.
  1215--1218.

\bibitem{Kotsiantis07}
S.~B. Kotsiantis, Supervised machine learning: {A} review of classification
  techniques, Informatica (Slovenia) 31~(3) (2007) 249--268.

\bibitem{ParkS20}
G.~Park, M.~Song, Predicting performances in business processes using deep
  neural networks, Decis. Support Syst. 129 (2020).

\bibitem{MetzgerN18}
A.~Metzger, A.~Neubauer, Considering non-sequential control flows for process
  prediction with recurrent neural networks, in: 44th Euromicro Conference on
  Software Engineering and Advanced Applications ({SEAA} 2018), Prague, Czech
  Republic, August 29 – 31, 2018, {IEEE} Computer Society, 2018, pp.
  268--272.

\bibitem{Breiman01}
L.~Breiman, Random forests, Mach. Learn. 45~(1) (2001) 5--32.

\bibitem{Dietterich2000}
T.~G. Dietterich, Ensemble methods in machine learning, in: J.~Kittler, F.~Roli
  (Eds.), Multiple Classifier Systems, First International Workshop, {MCS}
  2000, Cagliari, Italy, June 21-23, 2000, Proceedings, Vol. 1857 of LNCS,
  Springer, 2000, pp. 1--15.

\bibitem{HochreiterS96}
S.~Hochreiter, J.~Schmidhuber, {LSTM} can solve hard long time lag problems,
  in: M.~Mozer, M.~I. Jordan, T.~Petsche (Eds.), Advances in Neural Information
  Processing Systems 9, NIPS, Denver, CO, USA, December 2-5, 1996, {MIT} Press,
  1996, pp. 473--479.

\bibitem{Goodfellow-et-al-2016}
I.~Goodfellow, Y.~Bengio, A.~Courville, Deep Learning, MIT Press, 2016.

\bibitem{zhou2012ensemble}
Z.-H. Zhou, Ensemble methods: foundations and algorithms, Chapman and Hall/CRC,
  2012.

\bibitem{TamaCK20}
B.~A. Tama, M.~Comuzzi, J.~Ko, An empirical investigation of different
  classifiers, encoding, and ensemble schemes for next event prediction using
  business process event logs, {ACM} Trans. Intell. Syst. Technol. 11~(6)
  (2020) 68:1--68:34.

\bibitem{TaxVRD17}
N.~Tax, I.~Verenich, M.~L. Rosa, M.~Dumas, Predictive business process
  monitoring with {LSTM} neural networks, in: E.~Dubois, K.~Pohl (Eds.), CAiSE
  2017, Essen, Germany, June 12-16, 2017, Vol. 10253 of LNCS, Springer, 2017,
  pp. 477--492.

\bibitem{NavarinVPS17}
N.~Navarin, B.~Vincenzi, M.~Polato, A.~Sperduti, {LSTM} networks for data-aware
  remaining time prediction of business process instances, in: Symposium Series
  on Comp. Intelligence, Honolulu, USA, Nov 27-Dec 1, 2017, {IEEE}, 2017, pp.
  1--7.

\bibitem{schulman2017proximal}
J.~Schulman, F.~Wolski, P.~Dhariwal, A.~Radford, O.~Klimov, Proximal policy
  optimization algorithms, CoRR abs/1707.06347 (2017).

\bibitem{ShoushD22}
M.~Shoush, M.~Dumas, When to intervene? prescriptive process monitoring under
  uncertainty and resource constraints, in: {BPM 2022}, {Springer LNCS}, 2022.

\bibitem{boughorbel2017optimal}
S.~Boughorbel, F.~Jarray, M.~El-Anbari, Optimal classifier for imbalanced data
  using matthews correlation coefficient metric, PloS one 12~(6) (2017)
  e0177678.

\bibitem{MaaradjiDRO15}
A.~Maaradji, M.~Dumas, M.~L. Rosa, A.~Ostovar, Fast and accurate business
  process drift detection, in: H.~R. Motahari{-}Nezhad, J.~Recker, M.~Weidlich
  (Eds.), {BPM} 2015, Innsbruck, Austria, Vol. 9253 of LNCS, Springer, 2015,
  pp. 406--422.

\bibitem{LiuHC18}
N.~Liu, J.~Huang, L.~Cui, A framework for online process concept drift
  detection from event streams, in: 2018 Int'l Conference on Services
  Computing, {SCC} 2018, San Francisco, CA, USA, {IEEE}, 2018, pp. 105--112.

\bibitem{LeitnerHD13}
P.~Leitner, W.~Hummer, S.~Dustdar, Cost-based optimization of service
  compositions, {IEEE} Trans. Serv. Comput. 6~(2) (2013) 239--251.

\bibitem{WangKSLTMBKB17}
J.~Wang, Z.~Kurth{-}Nelson, H.~Soyer, J.~Z. Leibo, D.~Tirumala, R.~Munos,
  C.~Blundell, D.~Kumaran, M.~M. Botvinick, Learning to reinforcement learn,
  in: G.~Gunzelmann, A.~Howes, T.~Tenbrink, E.~J. Davelaar (Eds.), 39th Annual
  Meeting of the Cognitive Science Society, CogSci 2017, London, UK, 16-29 July
  2017, 2017.

\bibitem{Miller19}
T.~Miller, Explanation in artificial intelligence: Insights from the social
  sciences, Artif. Intell. 267 (2019) 1--38.

\bibitem{HuangMP21}
T.~Huang, A.~Metzger, K.~Pohl, Counterfactual explanations for predictive
  business process monitoring, in: M.~Themistocleous, M.~Papadaki (Eds.),
  Information Systems - 18th European, Mediterranean, and Middle Eastern
  Conference, {EMCIS} 2021, Vol. 437 of LNBIP, Springer, 2021, pp. 399--413.

\bibitem{FeitMP22}
F.~Feit, A.~Metzger, K.~Pohl, Explaining online reinforcement learning
  decisions of self-adaptive systems, in: E.~{Di Nitto}, I.~Gerostathopoulos
  (Eds.), Intl Conference on Autonomic Computing and Self-Organizing Systems,
  {ACSOS} 2022, {IEEE}, 2022.

\bibitem{Maggi_Dumas14}
F.~M. Maggi, C.~D. Francescomarino, M.~Dumas, C.~Ghidini, Predictive monitoring
  of business processes, in: M.~Jarke, {et al.} (Eds.), {CAiSE} 2014, Vol. 8484
  of LNCS, Springer, 2014, pp. 457--472.

\bibitem{FrancescomarinoDMT19}
C.~D. Francescomarino, M.~Dumas, F.~M. Maggi, I.~Teinemaa, Clustering-based
  predictive process monitoring, {IEEE} Trans. Serv. Comput. 12~(6) (2019)
  896--909.

\bibitem{DachraouiBC15}
A.~Dachraoui, A.~Bondu, A.~Cornu{\'{e}}jols, Early classification of time
  series as a non myopic sequential decision making problem, in: A.~Appice,
  P.~P. Rodrigues, V.~S. Costa, C.~Soares, J.~Gama, A.~Jorge (Eds.), Machine
  Learning and Knowledge Discovery in Databases - European Conference, {ECML}
  {PKDD} 2015, Porto, Portugal, September 7-11, 2015, Proceedings, Part {I},
  Vol. 9284 of Lecture Notes in Computer Science, Springer, 2015, pp. 433--447.

\bibitem{HuangALD11}
Z.~Huang, W.~M.~P. van~der Aalst, X.~Lu, H.~Duan, Reinforcement learning based
  resource allocation in business process management, Data Knowl. Eng. 70~(1)
  (2011) 127--145.

\bibitem{Silvander19}
J.~Silvander, Business process optimization with reinforcement learning, in:
  9th Intl. Symposium on Business Modeling and Software Design {BMSD} 2019,
  Vol. 356 of LNBIP, Springer, 2019, pp. 203--212.

\bibitem{BranchiFG22}
S.~Branchi, C.~D. Francescomarino, C.~Ghidini, D.~Massimo, F.~Ricci,
  M.~Ronzani, Learning to act: a reinforcement learning approach to recommend
  the best next activities, in: Business Process Management Forum - {BPM} Forum
  2022, Muenster, Germany, 2022.

\end{thebibliography}

\end{document}